\documentclass[11 pt]{article}
\pdfoutput=1 
\usepackage{jheppub} 
\usepackage{amsmath}
\usepackage{caption}
\usepackage{subcaption,longtable,stmaryrd}
\usepackage{slashed}
\usepackage{bigints}
\usepackage{cancel}
\usepackage{multicol}
\usepackage{blindtext}
\usepackage{tikz}
\usepackage{enumitem}
\usepackage[customcolors,shade]{hf-tikz}
\usepackage{graphicx}
\usepackage[bitstream-charter]{mathdesign}
\usepackage{cancel}
\DeclareSymbolFont{usualmathcal}{OMS}{cmsy}{m}{n}
\DeclareMathAlphabet\mathbfcal{OMS}{cmsy}{b}{n}
\DeclareSymbolFontAlphabet{\mathcal}{usualmathcal}
\DeclareSymbolFont{rmlargesymbols}{OMX}{mdbch}{m}{n}
\DeclareMathSymbol{\rmintop}{\mathop}{rmlargesymbols}{82}
\DeclareMathSymbol{\rmointop}{\mathop}{rmlargesymbols}{72}

\definecolor{mygray}{gray}{0.5}
\title{{\bf  3D $\mathcal{N}=1$ supergravity from Virasoro TQFT: 
 Gravitational partition function and Out-of-time-order correlator }}

\author[a]{Arpan Bhattacharyya,}
\author[a]{Saptaswa Ghosh,}
\author[b]{Poulami Nandi,}
\author[a]{Sounak Pal }
\affiliation[a]{\it Indian Institute of Technology, Gandhinagar, Gujarat-382055, India}
\affiliation[b]{David Rittenhouse Laboratory, University of Pennsylvania, 209 South 33rd Street, Philadelphia, PA
19104, USA}
\emailAdd{abhattacharyya@iitgn.ac.in}
\emailAdd{saptaswaghosh@iitgn.ac.in}
\emailAdd{pnandi@sas.upenn.edu}
\emailAdd{palsounak@iitgn.ac.in}
\abstract{ In this paper, we compute the partition functions of $\mathcal{N}=1$ SUGRA for different boundary topologies, i.e. punctured sphere and torus, using super-Virasoro TQFT. We use fusion and modular kernels of the super-Liouville theory to compute the necklace-channel conformal block and showcase formalism by proving that the inner product holds for superconformal blocks, defined as states in the Hilbert space. Finally, we compute the out-of-time-order correlator for the torus topology with superconformal primary insertions as matter using the tools of super-Virasoro TQFT.}
\makeatletter
\begin{document}
\maketitle
\flushbottom
\section{Introduction}
Quantum gravity has been a topic of interest for many years. However, the construction of tractable problems in quantum gravity is non-trivially challenging, with much development occurring in the lower dimensional cases \cite{Jackiw:1984je,Teitelboim:1983ux,Saad:2019lba}. Also, several insights came from the AdS/CFT correspondence \cite{Maldacena:1997re, Witten:1998qj}. In the last few years, many studies on computing the partition function for two-dimensional gravity and its deformation have been made \cite{Saad:2019lba,Turiaci:2023jfa,Iliesiu:2019lfc,Iliesiu:2020qvm,Iliesiu:2020zld,Collier:2023cyw,Ebert:2022gyn,Bhattacharyya:2023gvg}.
However, the portrait of a richer three-dimensional gravity is still not very clear. In three spacetime dimensions, pure Einstein gravity is non-dynamical, implying the non-existence of gravitational waves. This simplifies the gravity theory dramatically and provides a hint that gravity theory in three dimensions can be formulated in terms of topological quantum field theory (TQFT) \cite{Atiyah:1989vu,Mikhaylov:2017ngi},  where the edge modes are in the form of the graviton excitations in the asymptotic boundaries. However, full quantum treatment of three-dimensional gravity did not achieve much progress, unlike two-dimensional Jackiw-Titelboim gravity \cite{Saad:2019lba}.
Though the ensemble average description works well in the case of JT gravity, in the case of 3d gravity there are strong constraints of locality in terms of operator product expansion and modular transformations. The solution to the crossing equations in the holographic regime is rarely available to us. There have been multiple efforts of quantizing three-dimensional gravity \cite{Maloney:2007ud,Yin:2007gv, Witten:2007kt, Manschot:2007zb} \textcolor{black}{emerging} since the late 2000s based on AdS/CFT conjecture \cite{Maldacena:1997re}. Despite significant progress, the status of dual boundary CFT is not yet fully settled. Taking intuition from two-dimensional gravity \cite{Saad:2019lba}, it is expected that the boundary CFT is not just a family of 2D CFTs admitting a large-c limit but an ensemble of chaotic large-c CFTs. In spite of partial progress on this line \cite{Maloney:2020nni,Collier:2021rsn,Afkhami-Jeddi:2020ezh,Benjamin:2021wzr,Chandra:2022bqq,Belin:2020hea,Mertens:2022ujr,Eberhardt:2019ywk,Eberhardt:2018ouy}, there is no known unitary large-c chaotic CFT with a sparse light spectrum and Virasoro symmetry. Recently, the problem has been nicely tackled in \cite{Collier:2023fwi,Collier:2024mgv} showing the relation of $AdS_3$ quantum gravity with topological field theory, which relies on the bulk description of the problem, unlike the other conventional approaches. \textcolor{black}{This} opens up a new pathway to computing observables directly from the bulk description. These issues are further addressed in \cite{Dymarsky:2024frx,deBoer:2024kat,Jafferis:2024jkb}\footnote{In \cite{Chen:2024unp}, a simplicial 3D gravity model using BCFT data has been constructed.}.

\par 
In this paper, we aim to address the quantization of $\mathcal{N}=1$ supergravity \cite{Ducrocq:2021vkh}  via geometric quantization using super-Virasoro TQFT. The TQFT description for pure $AdS_3$ gravity has been studied for decades, starting from the matching at the classical level between $PSL(2,\mathbb{R})\times PSL(2,\mathbb{R})$ Chern-Simons theory and Einstein's gravity in $AdS_3$. However, the direct quantization $PSL(2,\mathbb{R})\times PSL(2,\mathbb{R})$ Chern-Simons theory is not the same as quantizing pure $AdS_3$ gravity. The fundamental principles in gravity need the metric to be non-degenerate, while the CS theory does not have such a requirement. 
The gravitational phase decomposes to two copies of Teichmuller space on a spatial slice (say $\Sigma$). Teichmuller space describes the shape of the initial value surface and becomes the universal covering of the moduli space of Riemann surfaces. For a Riemann surface $\Sigma_{g,n}$ of genus $g$ and $n$ punctures, the Teichmuller space $\mathcal{T}_{g,n}$ can also be understood as the space of all metrics on $\Sigma_{g,n}$ modulo small diffeomorphisms and Weyl rescalings \cite{Eberhardt:2023mrq},
\begin{equation}  \mathcal{T}_{g,n}=\frac{\text{Metrics on } \Sigma_{g,n}}{\text{Diff}_0 (\Sigma_{g,n}) \times \text{Weyl}(\Sigma_{g,n}) }.
\end{equation} 
The important fact here is that the Teichmuller space can be quantized on its own, and therefore, it is possible to assign a  Hilbert space $H_{\Sigma}$ having a well-defined inner product.
\textcolor{black}{Quantum Teichmuller theory has been an intriguing topic connecting various branches of mathematics and physics} \cite{Atiyah:1989vu,EllegaardAndersen:2011vps,Mikhaylov:2017ngi,Collier:2023fwi}. 
There are many generalizations possible for the Teichmuller TQFT, especially considering moduli spaces of super Riemann surfaces. Though the super Teichmuller space plays a fundamental role in superstring perturbation theory, geometric quantization is a relatively new direction. The main goal of the paper is to generalize the technologies of geometric quantization \cite{Eberhardt:2023mrq} for the supersymmetric case. Sometimes it is also referred to as  \textit{super-Teichmuller spin TQFT} \cite{Aghaei:2015bqi,Aghaei:2020otq} in the literature. In this paper, we compute the partition functions of different 3-manifolds. Here, the algorithm is based on surgery techniques \cite{Sen:2024nfd} of super-Riemann surfaces \cite{Witten:2012ga}. It turns out that the super-Teichmuller TQFT is connected to the three-dimensional Chern-Simons theory with complexifications of $OSp(1|2)$ and $SO(3)$ gauge groups.
\par
In the process of geometric quantization, we identified the states in the bulk Hilbert space with superconformal blocks. Conformal blocks \cite{Dolan:2000ut} are known to be the solution of the Casimir equation for the corresponding conformal group. For $\mathcal{N}=1$ super-Liouville theory, we systematically establish the inner product of conformal blocks following \cite{Eberhardt:2023mrq}. The inner product turns out to produce a Dirac delta function with the difference of two different Wilson line momenta. The $\mathcal{N}=1$ super-Liouville theory has contributions from fermions leading to the choice of different spin structures in the super-Riemann surface. Depending on the different boundary conditions of the super-torus geometry, we aim to perform different modular sums, which are given in terms of the famous Eisenstein series representation.
Apart from this, the spin structures using Kasteleyn orientations are known for $\mathcal{N}=1$ \cite{Aghaei:2020otq}. The isometry group OSp(1|2) acts on the upper half-plane $\mathbb{H}^{1|1}$ by general M\"obius transformations of the form,
\begin{align}
\begin{split}&x\rightarrow x'=\frac{ax+b+\rho \theta}{cx+d+\delta \theta},~~~~~
\theta\rightarrow \theta'=\frac{\alpha x+\beta + g\theta}{cx+d+\delta \theta}.
\end{split}
\end{align}
\noindent
We can define two types of conformal invariants, even Z and odd $\zeta$ under the general M\"obius transformation. The former is the superconformal generalization of the normal cross ratio in the super upper half-plane for points $P_j=(x_j|\theta_j)$, $j=1,\cdots 4$:
$$Z:=\frac{X_{14}X_{23}}{X_{12}X_{34}},$$
where $X_{ij}:=x_i-x_j-\theta_i\theta_j$. The other odd invariant cross ratio $\zeta$ for the collection of three points $P_j=(x_j|\theta_j)$, $j=1,\cdots 3$:
\begin{align}
    \zeta:=\pm \frac{x_{23}\theta_1+x_{31}\theta_2+x_{12}\theta_3-\frac{1}{2}\theta_1\theta_2\theta_3}{(X_{12}X_{23}X_{31})^{1/2}},
    \label{eqzeta}
\end{align}
where, $x_{ij}:=x_i-x_j$.
\par
\textcolor{black}{The field of quantum chaos \cite{Larkin1969QuasiclassicalMI,Haake:2010fgh}, developed in the early seventies, addresses the quantum manifestation of classically chaotic systems. There are many studies aiming to probe the quantum chaotic nature of quantum many-body  systems  \cite{Bohigas:1983er,Berry,DAlessio:2015qtq,Hosur:2015ylk,Nahum:2017yvy,vonKeyserlingk:2017dyr, Aleiner:2016eni, Garttner:2016mqj,syk,Maldacena:2016hyu,Cotler:2017jue,Chowdhury:2017jzb,khemani:2018sdn,Seshadri:2018yya,Hashimoto:2017oit,Ali:2019zcj,Bhattacharyya:2020art}\footnote{This list is by no means exhaustive. Interested readers are referred to references and citations of these papers.}. The study of quantum chaos in quantum gravity and holography has seen a renewed interest in recent years \cite{Sachdev1992GaplessSG,DiFrancesco:1993cyw,Roberts:2014ifa,Maldacena:2015waa,Maldacena:2016hyu,Kitaev:2017awl,syk,Garcia-Garcia:2016mno,Papadodimas:2015xma,Sarosi:2017ykf,Cotler:2016fpe,Balasubramanian:2016ids,Krishnan:2016bvg,Dyer:2016pou}\footnote{Again this list is by no means exhaustive. Interested readers are referred to this review \cite{Jahnke:2018off} and the thesis \cite{NHJ} for more references.}. The gap between the two apparently distinct fields has been bridged in  \cite{Roberts:2014ifa,Altland:2022xqx}, focusing on the 2D quantum gravity.} Over the years, several diagnostics have been developed to probe quantum chaos \cite{Bhattacharyya:2019txx,Kudler-Flam:2019kxq}. In particular, one computes the out-of-time-order correlator (OTOC) to probe quantum information scrambling and chaotic dynamics \cite{Zhou:2021syv, Swingle:2018ekw,Shenker:2013pqa,Roberts:2014ifa,Maldacena:2015waa,Stanford:2015owe, Mertens:2022irh}\footnote{Interested readers are referred to \cite{Xu:2022vko,Garcia-Mata:2022voo} for more references.}. One of the primary focuses of our study in this paper, is the OTOC of the form,
$$\langle \mathcal{O}_A\mathcal{O}_B\mathcal{O}_A\mathcal{O}_B\rangle_\beta$$ on a torus. Here $\beta$ is the inverse temperature. $\mathcal{O}_A$ and $\mathcal{O}_B$ are approximately local insertions smeared over the thermal circle. The behaviour of such functions should be contrasted with the behaviour of correlation functions of the following type,
$$\langle \mathcal{O}_A(t)\mathcal{O}_A(t)\mathcal{O}_B\mathcal{O}_B\rangle_\beta , ~~~~\langle\mathcal{O}_A\mathcal{O}_B(t)\mathcal{O}_B(t)\mathcal{O}_A\rangle_\beta $$
which tends to a $\mathcal{O}(1)$ value $\langle \mathcal{O}_A\mathcal{O}_A\rangle_\beta\langle \mathcal{O}_B\mathcal{O}_B\rangle_\beta$ as $t$ increases. All of the above four-point configurations can be obtained from the Euclidean four-point function. The difference in behaviour arises due to analytic continuation and multivaluedness of a function. OTOCs have been well studied in the literature, particularly for 2D CFTs \cite{Roberts:2014ifa, Kusuki:2019gjs,Caputa:2017rkm,Caputa:2016tgt,Craps:2021bmz,Das:2022jrr}.
\textcolor{black}{One can extract the quantum Lyapunov exponent from the exponential behaviour of the OTOC, and for certain holographic theories (at finite temperature), it satisfies the Maldacena-Shenker-Stanford (MSS) bound \cite{Maldacena:2015waa}.  Also, by investigating the early-time exponential behaviour of OTOC, one can extract the information about scrambling time \cite{Hosur:2015ylk}. It is worth noting that black holes are fast scramblers, i.e scrambling time scales logarithmically in the number of degrees of freedom of the system, $t_{Sc}\sim \log\, N$ \cite{Sekino:2008he}. Motivated by these, in this paper, to probe the chaotic nature of the bulk gravitational theory,  we compute the leading contribution to the OTOC from the superconformal matter insertions on the super-torus using the formalism of the Virasoro-TQFT. We will restrict ourselves to the large-$c$ regime in $\mathcal{N}=1$ superconformal field theory to connect with the gravitational case.}
\par
\medskip
\noindent
Our paper is organized as follows.
In Sec.~(\ref{sec 2}), we start by reviewing $\mathcal{N}=1$ super-Liouville theory, and subsequently,  we describe the ingredients for computing the partition function for wormholes with different number of boundaries and topologies using super-Teichmuller TQFT. Furthermore, we discuss the process of geometric quantization for super-Liouville theory and the inner product of superconformal blocks. In Sec.~(\ref{sec 3}),  we calculated the (super-) VTQFT partition and the gravitational partition function by taking some specific examples of topologies. Then, in Sec.~(\ref{sec 4}),  we put the calculation of 4-point superconformal blocks on the torus, which is required to compute the OTOC.
Finally, in Sec.~(\ref{sec 5}), to probe the regime of chaos, we initiate a computation of the out-of-time-order correlation function with super-Liouville matter insertions. To get an analytic handle on it, we take some specific but relevant limits. We use braiding and modular transformations to investigate the OTOC in both channels, i.e., the necklace and the OPE channel. In Sec.~(\ref{sec 6}), we conclude by summarising the main findings and giving some future outlooks. Last but not least, we have added a few appendices to give the details of some of the computations.   
\section*{Some Definition:}
\textbullet $\,\,$Modular parameter $\tau$ is related to $q$ as $q=e^{2\pi i \tau}\,.$\\
\textbullet $\,\,$$\,\Theta(\tau)$ are Jacobi theta functions\,. \\ 
\textbullet $\,\,$ $\chi_{NS}$ are super-virasoro characters\,.\\
\textbullet $\,\,$ $\xi(z,\theta)$ are super-elliptic functions\,.\\
\textbullet $\,\,$ $(a)_m$ denotes the Pochhammer symbols, $(a)_m=:\frac{\Gamma(a+m)}{\Gamma(a)}\,.$\\ 
\textbullet $\,\,$ $s\mathcal{T}\equiv$ super-Teichmuller\,.


\section{Geometric quantization and inner product of super conformal blocks}
\label{sec 2}
In this section, we will discuss how to compute the gravitational partition function using the super-Teichmuller spin TQFT, the boundary theory being the $\mathcal{N}=1$ super-Liouville theory. The gravitational partition function is found from the character of the $\textrm{{\bf Osp}}(1|2)$ algebra.
The gravitational partition function can also be given by the square of the super-Liouville partition function as shown in \cite{Eberhardt:2023mrq}. We claim that the inner product of the super-Liouville zero-point conformal block on the torus is,
\hfsetfillcolor{gray!8}
\hfsetbordercolor{white}
\begingroup\small
\begin{align}
\begin{split}\tikzmarkin[disable rounded corners=true]{e22}(0.5,-0.5)(-0.1,0.8) 
\langle \mathcal{F}^{s(0)}_{(1,0)}(\vec{P}_1)|\mathcal{F}^{s(0)}_{(1,0)}(\vec{P}_2)\rangle=\int_{\mathcal{T}} d(\textrm{sWP})~{\textrm{sdet}}(\hat{\mathcal{P}}_1^{\dagger}\hat{\mathcal{P}}_1)^{1/2}~Z_{\text{timelike SL}}\,\mathcal{F}^{s(0)}_{(1,0)}(\vec P_2)\overline{\mathcal{F}^{s(0)}_{(1,0)}}(\vec P_1)\propto{ \delta(\vec{ P_1}-\vec{P_2})}.\tikzmarkend{e22}
\end{split}
\label{eqclaim}
\end{align}
\endgroup
where, $\mathcal{F}^{s(n)}_{(g,n)}(\vec{P})$ is the $n$-point super-conformal block for Riemann surface with $g$-genus and $n$-puncture, $P$ denotes the Liouville momenta , $\textrm{sWP}$ is called the {\bf super-Weil-Peterson volume} and \textcolor{black}{$\textrm{sdet}$ is the super-determinant.}
In this section, we wish to prove our claim. However, before we get into the details of our proof, let us take a detour to briefly review the necessary concepts utilised in this section.

\section*{A brief review of super-Liouville Theory}
Let us review the basics of the $\mathcal{N}=1$ super-Liouville theory. The theory is defined via following Lagrangian density \cite{Poghossian:1996agj},
\begin{align}\mathcal{L}=\frac{1}{2}g_{ab}\partial_a\varphi\partial_b\varphi+\frac{1}{2\pi}(\psi\bar\partial\psi+\bar\psi\partial\bar\psi)+2i\mu b^2\bar\psi\psi e^{b\varphi}+2\pi\mu^2b^2e^{2b\varphi}+\frac{1}{2\pi}Q R \phi\,.\label{4.1 p}\end{align}
The energy-momentum tensor and the supercurrent is given by,
\begin{align}
\begin{split}
    &T=-\frac{1}{2}(\partial\varphi\partial\varphi-Q\partial^2\varphi+\psi\partial\psi)\,,\\&
    G=i(\psi\partial\varphi-Q\partial\psi)\,.\end{split}\end{align}

\smallskip
\noindent

\noindent
The superconformal algebra is given by,
\begin{align}
\begin{split}
&[L_m,L_n]=(m-n)L_{m+n}+\frac{c_{SL}}{12}m(m^2-1)\delta_{m+n}, \\&
[L_m,G_k]=\frac{m-2k}{2}{G_{m+k}}, \\&
\{G_k,G_l\}=2L_{l+k}+\frac{c}{3}(k^2-\frac{1}{4})\delta_{k+l}\,.
\end{split}
\end{align}
Here the central charge is given by $c_{SL}=\frac{3}{2}+3Q^2$ where $Q=b+\frac{1}{b}.$ $k,l$ takes integer values for the Ramond algebra, and half-integer ones for Neveu-Schwarz (NS) algebra. $Q$ is known as the  total {\bf background charge}.\par

\smallskip
\noindent
For the \textcolor{black}{super Liouville theory}, one-point function of  the vertex operator, \textcolor{black}{$V_a=e^{a {\Phi}_{\textrm{SL}}}$  with ${\Phi}_{\textrm{SL}}=\varphi+\theta\psi-\bar\theta\bar\psi$,}   on a torus is given by \cite{Artemev:2022sfi},
$$\langle V_a\rangle_{\tau}=\int_{\gamma} \frac{dP}{4\pi}C^{(L)Q/2+iP}_{a,Q/2+iP}(q\bar{q})^{-1/24+P^2}\times |F_L(\Delta_a^L,\Delta^L_{Q/2+iP},q)|^2$$
where the structure constant is given by,\textcolor{black}{\begin{align}C^{(L)Q/2+iP}_{a,Q/2+iP}=(\pi \mu \gamma (b^2)b^{2-2b^2})^{-a/b}\frac{\Upsilon_{NS}(b)\Upsilon_{NS}(2a)\Upsilon_{NS}(2iP)\Upsilon_{NS}(-2iP)}{\Upsilon^2_{NS}(a)\Upsilon_{NS}(a+2iP)\Upsilon_{NS}(a-2iP)}\end{align}}
and $F_L$ is the conformal block. The  $\Upsilon(x)$ functions are defined as,
\begin{align}
\begin{split}
&\Upsilon_b(x)=\frac{1}{\Gamma_b(x)\Gamma_b(Q-x)},~~~~~~~~S_b(x)=\frac{\Gamma_b(x)}{\Gamma_b(Q-x)}\,,\\ &
\Gamma_{NS}(x)=\Gamma_b(x/2)\,\Gamma_b(\frac{x+Q}{2}),~~~\Gamma_R(x)=\Gamma_b(\frac{x+b}{2})\Gamma_b(\frac{x+b^{-1}}{2})\,,
\\ &
\Upsilon_{NS}(x)=\Upsilon_b(x/2)\,\Upsilon_b(\frac{x+Q}{2})\,, ~~\Upsilon_R(x)=\Upsilon_b(\frac{x+b}{2})\Upsilon_b(\frac{x+b^{-1}}{2}).
\label{8.8p}
\end{split}
\end{align}

\noindent
Using the following definition of the superfields \cite{Poghossian:1996agj},
\begin{align}
\begin{split}
\textcolor{black}{V_a}\equiv\Phi_a&=\phi_a(z,\bar z)+\theta \Lambda_a(z,\bar z)-\bar\theta \bar\Lambda_a(z,\bar z)+\theta \overline \theta F(z,\bar z)\,\sim e^{a \Phi_{SL}}\,,\label{2.25nnm}\end{split}
\end{align}
\textrm{where}
\begin{align}
\begin{split}
\phi_a\sim e^{a\varphi},\hspace{2 cm}
\Lambda_a\sim a e^{a\varphi}\psi.\label{2.25nm}\end{split}\end{align}
and $\Phi_{SL}=\varphi+\theta\psi-\bar\theta \bar\psi+\theta \bar\theta F$.
Now we can write the two-point and three-point functions in the following way,
\begin{align}
\langle \Phi_1(Z_1,\bar Z_1)\Phi_2(Z_2,\bar Z_2)\rangle=\frac{c_{12}}{Z_{12}^{2\Delta}\bar Z_{12}^{2\bar \Delta}}, \,\,\,\,\,\,\, \Delta=\Delta_1=\Delta_2, \,\,\,\bar\Delta=\bar\Delta_1=\bar\Delta_2\,.
\end{align}
Here $c_{12}$ is a constant, and 
\begin{align}
\begin{split}
\langle \Phi(z_3,\theta_3;\bar z_3,\bar \theta_3)\Phi(z_2,\theta_2;\bar z_2,\bar \theta_2)\Phi(z_1,\theta_1;\bar z_1,\bar \theta_1)\rangle &=  Z_{32}^{\beta_1}Z_{31}^{\beta_{2}}Z_{21}^{\beta_{3}}\\&\,\langle\Phi(\infty,0;\infty,0)\Phi(1,\Theta;1,\bar \Theta)\Phi(0,0;0,0)\rangle\label{B.4 m} 
\end{split}
\end{align}
where, $$\beta_i=\Delta_i-\Delta_j-\Delta_k,~ Z_{12}=z_1-z_2-\theta_1\theta_2,~  \Theta=\frac{1}{\sqrt{z_{12}z_{13}z_{23}}}\Big( \theta_1z_{23}+\theta_2z_{31}+\theta_3z_{12}-\frac{1}{2}\theta_1\theta_2\theta_3\Big)\,.$$


We used conformal invariance to write the three-point function (\ref{B.4 m}).

\section*{Geometric quantization in Liouville theory}
Before proceeding to prove (\ref{eqclaim}), we briefly review the geometric quantization technique in the context of the Liouville theory.
\noindent
To carry out the path integral over metric, we parametrize the space of metrics by $g=\textcolor{black}{\exp(\delta v)e^{2\sigma }\hat g}$. Here $\hat g$ is transversal to the orbits of $\textrm{Weyl} (M)$ and $\textrm{Diff}_0 (M)$. The joint action of the reparametrizations and Weyl transformations is given by \cite{DHoker:1988pdl},
\begin{align}
\delta g_{mn}=(2\delta \sigma + \nabla^p\delta v_p)g_{mn}+(\mathcal{P}_1\delta v )_{mn}.
\end{align}
If the diffeomorphisms are generated by a vector field $v$, then 
\begin{align}
(\mathcal P_1 v )_{mn}=\nabla_m v_n + \nabla_nv_m-(\nabla_c v^c)g_{mn}\,.
\end{align}
\noindent
$\mathcal P_1$ generates symmetric and traceless variations. Deformations $\delta g^{\perp}$
that are elements of $\textrm{Ker}\mathcal P_1^{\dagger}$ (satisfying $\mathcal P_1^{\dagger}\delta g^{\perp}=0$) are orthogonal to those generated by conformal transformations and diffeomorphisms.
\begin{align}
\langle\delta g^{\perp},\mathcal{P}_1\vec{v}\rangle=\langle \mathcal{P}_1^{\dagger}\delta g^{\perp},\mathcal{P}_1\vec{v}\rangle=0\,.
\end{align}
Therefore, only those that are not achieved by diffeomorphism and Weyl transformation are inside the ($ \textrm{Range} \mathcal{P}_1)^{\perp}$.
So any deformation can be decomposed into the following,
\begin{align}
\Big\{\delta g_{mn}\Big\}=\Big\{{\bf\delta \sigma} g_{mn}\Big\}\bigoplus \textrm{Range} \,{\bf \mathcal{P}_1}\bigoplus Ker {\bf \mathcal{P}_1^{\dagger}},
\end{align}
where the action of $\mathcal{P}_1^{\dagger}$ on the symmetric tensors are given by,
\begin{align}
(\mathcal{P}_1^{\dagger}\delta g)_m=-2\nabla^n\delta g_{mn},
\end{align}
and we have the following identification 
\begin{align}
(\textrm{Range} \,\mathcal{P}_1)^{\perp}=\textrm{Kernel}\, \mathcal{P}_1^{\dagger}\,.
\end{align}

\noindent
To determine the number of zero modes of the operators, one needs to resort to an index theorem, which gives the difference between the number of zero modes of the operator and its adjoint, expressing it in terms of a topological invariant.
\noindent
The index theorem reduces to the \textit{\bf Riemann-Roch theorem} \cite{BSMF_1958__86__97_0}, which says;
\begin{align}
\dim(\textrm{Ker}\mathcal{P})-\, \dim (\textrm{Ker}\mathcal{P}^{\dagger}) =3\chi(M)\,.
\end{align}
In the absence of the anomalies, the path integrals should reduce over the space of inequivalent metrics under the Weyl and diffeomorphism symmetries \cite{DHoker:1988pdl}.
\noindent
Now, we proceed to write the path integral measure using the formulas for the bosonic string theory,

\begin{align}
\begin{split}
  &\Bigg(\frac{\textrm{det} \mathcal{P}_1^{\dagger}\mathcal{P}_1}{\textrm{det}\langle \phi_j |\phi_k\rangle}\Bigg)^{1/2}=\Bigg(\frac{\textrm{det} \hat{\mathcal P}_1^{\dagger}\hat{\mathcal P}_1}{\textrm{det}\langle \hat\phi_j |\hat\phi_k\rangle}\Bigg)^{1/2}e^{-26 S_L(\sigma)}\,,\\&
\Bigg(\frac{8\pi^2 \textrm{det}'\Delta_{g}}{d^2\xi \sqrt{g}}\Bigg)^{1/2}=\Bigg(\frac{8\pi^2 \textrm{det}'\Delta_{\hat g}}{d^2\xi \sqrt{\hat g}}\Bigg)^{1/2}e^{-S_L(\sigma)} \,.
\end{split}
\end{align}
\noindent
Now to compute the measure due to the change of variable from $g\rightarrow\{\sigma , v,\hat g\},$ we need to compute the Jacobian in the following way,
\begin{align}Dg_{mn}= \textrm{Vol}_g(g\delta \sigma ,P_1\delta v,f_j)\, D\sigma \, Dv \, dm_j\,.
\label{6.2 w}\end{align} 
In equation (\ref{6.2 w}) we have used the fact,
\begin{align}
\delta g(m)=\sum_{j=1}^{3h-3} \delta m_j\hat{f}_j\,.
\end{align}
 Here, the coordinates along $\hat{S}$ are $\delta \sigma \hat g ,\hat{\mathcal{P}}_1 \delta v$ and $f_j$.
\noindent
Using the orthogonality, the volume can be decomposed as \cite{DHoker:1988pdl},
\begin{align}
Dg_{mn}=\text{Vol}_g(g\delta \sigma) \,\text{Vol}_g(\mathcal{P}_1\delta v)\, \text{Vol}_g(f_j\perp \textrm{Proj. Ker} \,\mathcal{P}_1^{\dagger})\,D\sigma Dv dm_j
\end{align}
where one can define, $\text{Vol}_g(\mathcal{P}_1\delta v):={(\textrm{det}(\mathcal {P}_1^{\dagger}P_1))^{1/2}}$ and $\text{Vol}_g(f_j\perp \textrm{Proj}. \textrm{Ker} \mathcal{P}_1^{\dagger}):=\frac{\textrm{det} \langle f_j | \phi_k\rangle_g}{(\textrm{det} \langle \phi_j | \phi_k\rangle)^{1/2}_g}.$\\
Using ultralocality, the first term can be ignored.
Again, $$\langle f_j|\phi_k\rangle_g=\langle \hat{f}_j|\phi_k\rangle_g=\langle \mu_j|\phi_k\rangle_g,$$
\noindent
where, $\mu^{z}_{j \bar z}=\hat{g}^{z\bar z}\hat{f}_{j z\bar z}$ are the \textit{Beltrami differentials} \cite{Baulieu:1987jz} corresponding to the moduli deformations along the slice $\hat{S}$.
Now we proceed to write the inner product of the states in the Teichmuller space using the vierbein formalism $ds^2 = e^{+}\bigotimes e^{-}$.
The zweibein $(e^+,e^-)$ are respresented as $$e^+=e^{\xi(z,\bar z)}|dz+ \mu(z,\bar z) d\bar{z}|\,\,\,\,\,\,, \,\,\,\,\,\,\, e^-=\overline{e^{+}}\,.$$
Now, following \cite{VERLINDE1990652}, we achieve the following definition of the inner product,
\begin{align}
\begin{split}\langle \mathcal{F}^{s(0)}_{(1,0)}(\vec{P}_1)|\mathcal{F}^{s(0)}_{(1,0)}(\vec{P}_2)\rangle=&\int_{\mathcal{T}} d(WP) \; \mathcal{D}\sigma \; \textcolor{black}{\mathcal{D} v}\,(\textrm{det}(\hat{ \mathcal{P}_1}^{\dagger}\hat{ \mathcal{P}_1}))^{1/2}e^{(c_m-26) \, S_L(\sigma)} \overline{\mathcal{F}^{s(0)}_{(1,0)}(\vec{P}_1)}\mathcal{F}^{s(0)}_{(1,0)}(\vec{P}_2)\,,\\&
=\int_{\mathcal{T}}d(WP)Z_{T.L}\textrm{det}(\hat {\mathcal{P}}_1^{\dagger}\hat P_1))^{1/2} \overline{\mathcal{ F}^{s(0)}_{(1,0)}(\vec{P}_1)}\mathcal{F}^{s(0)}_{(1,0)}(\vec{P}_2)\,.
\end{split}
\end{align}
In the first line, we have used $d(WP)=\prod_{j=1}^{6h-6}dm_j \frac{\textrm{det} \langle f_j | \phi_k\rangle _g}{(\textrm{det} \langle \phi_j | \phi_k\rangle)^{1/2}_g}\,.$
\medskip
\noindent

Now we are ready to proceed to compute the inner product of the (super-)conformal blocks in the subsequent section.

\section*{Towards the inner product of super-Liouville torus conformal blocks }
With the necessary machinery at hand, we now proceed to prove our claim \eqref{eqclaim}. Though this seems a straightforward generalization of the Liouville inner product, a few subtleties are involved here,  at $\mathcal{N}=1$ super-Liouville theory. In this case, we need to consider the super-Teichmuller space, which has two types of spin structures, even and odd, according to the triangulations of the super-Riemann surface(SRS).
\noindent
Also, the generalization of the ghosts in the Liouville theory are Super-ghosts\footnote{\textcolor{black}{One also uses the convention of $\rho ,\sigma$ ghosts. The $\rho$ ghost cancels the contribution of two pairs of twisted fermions in $\mathcal{N}=4$ supersymmetric theory and $\sigma$ cancels the contribution of pair of bosonic oscillators \cite{Eberhardt:2018ouy}. }} (b,c,$\beta,\gamma$). The super weil-Peterson volumes are given by \cite{DHoker:1988pdl},
\begin{equation}
   d(sWP)= \frac{\textrm{sdet}\langle \mu_j |\hat \Phi_k\rangle}{\textrm{sdet}\langle \hat \Phi_j |\hat \Phi_k\rangle}\prod_J dm_J\,.
\end{equation}
The odd(C) and even$(B)$ superfields, where  $C=c+\theta\gamma,\, B=\beta+\theta b$ satisfy,\footnote{Note that, the ghost superfields can have different definitions in different gauges (Wess-Zumino supergauge).}

\begin{equation}
 \int D(B'\overline{B'} C\overline{C'})e^{-I_{sgh(C,B')}}=(\textrm{sdet}(P_1^{\dagger}P_1))^{1/2}\,.
\end{equation}
\noindent
Hence, the inner product in the super-Teichmuller can be written as \footnote{
The super-Virasoro character function is given by \cite{Alkalaev:2018qaz},
\begin{align}
\chi_{H.C}(q)=\prod_{n=0}^\infty \frac{1+q^{n+1/2}}{1-q^{n+1}}\,.
\end{align}
For type-A contraction of the superalgebra, the character is trivial and $\chi_{F}(q)=1 $.} ,
\begin{align}
\begin{split}
&\langle \mathcal{F}^{s(0)}_{(1,0)}(\vec{P}_1)|\mathcal{F}^{s(0)}_{(1,0)}(\vec{P}_2)\rangle=\int_{\mathcal{T}}d(sWP)\,\textrm{sdet}(\hat{\mathcal{P}}_1^{\dagger}\hat{\mathcal{P}}_1)^{1/2}Z_{\text{timelike SL}}\,\mathcal{F}^{s(0)}_{(1,0)}(\vec{P}_2)\overline{\mathcal{F}^{s(0)}_{(1,0)}(\vec{P}_1)},
\label{6.6 m}
\end{split}
\end{align}
which proves the claim that we have made in \eqref{eqclaim}.
\medskip

Now, let us calculate the inner product explicitly to show the orthogonality of the superconformal blocks on the super-torus. To proceed further and compute the action on the torus, we define the super-coordinates $Z=(z,\theta)$ along with the super-derivatives as,
\begin{align}
D_{\alpha}=\partial_{\theta}+\theta \partial{z},\hspace{2 cm}{D}_{\bar\alpha}=\partial_{\bar \theta}-\bar \theta\partial{\bar z}. \end{align}
\textcolor{black}{Before going to the details of the inner product, the first task is to compute the torus partition function in super-Liouville theory. We consider the super-Liouville action with an even spin structure. In principle, the partition function can be computed perturbatively in the cosmological constant $\mu$. But we know that the VTQFT formalism works for Liouville theories with central charge $c_{L}\ge 1$ \cite{Collier:2023fwi,Collier:2024mgv}. For simplicity, in our case, we take the large charge limit (and we do not expand in $\mu$), i.e. $c_{SL}\to \infty$ (semiclassical limit) and correspondingly, the semiclassical partition function reduces to $(\textrm{free bosonic})\otimes (\textrm{free fermionic})$ partition function along with some prefactor depending on the cosmological constant $\mu$. It can be shown that in this limit, $\mu$ acts as a regulator for the zero mode integral. An explicit derivation of the claim for the Liouville theory is given in Appendix~(\ref{AppE}). The same extends to super-Liouville straightforwardly. }\par
In the large charge limit, the action reduces to\footnote{A non-propagating top form $\theta\bar\theta  F$ in $\Phi_{SL}$ is ignored as it can be integrated out and contributes only an overall \(\tau\)-independent normalization.} \cite{Poghossian:1996agj},
\begin{align}
\begin{split}
S_{\textrm{sl}}&=\int_{s\mathcal{T}}d^2zd^2\theta D_{\alpha}\Phi_{SL} D_{\bar\alpha}\Phi_{SL}\,\\&
=\int_{s\mathcal{T}}d^2zd^2\theta\,(\bar\psi \psi+\bar \theta \psi \bar\partial \varphi-\bar\theta\theta \psi\bar\partial\psi-\theta\bar{\psi}\partial \varphi-\theta \bar{\theta} \partial\varphi\bar\partial\varphi+\theta \bar{\theta}\bar\psi\partial\bar{\psi})\,,\\&
=\int_{\mathcal{T}} d^2z\, (\partial{\varphi}\bar\partial{\varphi}+\psi\bar{\partial}\psi+\bar{\psi}\partial\bar{\psi})\,.
\end{split}
\end{align}
Now $d^2\theta=d\bar\theta d\theta$;
The super-Liouville partition function on the super-Torus  is given by,
\begin{align}
    \begin{split}
Z_{\textrm{sl}_{\textrm{even}}}&=\int [\mathcal{D}\varphi\mathcal{D}\psi\mathcal{D}\bar\psi]e^{-S_{\textrm{sl}}}\,,\\&={Z_{\textrm{Bosonic}}}\,\,\,\,\times {Z_{\textrm{Fermionic}}}\,,\\&
       =\textrm{det}(\nabla^2)^{-1/2}\times Pf(\partial)Pf(\bar\partial)\,,\\&
       =\frac{1}{\sqrt{\tau_2}\,|\eta(\tau)|^2}\, \Bigg(\Bigg|\frac{\Theta_3(z=0|\tau)}{\eta(\tau)}\Bigg|+\Bigg|\frac{\Theta_4(z=0|\tau)}{\eta(\tau)}\Bigg|+\Bigg|\frac{\Theta_2(z=0|\tau)}{\eta(\tau)}\Bigg|\Bigg)\,.
    \end{split}
\end{align}
The $Z_{\textrm{sl}_{\textrm{even}}}$ turns out to be modular invariant as it is a product of the free bosonic and fermionic modular invariant partition functions on the torus. The Pffafian $(Pf(A))$ is defined as,
\begin{equation}
    Pf(A)=\sqrt{\textrm{det} A} \, .
\end{equation}
\noindent
Therefore the inner product defined in (\ref{6.6 m}) becomes,
\begin{equation}
\langle \mathcal{F}^{s(0)}_{(1,0)}(\vec{P}_1)|\mathcal{F}^{s(0)}_{(1,0)}(\vec{P}_2)\rangle=\int_{\mathcal{T}}d(sWP)\,\,{\textrm{sdet}}_{(1,n=0)}(\hat{\mathcal{P}}_1^{\dagger}\hat{\mathcal{P}}_1)^{1/2}\,\,Z_{\text{timelike sl}}\,\,\mathcal{F}^{s(0)}_{(1,0)}(\vec{P}_2)\overline{\mathcal{F}^{s(0)}_{(1,0)}(\vec{P}_1)} \label{Int}
\end{equation}
\noindent
where, $d(sWP)=\frac{d^4 \tau}{Y},~Y=\textrm{Im}(z)+\frac{\theta\bar\theta}{2}$  and $\textrm{sdet}_{(1,n=0)}(\hat{\mathcal{P}}_1^{\dagger}\hat{\mathcal{P}}_1)^{1/2}=\frac{|\eta(\tau)|^6}{\Theta^2[a,b](0,\tau)}$ \cite{Grosche:1989he} \footnote{We can think that we are working with "Non-critical strings". Then, we can identify $ Z_{{\bf \textrm{super-ghost}}}^{\mathbb{T}^2}=\textrm{sdet}_{(1,n=0)}(\hat{\mathcal{P}}_1^{\dagger}\hat{\mathcal{P}}_1)^{1/2}$ expressions of which follows from \cite{DHoker:1988pdl}. }.

\medskip
\medskip
\noindent
For the odd-spin structure, it turns out to be more subtle. In this case, the odd modular parameter also contributes to the fermionic partition function as shown in \eqref{3.24a}.
The superconformal blocks on the torus are just zero-point super-Virasoro characters. For simplicity, we will discuss the even-spin sector. As the super-Virasoro conformal block on the torus are super-Virasoro characters \cite{Poghossian:1996agj}, these are given by,\begin{equation}\mathcal{F}^{s(0)}_{1,0}=\chi_{\textcolor{black}{\textrm{even}}}=\sqrt{\frac{\Theta[a,b](0,\tau)}{\eta(\tau)}}\frac{q^{P^2}}{\eta(\tau)}\,.\end{equation}

\smallskip
\noindent
Before going further, let us take a moment to remind the reader of the definition of the Theta and Dedekind's eta functions. Few modular properties of theta functions are given in Appendix~(B) of \cite{Dei:2024sct}.  
\begin{align}
\begin{split}
 &\Theta_2(\tau)=2q^{1/8}\prod_{n=1}^\infty(1-q^n)(1+q^n)^2,\\
&\Theta_3(\tau)=\prod_{n=1}^\infty(1-q^n)(1+q^{n-1/2})^2,\\
&\Theta_4(\tau)=\prod_{n=1}^\infty(1-q^n)(1-q^{n-1/2})^2.
\end{split}
\end{align}



\noindent
The above theta functions have simple transformation properties under the action of the modular group. The Dedekind's eta ($\eta$) function is defined to be \cite{Ginsparg:1988ui}{\footnote{A few known $\Theta[a,b](z,\tau) $ are defined as follows.
\begin{align}
\begin{split}
&\Theta[a,b](z,\tau)=\sum_n e^{i\pi[(n+a)^2 \tau+2(n+a)(z+b)]}
\end{split}
\end{align}
\text{along with some additional definitions}\begin{align}
\begin{split}
\Theta_1=\Theta[1/2,1/2],\,\,\,\,\,\,\Theta_2=\Theta[1/2,0],\,\,\,\,\,\,\Theta_3=\Theta[0,0],\,\,\,\,\,\,\Theta_4=\Theta[0,1/2] .\label{2.35r}
\end{split}
\end{align}

\textbullet \textcolor{black}{ \,$NS,\widetilde{NS},R$ sector are the even torus boundary conditions and $\widetilde{R}$ is the odd torus boundary condition.
}} ,
\begin{align}
\eta(\tau)=q^{1/24}\prod_{n=1}^{\infty}\,\,(1-q^n)\,\label{2.39p}.
\end{align}}
\par 
\begin{table}
    \centering
    \begin{tabular}{|c|c|c|}
    \hline
       AA  & $Tr_{NS}q^{L_0-c/24}$ & $\sqrt{\frac{\Theta_3(\tau)}{\eta(\tau)}}$\\
       \hline
       AP  & $Tr_{NS}(-1)^Fq^{L_0-c/24}$ & $\sqrt{\frac{\Theta_4(\tau)}{\eta(\tau)}}$\\
              \hline
       PA  & $Tr_{R}q^{L_0-c/24}$ & $\sqrt{\frac{\Theta_2(\tau)}{\eta(\tau)}}$\\
              \hline
        PP &  $Tr_{R}(-1)^Fq^{L_0-c/24}$&$\sqrt{\frac{\Theta_1(\tau)}{\eta(\tau)}}$ \\
               \hline
    \end{tabular}
    \caption{Table of a few super-characters corresponding to different boundary conditions in the Neveu-Schwarz and Ramond sectors. The "PP"(Periodic-Periodic) structure is called the `odd', and the other three types are called `even' spin structures.}
    \label{tab}
\end{table}
After computing the integral in (\ref{Int}), we get the inner product to be proportional to the delta function. For $P_1, P_2 \geq 0 $, the inner product of conformal blocks turns out to be,
\begin{align}
\langle\mathcal{F}^{s(0)}_{(1,0)}(\vec{P}_1)|\mathcal{F}^{s(0)}_{(1,0)}(\vec{P}_2)\rangle\propto \,\mathcal{N} \,\delta(\vec{P}_1-\vec{P}_2)\end{align}
where, $\mathcal{N}$ is the divergent constant.


\section{\textcolor{black}{ Supergravity partition function from VTQFT} }
\label{sec 3}

We begin this section by interpreting the connection between partition function of the super-Liouville theory and that of the supergravity, thereby extending the proposal of \cite{Collier:2023fwi}, i.e \textcolor{black}{equating the gravity partition function on a manifold with a fixed topology to the Virasoro TQFT partition function,} to the case of $\mathcal{N}=1$ super-Liouville theory with super-Teichmuller geometric quantization. We propose that,

\hfsetfillcolor{gray!8}
\hfsetbordercolor{white}
\begin{align}\tikzmarkin[disable rounded corners=true]{esp}(0.5,-0.7)(-0.9,0.8)Z_{\text{sugra}}({M})=\frac{1}{|\text{Map}(M,\partial M)|}~\sum_{\gamma\in \text{Map}(\partial M)}|Z_{\text{superVir}}(M^{\gamma})|^2\tikzmarkend{esp} \label{3.1 m}\end{align}
which can also be expressed alternatively as 
\footnote{Bulk $|\text{Map}(M,\partial M)|$ can be infinite. However, the bulk mapping class group transformation for the $3d$ manifold gives a boundary mapping class group transformation, mapping $\text{Map}(M,\partial M) \to \text{Map}(\partial M)$. For, $\gamma\in \text{Map}(M,\partial M) \subset \text{Map}(\partial M) $, \textcolor{black}{ we generalize} $Z_{\text{superVir}}(M^\gamma)=Z_{\text{superVir}}(M)$.  },
\begin{align}
\begin{split}
Z_{\text{sugra}}(M)=\frac{1}{|\text{Map}_0(M,\partial M)|}~~\sum_{\gamma\in \frac{\text{Map}(\partial M)}{\text{Map}(M,\partial M)}}|Z_{\text{superVir}}(M^{\gamma})|^2.
\end{split}
\end{align}
 The LHS of \eqref{3.1 m} can be thought as sum over different topologies. The simplest one is the $\Sigma_{g,n}\times \mathbb{I}\,.$  We take the the subclass, $\Sigma_{\{0,1\},n}\times \mathbb{I}$.
Also in \eqref{3.1 m}, the $\text{Map}(M,\partial M)$ and $\text{Map}(\partial M)$ are the bulk and boundary mapping class group respectively \cite{Collier:2023fwi}. 
 \textcolor{black}{The large diffeomorphisms in a gravity theory are not gauge transformations from the TQFT side. Hence, we coset the set of all diffeomorphisms by small diffeomorphisms, denoted by $\textrm{Diff}_0(M,\partial M)$ \cite{Eberhardt:2022wlc}.} These are described by the mapping class groups 
\begin{equation}
    \text{Map}(M,\partial{M})\equiv \frac{\textrm{Diff}(M,\partial M)}{\textrm{Diff}_0(M,\partial M)},~~~~~ \text{Map}(\partial{M})\equiv \frac{\text{Diff}(\partial M)}{\textrm{Diff}_0(\partial M)}.
\end{equation}
The mapping class group prevents the theory of quantum gravity from following basic axioms in QFT, e.g., factorization in amplitudes. Also, the part of the bulk mapping class group that acts trivially on the boundary is given by,
\begin{equation}\text{Map}_0(M,\partial M)=\frac{\text{Map}(M,\partial M)}{\textrm{Ker}(\text{\text{Map}}(M,\partial M)\rightarrow \text{Map}(\partial M))}.\end{equation}
\subsection*{For two boundary wormholes with genus zero}
We begin to test our claim \eqref{3.1 m} with a warm-up exercise considering the case of two boundary wormholes before moving on to the more complicated examples. We calculate the partition function of the simplest topology i.e a wormhole ($g=0$) with two boundary three-punctured spheres $(n=3)\,.$ Assuming the spheres are at the boundary and Wilson lines are non-spinning, we get from the inner product of conformal blocks,
\begin{align}
\begin{split}
{Z}_{\text{superVir}}^{\text{Wormhole}}\hspace{- 0.03 cm}\bigg(\begin{minipage}[h]{0.17\linewidth}
	\vspace{4pt}
	\scalebox{1.9}{\includegraphics[width=\linewidth]{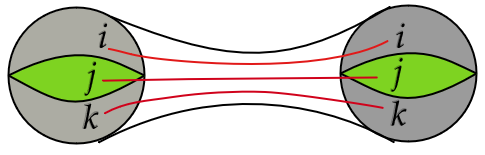}}
   \end{minipage} \hspace{2.3cm}\bigg)=\mathcal{N}\langle \mathcal{F}_{0,3}|\mathcal{F}_{0,3}\rangle=\frac{\mathcal{N}}{C_{NS}(i\hat P_i,i\hat P_j,i\hat P_k)} \, .\label{eq1}
  \end{split}
  \end{align}
  Hence\footnote{with,
     ${\mathcal{N}}=C_{NS}(P_i,P_j,P_k)^2\,.$}, 
  \begin{equation}
      Z_{\textcolor{black}{\textrm{sugra}}}=|Z_{\textrm{superVir}}|^2=\textcolor{black}{{|C_{NS}(i\hat P_i,i\hat P_j,i\hat P_k)|^2}} \, .
  \end{equation}
 As argued in \cite{Collier:2023fwi}, for the general case (i.e for arbitrary $g$ and $n$) we define,
\begin{align}
\begin{split}
\langle\mathcal{F}^{s\,\,\mathcal{C}}_{g,n}(\vec{P}_1)|\mathcal{F}^{s\,\,\mathcal{C}}_{g,n}(\vec{P}_2)\rangle=\frac{\delta^{3g-3+n}(\vec P_1-\vec P_2)}{\rho^{s\,\,\mathcal{C}}_{g,n}
(\vec P_1)} \,\label{3.7u} .
  \end{split}
  \end{align}
  In particular, extending \cite{Collier:2023fwi} for the supersymmetric case, we have \footnote{In Eqs~(\ref{eq1})-(\ref{eq2}) we only consider the NS (leaving the Ramond vertex insertions for future) insertions and restrict it to the bosonic component of the NS supermultiplet and use the even NS conformal-block sector (the $i=j=1$ component c). For the full NS supermultiplet, one must retain different external components of the superprimary operator together with both even and odd blocks (labeled by $i,j$ in (\ref{eq4})). The resulting data are fixed by the two independent NS structure constants ((\ref{eq5}) and (\ref{eq6})), superconformal Ward identities, and picture-changing operators that mix the corresponding component structures. Hence, for a general pants decomposition, $\prod_{\text{pairs of pants}}C_{NS}$ in (\ref{eq2}) should be replaced by a sum over the allowed components and even/odd intermediate-channel labels, with the corresponding super-Virasoro two-point pairings and fermionic signs along each internal NS cuff.  The no-insertion results, including (\ref{eqclaim}) and the unpunctured torus sector, remain unaffected.}
  \textcolor{black}{\begin{align}
  \begin{split}
  \rho^{s\,\,\mathcal{ C}}_{g,n}(\vec{P})=\prod_{\textcolor{black}{\text{cuffs a}}} \Omega_{0\,\,s}( P_a) \prod_{\text{pairs of pants}}C_{NS}( P_i, P_j, P_k),
  \end{split} \label{eq2}
  \end{align}}
 where $\Omega_{0\,\,s}(\alpha_a)$\footnote{$\alpha_a =Q/2+i P_a.$ } is defined to be,

\begin{align}
\Omega_{0\,\,s}(\alpha_a)=\frac{W_{NS}(\alpha_a)}{W_{NS}(Q-\alpha_a)},
\end{align}
where $W_{NS}$ has been defined in Appendix~(\ref{AppB}).\\ 

 
\begin{center}
\begin{figure}
\centering
\scalebox{0.08}{\includegraphics{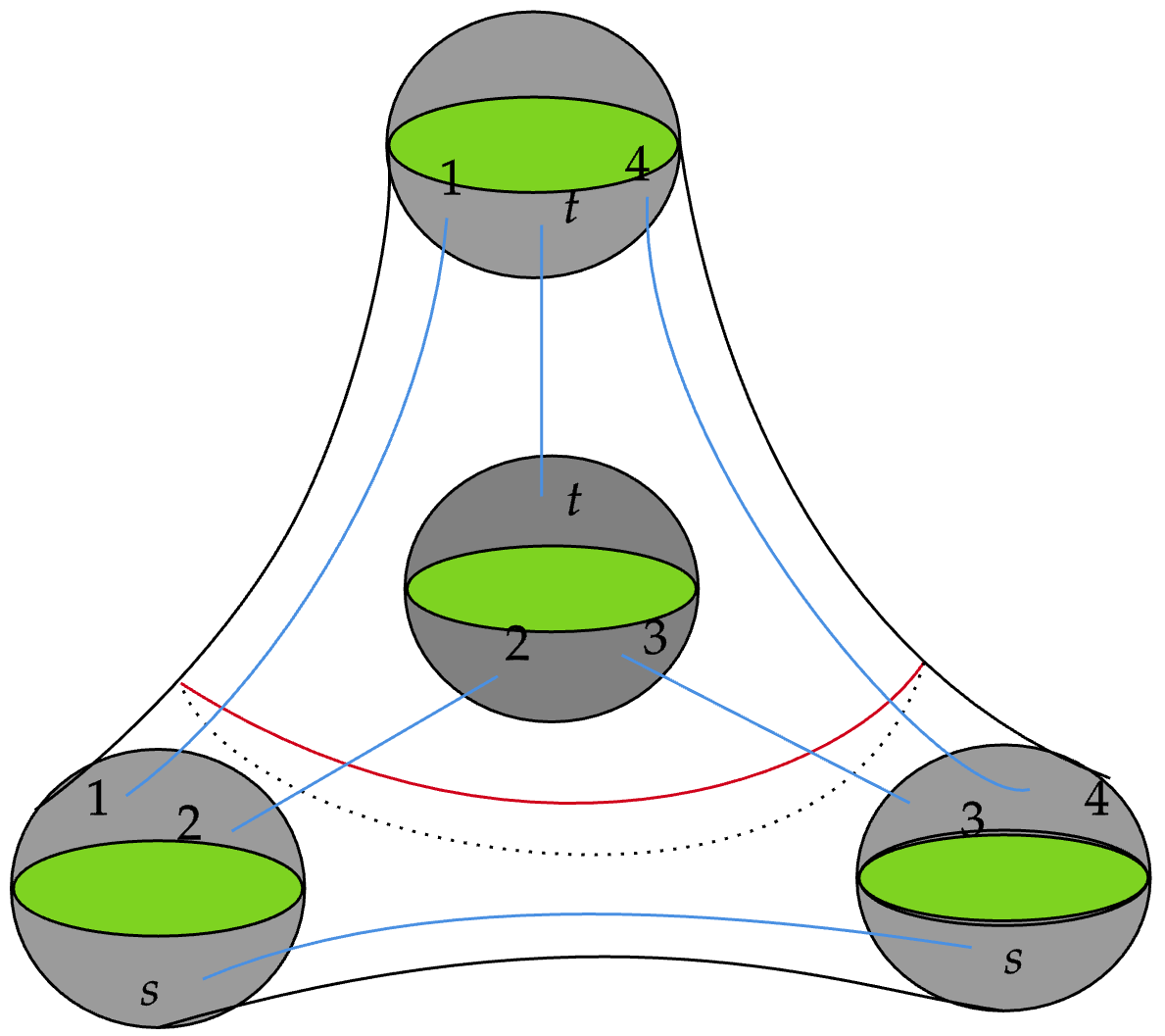}}
\scalebox{0.08}{\includegraphics{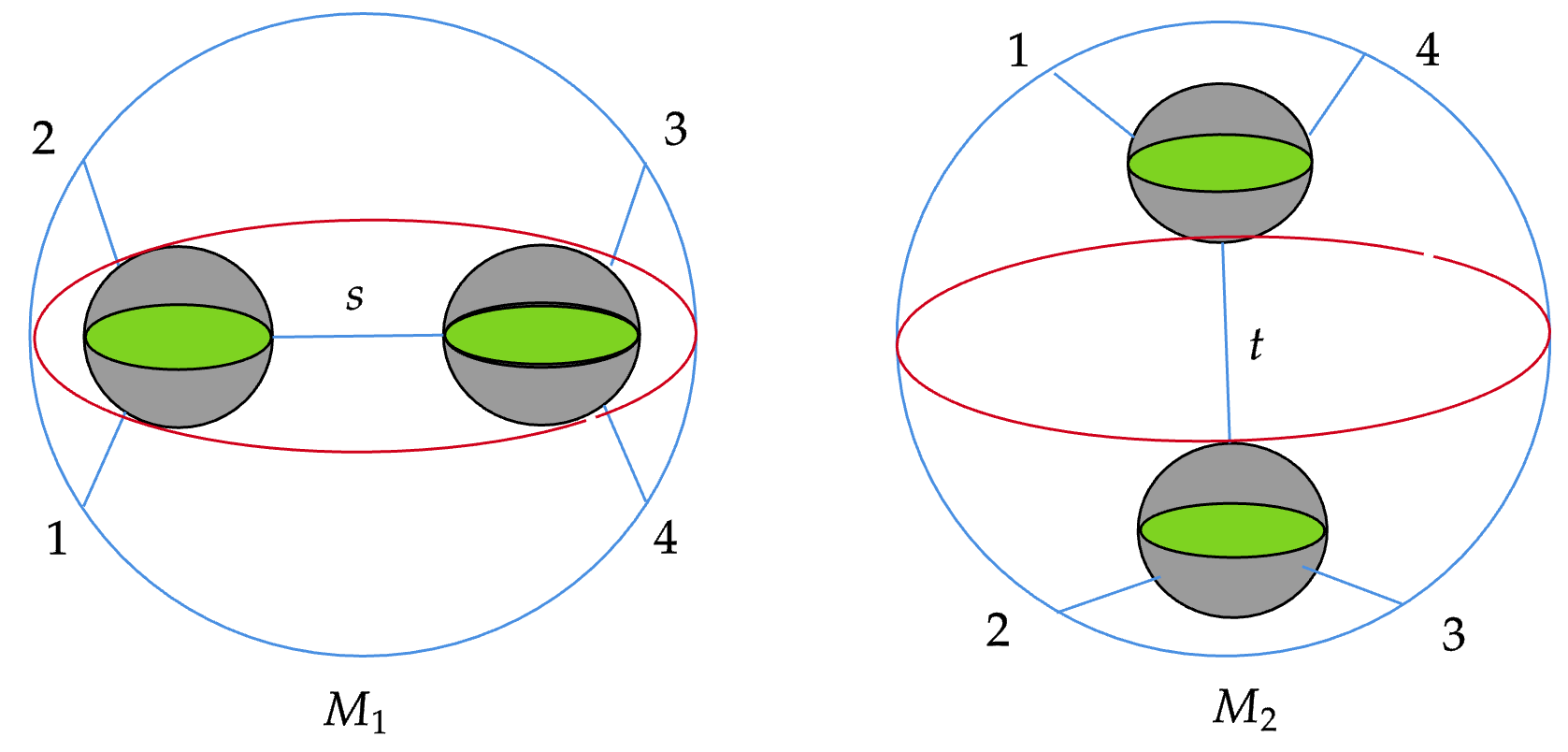}}
\caption{Figure depicting the four three-punctured sphere wormholes in super-Liouville theory with non-spinning Liouville momenta. The figure on the right side shows the Heegaard splitting. }
\label{fig h}
\end{figure}
\end{center}
\vspace{-2cm}
\subsection*{Four boundary genus zero wormhole}
 For a four-boundary wormhole, to find the gravitational partition function after Heegaard splitting into the basis of compression bodies ($M_1 \mbox{and }  M_2$ in Fig.~(\ref{fig h})), we need to use the super-Virasoro fusion kernel to convert the $s$ channel block to $t$ channel block. We use the three-point super-block for the three punctured spheres, which is similar to the DOZZ structure constant in Liouville theory, apart from a little modification \cite{He:2017lrg}.
\noindent
The four-boundary super-Virasoro sphere wormhole partition function can be written as\footnote{with,
     ${\mathcal{N}}=C_{NS}(P_1,P_2,P_s)C_{NS}(P_3,P_4,P_s)C_{NS}(P_t,P_2,P_3)C_{NS}(P_1,P_t,P_4)\,.$},
 \begin{align}
\begin{split}
{Z}_{\text{superVir}}^{\text{wormhole}}\hspace{- 0.03 cm}\left(\begin{minipage}[h]{0.1\linewidth}
	\vspace{4pt}
	\scalebox{1.3}{\includegraphics[width=\linewidth]{wormhole_3.png}}
   \end{minipage} \hspace{0.7cm}\right)=\frac{\,\mathcal{N}\,\,\mathbb{F}_{P_{\{s\}},P_{\{t\}}}\begin{bmatrix}
       P_1& P_4\\
       P_2& P_3
   \end{bmatrix}}{\Omega_{0\,\,s}(P_t)C_{NS}(P_1,P_4,P_t)C_{NS}{(P_2,P_3,P_t)}}\,.
  \end{split}
  \end{align}
  \noindent

  In evaluating the above, we used the Ponsot-Teschner fusion kernel \cite{Ponsot:1999uf}, which expands $t$-channel block $\{ t \}$ in terms of the complete basis of $s$-channel block $\{s\}$. The SLFT fusion kernel is given in Appendix~(\ref{C.1m}). The main advantage of using such a formula is even without knowing the $\{s\}$ or $\{t\}$ channel conformal blocks, we can use the closed form of the fusion kernel to obtain the OPE density, which is one of the important ingredients to calculate/check quantities, e.g. partition function, ETH \cite{Srednicki:1995pt,PhysRevA.43.2046,Rigol2007ThermalizationAI}.
  \subsection*{Gravity partition function of a torus}
  Usually, the partition function for a torus is written in the following way as the sum of the vacuum characters \cite{Benjamin:2019stq}.
\begin{align}
\begin{split}
{Z}_{\text{Vir}}^{\textrm{torus}}\hspace{- 0.03 cm}\left(\begin{minipage}[h]{0.17\linewidth}
	\vspace{4pt}
	\scalebox{0.8}{\includegraphics[width=\linewidth]{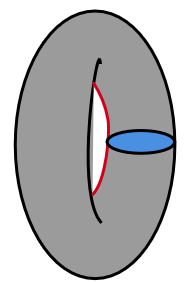}}
   \end{minipage} \hspace{-0.47 cm}\right)=\chi_0(q),\,\,\textrm{with,}\,\, \chi_0(q)=(1-q)\frac{q^{-\frac{c-1}{24}}}{\eta(q)}.
  \end{split}
  \end{align}
\noindent
\textcolor{black}{Hence, the gravity partition function of the torus is given by},

\textcolor{black}{\begin{align}
    \textcolor{black}{Z_{\textrm{gravity}}^{\textrm{torus}}=\sum_{\gamma\in PSL(2,\mathbb{Z})} |\chi_0(\gamma\cdot \tau)|^2}
\end{align}}
\noindent
\textcolor{black}{The sum is given by a divergent {\bf Poincare series} and can be written as,}
\textcolor{black}{
\begin{align}
Z_{\textrm{gravity}}^{\textrm{torus}}(\tau)=\frac{1}{\sqrt{\textrm{Im}(\tau})|\eta(\tau)|^2}\sum_{c,d}\,\Bigg(\sqrt{\textrm{Im}(\tau)} \,|\bar{q} q|^{-c+1/24}|1-q|^2\Bigg)\Bigg|_{\gamma}\,.
\end{align}}
\textcolor{black}{
Now, this {\bf Poincaré series} can be written as the special case of the Eisenstein series,
\begin{align}
E(\tau;n,m)=\sum_{c,d}\Bigg(\sqrt{\textrm{Im}(\tau)}\,q^{-n}\bar{q}^{-m}\Bigg)\Bigg|_\gamma,
\end{align}}
with $n-m$ becoming 0 or $\pm 1$. Using the modular weight of modular forms properly, the final result in terms of the Eisenstein Series can be written as \footnote{One can find the regularized sum for these divergent series in \cite{Maloney:2007ud,Fleig:2015vky}.},
\begingroup
\begin{align}
\begin{split}
\hspace{-0.5 cm}&Z_{\textrm{gravity}}^{\textrm{torus}}(\tau)=\frac{1}{\sqrt{Im\tau}|\eta(\tau)|^2}\Bigg(E(2k-1/12,0)+E(2k+2-1/12,0)\\&\hspace{ 3cm}-E(2k+1-1/12,1)-E(2k+1-1/12,-1)\Bigg)\,.
\end{split}
\end{align}\\
\endgroup
\noindent
 Now the non-degenerate and the degenerate primary characters transform under modular S-transformation as \cite{Benjamin:2019stq}\footnote{The vacuum character is the degenerate primary $\chi_{vac}=\chi_{1,1}$.} , \begin{align}
\chi_P\Big(-\frac{1}{\tau}\Big)=\int_0^\infty dP'S_{P}^{p'} \chi_{p'},\,\,\,\chi_{m,n}\Big(-\frac{1}{\tau}\Big)=\int_0^\infty dp'S_{m,n}^{P'} \chi_{P'}.
\end{align}
\noindent
with $S_{m,n}^{P'}=2\sqrt{2}\sinh(\pi m P' b^{-1})\sinh(\pi n P' b)$ and $S_{P}^{P'}=\sqrt{2}\cos(\pi P P')$.
\textcolor{black}{
Now the Cardy density of states for Liouville CFT has the asymptotic form given by,}
 \begin{align}
  \rho(P,\bar P)=\mathbb{S}_{P\mathbb{I}}[\mathbb{I}] \mathbb{S}_{\bar P\mathbb{I}}[\mathbb{I}] +\sum_i\mathbb{S}_{PP_i}[\mathbb{I}]\mathbb{S}_{\bar P\bar{P}_i}[\mathbb{I}],
 \end{align}
 where $\mathbb{S}_{P\mathbb{I}}[\mathbb{I}]$ is the modular S-matrix.  Asymptotic spectrum of the CFT is quite generally given by\footnote{with $\rho(P)\sim\rho_0(P)\rho_0(\bar P)$ as $P,\bar P\rightarrow \infty\,.$},
 \begin{align}
 \rho_0(P)=\mathbb{S}_{P\mathbb{I}}[\mathbb{I}]=4\sqrt{2}\sinh(2\pi b P )\sinh(2\pi b^{-1}P)\,.
 \end{align}
Considering analytic continuation of the Liouville OPE for $\alpha\slashed{\in}[0,Q]$, we can obtain the following representation of fusion, 
\begin{align}
\mathbfcal{V}_{\alpha_A}\times \mathbfcal{V}_{\alpha_B}:=\sum_{\substack{\alpha_{n,m}<Q/2 \\ n,m \in \mathbb{Z}_{\geq0}}}\mathcal{V}_{\alpha_{n,m}} + \int_{Q/2+0}^{Q/2+i\infty}d\alpha \mathcal{V}_\alpha,
\end{align}
where $\mathbfcal{V}_{\alpha}$ is a primary with associated conformal dimension $\Delta=\alpha(Q-\alpha)\,.$ In calculating OTOC we consider $\alpha>Q/2$, so we only have the contribution from the integral. \par
 \subsection*{SUGRA partition function of an even torus}
By continuing the same argument, the super Virasoro partition function takes the form \eqref{2.35r},
\begin{align}
    \begin{split}
        Z_{\textrm{superVir}}^{\textrm{torus}}=\chi_{\textrm{even}}(\tau)=\sqrt{\frac{\Theta_i(0,\tau)}{\eta(\tau)}}\frac{q^{P^2}}{\eta(\tau)},\,\forall i=2,3,4,
    \end{split}
\end{align}
 Correspondingly, the supergravity partition function is given by,
\begin{align}
    \begin{split}
        Z_{\textrm{sugra}}^{\textrm{torus}}=\sum_{\gamma}|\chi_{\textrm{even}}(\gamma\cdot \tau)|^2.
    \end{split}
\end{align}\\
Having all these ingredients in hand, we proceed to calculate the partition function for different subdominant topologies.

\subsection*{Calculation of partition function Torus Wormhole:}
After calculating the partition function of multiboundary sphere wormholes, we now proceed to calculate the partition function of the two boundary wormholes on the super-torus. The computation of the partition function is similar to the Liouville case, and the modular sum is the same as the usual (non-supersymmetric) Liouville theory. The main reason behind this is that for the even spin structure, the superspace generalisation of the torus does not contain the $\textrm{Im}(\tau)$ factor or any other extra modular parameter to change the modular sum. The modular sum changes substantially when we consider the odd-spin structure on the super-torus.

\smallskip
\noindent
\textbf{ Even spin structure:}
The two boundary torus wormhole partition function is given by,

\begin{align}
\begin{split}
&{Z}_{\text{super-Liouville}}^{\text{torus}}\hspace{- 0.03 cm}
   =\frac{1}{\sqrt{-i(\tau-\bar \tau)}|\eta(\tau)|^2}\times\Bigg(\Bigg|\frac{\Theta_3(z=0|\tau)}{\eta(\tau)}\Bigg|+\Bigg|\frac{\Theta_4(z=0|\tau)}{\eta(\tau)}\Bigg|+\Bigg|\frac{\Theta_2(z=0|\tau)}{\eta(\tau)}\Bigg|\Bigg)\,.\label{3.16 y}
  \end{split}
  \end{align}
\noindent
The $\Theta$ functions appearing in \eqref{3.16 y} are chosen according to the boundary conditions mentioned in Table~\eqref{tab}. 
Now replacing $\tau\rightarrow\tau_1$ and $-\bar\tau\rightarrow \gamma \cdot\tau_2$, where the usual modular transformation $\gamma \cdot\tau=\frac{a\tau+b}{c\tau+d}$, with $ad-bc\neq 0$, the gravity partition function is written as,\\
\begingroup
\begin{align}
\begin{split}
\hspace{-0.5 cm}Z_{\textrm{sugra}}(T^2\times [0,1])&=\left(\begin{minipage}[h]{0.17\linewidth}
	\vspace{4pt}
	\scalebox{1.3}{\includegraphics[width=\linewidth]{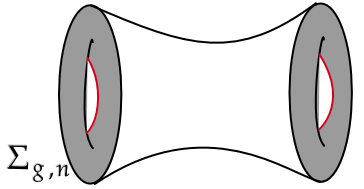}}
   \end{minipage} \hspace{0.7cm}\right)\\&\equiv Z^{\textrm{torus-wormhole}}_{\textrm{superVir}}(\tau_1)Z^{\textrm{torus-wormhole}}_\textrm{{superVir}}(\tau_2)\sum_\gamma\frac{\sqrt{\textrm{Im}(\tau_1)\,\textrm{Im}(\gamma\cdot \tau_2)}}{|\tau_1+\gamma\cdot \tau_2|}\,.
   \end{split}
\end{align}\\
\endgroup
However, it should be noted that there exists an issue with the calculated partition function in $\mathcal{N}=1$ super-Liouville theory with the modular sum, which is similar to \cite{Eberhardt:2023mrq}. 
For the gravitino field, there is at least one cycle around which the gravitino changes sign due to the boundary condition, and in such cases, there are no zero modes.

\textcolor{black}{At this point, we note that while computing the inner product between different conformal blocks of the same spin structure but with different boundary conditions in different cycles of the torus, one should, in principle, introduce the element of spin mapping class group ($U(\gamma)$) sandwiched between the states of the Hilbert space i.e the superconformal blocks, schematically,  $\langle\mathcal{F}(\vec P_1)|U(\gamma))|\mathcal{F}(\vec P_2)\rangle$.  While the \textbf {odd spin structure} cannot be obtained from an \textbf{even} one by applying a S (modular S transform) or T (T-transform) move, the inner product is ambiguous between even and odd spin structures as one can not construct the mapping torus because of the absence of suitable spin mapping class group action, $U(\gamma)$, between them. \footnote{\textcolor{black}{Mapping torus is obtained for a $\Sigma\times S^1$ topology as,
$$M.T(\Sigma\times S^1)=(\Sigma\times S^1)_{[\phi]}:=(\Sigma \times [0,1])/\sim,$$ where the equivalence class$\sim$ is given by,
$(x,0)\sim(\phi(x),1)$ for $x\in \Sigma$ and $\phi\in \textrm{Aut}(\Sigma)$.}} Hence,  at this point, it is quite unclear to us how to define the super-Virasoro partition function for this. We leave this for detailed future investigation. Now, the action of the spin mapping class group changes the boundary condition (within an even/odd spin structure) from one to the other depending on the action of the S or T moves, as shown in \cite{Aghaei:2020otq}. Hence the ``mapping torus''  has a fundamental group, \begin{align}
    \pi_1(M)=\pi_1(\Sigma)\rtimes_{\phi_*} \mathbb{Z},
\end{align}
where $\phi_*:\pi_1(\Sigma)\rightarrow \pi_1(\Sigma) $ is the induced map from $\phi\in \textrm{Aut}(\Sigma)$. Here, M is the 3-manifold with spin-2 manifolds $(\Sigma)$ on the boundary. Now, to compute the gravity partition function, one should, in principle, sum over the spin mapping class group. In general, the representation of the spin mapping class group ($U(\gamma)$) will not be proportional to identity, e.g., for the case of a two-boundary torus ($\mathbb{T}^2$) with the even spin structure at the two boundaries but with different boundary conditions. But in this paper, we have chosen the same spin structure on $\Sigma$  with the same boundary condition; hence, for our case, $U(\gamma)$ is always the identity. We leave a more general study of the partition function of two boundary tori with the same (even) spin structure but different boundary conditions for the future.}
\\\\
\textbf{Odd torus-Wormhole partition function:}
For the odd spin structure, we identify the points,
\begin{align}
\begin{split}
(z,\theta)\sim (z+1,\theta); \hspace{2 cm}(z,\theta)\sim(z+\tau+\theta\delta,\theta+\delta)\,.
\end{split}
\end{align}
\noindent
Clearly, due to this, a function (for odd spin structure) $\xi(z,\theta)$ satisfies  
\begin{align}
\begin{split}
&\xi(z,\theta)=\xi(z+\tau+\theta\delta,\theta+\delta)\,,\\ &
\xi(z,\theta)=\xi(z+1,\theta)\,.\label{4.7a}
\end{split}
\end{align}
Physically, the \textcolor{black}{odd($\delta$)}  and the even ($\tau$) are the modular periods of  $(\theta,z)$, corresponding to zero modes for gravitino and the graviton (these constant modes cannot be gauged). These type of functions in \eqref{4.7a} are also known as { `Superelliptic'}
functions. 
To obtain the partition function\footnote{Generally, the partition function for the odd sector is 0, if one does not consider the coupling of $\psi$ and $\bar\psi$.}, we have the following action as \cite{Myung:1989bg},
\begin{align}
\begin{split}
A'=\frac{g}{2\pi}\int_{\mathcal{T}} d^2\omega&\Bigg[\partial_{\omega}\varphi_1\partial_{\bar{\omega}}\varphi_1+\bar{\psi}_1\partial_{\omega}\bar{\psi}_1-\psi_1\partial_{\bar{\omega}}\psi_1+\frac{\delta\bar{\delta}}{2\tau_I^2}\bar\psi_1\psi_1-F_1^2-\frac{i\delta}{\tau_I}\psi_1\partial_{\omega}\varphi_1-\frac{i\bar\delta}{\tau_I}\bar\psi_1\partial_{\bar\omega}\varphi_1\\&
+g\frac{\delta\bar \delta}{4\pi \tau_I}\bar\psi_0\psi_0-\frac{g\tau_IF_0^2}
{2\pi}\Bigg]\,.\label{3.14 r}
\end{split}
\end{align}
Here the new coordinates $W=(\omega,\hat\theta)$ is related with $Z=(z,\theta)$ in the following way,

\begin{align}
z=\omega+\hat\theta\delta \frac{\omega_I}{\tau_I},\,\,\,\,\,\, \theta=\hat\theta+ \delta \frac{\omega_I}{\tau_I}\,.
\end{align}
$\omega_I$ and $\tau_I$ denotes the imaginary part of $\omega,\tau$. The action (\ref{3.14 r}) comes from the expansion of \begin{align}A'=\frac{g}{2\pi}\int_{s\mathcal{T}}d^4z D_{\theta}JD_{\bar\theta}J\end{align}
with \begin{align}J(W,\bar{W})=\phi(\omega,\bar\omega)+\hat\theta \psi(\omega,\bar\omega)-\bar{\hat\theta}\bar\psi(\omega,\bar\omega)+\bar{\hat\theta}\hat\theta F(\omega,\bar\omega)\,.\end{align}

The auxiliary field F can be removed by using the equation of motion.
The total partition function is given by \cite{Myung:1989bg},
\begin{align}
\begin{split}
Z_{\textrm{sl}_{\textrm{odd}}}={\int \mathcal D\varphi_1 e^{-A_{\varphi_1}}}~ {\int \mathcal D S_0 \,\mathcal D[\psi_1,\bar\psi_1]
e^{-A_r}}=Z_0\cdot Z_1,\label{3.24a}
\end{split}
\end{align}
where,
\begin{align}
    \begin{split}
        &Z_{0}=\Bigg(\frac{g}{2\tau_I}\Bigg)^{1/2}\frac{1}{|\eta(\tau)|^2}\,\,,\quad
        Z_1=\frac{\bar\delta \delta}{\tau_I}|\eta(\tau)|^2,
        \label{3.26m}
    \end{split}
\end{align}
where, $\mathcal D S_0=C dF_0 d\psi_0d\bar{\psi_0}$ and  $A_r=A'-A_{\Phi_1}$.      $Z_0$ comes from the contribution of zero modes of action, and $Z_1$ comes from the path integral over $\Phi_1$.
In an odd super-torus, the modular transformation works as,
\begin{align}
\tau'=\frac{a\tau+b}{c\tau+d},\,\,\,\,\,\, \,\,\,\,\,\, \delta'=\frac{\delta}{(c\tau+d)^{3/2}},
\end{align}
\textcolor{black}{where $\tau,\delta$ belongs to the super-torus} and $\begin{bmatrix}
    a & b \\
    c & d
\end{bmatrix}
\in SL(2,\mathbb{Z})\,.$ 
\textcolor{black}{The modular sum (which gives the gravitational partition function) vanishes as the partition function is trivially zero (as the odd R-R sector super-Virasoro character trivially vanishes for decoupled $\psi$ and $\bar\psi$), though we have a non vanishing super-Liouville partition function $Z_{sl_{odd}}$ given by the product of $Z_0$ and $Z_1$ in (\ref{3.26m}) due to the coupling of $\psi$ and $\bar{\psi}$ via odd modular parameter in zero modes.
Hence, in the odd sector, the supergravity partition function becomes,}
\begin{align}
Z^{\textrm{odd}}_{\textrm{sugra}}(T^2\times [0,1])=\sum_{\substack{\gamma\in \frac{a\tau+b}{c\tau+d}\\ \gamma'\in \frac{\delta}{(c\tau+d)^{3/2}} }}\Bigg|Z_{\textrm{superVir}}^{\textrm{odd}}(\gamma\cdot\tau,\gamma'\cdot \delta)\Bigg|^2.
\end{align}

\textcolor{black}{We will leave performing the modular sum and its extensions for future work}. \textcolor{black}{For this case, the boundary conditions are periodic in both cycles of the torus. Hence, the gravitino field does not change its sign, and therefore, there are zero modes appearing in theory. Due to this, the Laplacian operator does not yield an inverse. This is one of the important subtleties involved in the "PP" boundary condition. So, we perform the calculation where the fermionic mode is at least antiperiodic in one of the cycles of the torus (i.e. for an even spin structure), leaving the complete analysis for the odd spin case for future studies. }\par
\textcolor{black}{In this section, we focused on the computation of the partition function. Now, a natural question arises: What are the gravitational descriptions of higher point functions? The answer to this question is a bit tricky. The direct answer is that, although there is a general holographic correspondence between 3D gravity and 2D CFT working for ensemble averages of large charge CFTs, they reproduce the gravitational counterparts in very few cases. For the Gaussian ensemble, it reproduces the gravitational result upto two boundary wormholes \cite{deBoer:2024kat}. However, for wormholes with more than two boundaries, non-Gaussianities \cite{Belin:2023efa} are important, and Gaussian contractions lead to wrong results. Partial progress has been made by considering non-Gaussian contractions to reproduce the gravitational counterparts in \cite{deBoer:2024kat}.}
\textcolor{black}{In VTQFT formalism, one can define states in super-Virasoro CFT as four-punctured torus surfaces, and one can calculate the gravitational counterpart by squaring and performing the modular sum. Generally, the one-punctured torus wormholes correspond to carving out a solid torus from the interior and gluing along the inner torus boundary. These types of wormholes are called Maldacena-Maoz wormholes \cite{Maldacena:2004rf}. Furthermore, one can think of the higher point functions as conical defects propagating in the bulk of such a wormhole.}\\ \par 
\textcolor{black}{In our paper, we focus on calculating the fount-point OTOC on torus (leading geometry), taking resort of (super-) VTQFT {\bf braiding transformations}. One can calculate the contribution to the OTOC in the bulk side coming from the torus-wormhole by taking the inner product of four-point conformal blocks on the torus. In the next section, we present the details of the computation of OTOC.}
\section{Four-point Conformal blocks on $T^2$}
\label{sec 4}
In this section, we proceed to evaluate the torus ($T^2$) four-point block in the large-charge ($c\rightarrow \infty$) regime. One way to evaluate this is to use the monodromy method for super-torus using super-Ward identities. However, we use the procedure mentioned in \cite{Alkalaev:2017bzx} to calculate the large charge conformal blocks using the recursive approach. The calculated conformal blocks are Necklace channel conformal blocks. We showcase the formalism to calculate the OPE channel blocks from the necklace channel blocks using some basic fusion and modular transformations.
To begin with, we first review what is the torus one-point function in this approach in Liouville CFT.
\subsection*{NS Vertex operators}
Super-descendants $\Phi_{\Delta,\bar{\Delta}}(\xi,\bar \xi|Z,\bar Z)$ of the super-primary field $\Phi_{\Delta,\bar\Delta}(Z,\bar Z)=\Phi_{\Delta,\bar{\Delta}}(\nu,\bar \nu|Z,\bar Z)$ are given by the following relation,
\begin{align}
\begin{split}
&\Phi_{\Delta,\bar{\Delta}}(L_{-m}\xi,\bar \xi|Z,\bar Z)=\oint \frac{dW}{2\pi i} (W-Z)^{1-m}T(W)\Phi_{\Delta,\bar{\Delta}}(\xi,\bar \xi|Z,\bar Z)\,\,\,\,\,\,\,\,\,\,\,\,\,\,\, m\in \mathbb{N}\,,\\ &
\Phi_{\Delta,\bar{\Delta}}(G_{-k}\xi,\bar \xi|Z,\bar Z)=\oint \frac{dW}{2\pi i} (W-Z)^{1/2-k}T(W)\Phi_{\Delta,\bar{\Delta}}(\xi,\bar \xi|Z,\bar Z)\,\,\,\,\,\,\,\,\,\,\,\,\,\,\,k\in \mathbb{N}-1/2\,.
\end{split}
\end{align}
Few conformal ward identities can be found in \cite{Suchanek:2010kq,Hadasz:2006qb}.

\subsection*{Liouville Torus one-point function:}
The torus one-point function $\phi_{\lambda,\bar{\lambda}}$ with the modular parameter $\tau$ can be casted as \cite{Suchanek:2010kq},
\begin{align}
\begin{split}
\langle \phi_{\lambda,\bar\lambda}\rangle &=(q\bar q)^{-c/24}\sum_{\Delta,\bar\Delta} \,\,\sum_{f,\bar f\in \frac{1}{2}\bigcup\{0\}}q^{\Delta+f}\bar{q}^{\bar \Delta +\bar f }
\sum_{\substack{f=|M|+|K|=|N|+|L|\\ \scriptstyle{\bar{f}=|\bar M|+|\bar K|=|\bar N|+|\bar L|}}} [B^f_{\Delta }]^{MK,NL}[B^{\bar f}_{\bar\Delta}]^{\bar M \bar K,\bar N\bar L}\\&\times\langle \langle\Delta|_{MK}\bigotimes \langle\bar\Delta|_{\bar M\bar K}|\phi_{\lambda,\bar \lambda}(1,1)||\Delta\rangle_{NL}\bigotimes|\bar\Delta\rangle_{\bar N\bar L} \rangle
\end{split}
\end{align}
where $q=\exp(2\pi i \tau )$ and the sum runs over the whole supermodule of the Neveu-schwarz(NS) sector. The matrices $[B^f_{\Delta}]^{MK,NL}$,\quad $[B^f_{\bar \Delta}]^{\bar M\bar K,\bar N\bar L}$ are the inverse of the Gram matrices, and they are given by,

$$[B^f_{\Delta}]_{MK,NL}=\langle \langle\Delta|_{MK}\Big||\Delta\rangle_{NL}\rangle ,~~~~~~~~~ [B^{\bar f}_{\bar\Delta}]_{\bar M\bar K,\bar N\bar L}=\langle\langle\bar\Delta|_{\bar M\bar K}\Big||\bar\Delta\rangle_{\bar N\bar L}\rangle  $$
\noindent
calculated in the standard NS sector. Apart from this, we have,

$$|\Delta\rangle_{MK}=L_{-M}G_{-K}|\Delta\rangle=L_{-m_j}\cdots L_{-m_1}G_{-k_i}\cdots G_{-k_1}|\Delta\rangle$$
where $k_i>\cdots>k_1$ and $k_s\in \mathbb{N}-1/2$ alongwith $m_j\geq\cdots\geq m_1,m_r\in\mathbb{N}.$
\noindent
For $K,L \mbox{ and } \bar K,\bar L$ having the same parity one has,
\begingroup\footnotesize
\begin{align}
\begin{split}
\langle \langle\Delta|_{MK}\bigotimes {\langle\bar\Delta|_{\bar M\bar K}}\Big|\phi_{\lambda,\bar \lambda}(1,1)\Big||\Delta\rangle_{NL}\bigotimes|\bar\Delta\rangle_{\bar N\bar L} \rangle
=\rho_{NN}(\Delta_{MK},\Delta_{\lambda},\Delta_{NL})\rho_{NN}({\bar\Delta_{\bar M\bar K}},\bar\Delta_{\bar\lambda},{\bar\Delta_{\bar N \bar L}})C^{\lambda ,\bar \lambda }_{\Delta,\bar\Delta},\end{split}
\end{align}
\endgroup
\noindent
where the quantity $C^{\lambda ,\bar \lambda }_{\Delta,\bar\Delta}$ is the primary three-point structure constants. The quantity $\rho$
can be computed by taking the product of fusion polynomials.\footnote{Fusion polynomials are written as $$P^{rs}\begin{bmatrix}
\Delta_2 \\
\Delta_1
\end{bmatrix}=\mathcal{X}^{rs}_e(\lambda_1+\lambda_2)\mathcal{X}^{rs}_e(\lambda_1-\lambda_2)\,,$$ with $$\mathcal{X}^{rs}_e(\lambda)=\prod_{p=1-r}^{r-1} \prod_{q=1-s}^{s-1}(\frac{\lambda-pb -qb^{-1}}{2\sqrt{2}})$$ and the products run over $p=1-r+2k , \,\,\,\,\,\,\,\,\,\,\,\,\,\,\,\,\,\,q=1-s+2l,\,\,\,\,\,\,\,\,\,k+l\in 2\mathbb{N}\bigcup \{0\}.$}\par
\noindent
 In the next subsection, we move on to the computation of the torus four-point conformal block, which is relevant for computing the OTOC \cite{Kusuki:2019gjs,Roberts:2014ifa}.
\section*{ Super-Liouville four-point function on torus: }\par
Now, we proceed to compute the torus four-point block in $s$-channel. The $t$-channel block can be computed by using the fusion kernel method or by taking the OPE directly in the different channels. Then, we will show how to achieve the global blocks in the $c\rightarrow \infty $ limit.
\noindent
Recalling that $\langle\tilde\Delta_m|\phi_k(z_k)|\tilde\Delta_l\rangle= C_{\tilde\Delta_m\Delta_k\tilde\Delta_l}z_k^{\tilde\Delta_m-\Delta_k-\tilde\Delta_l}$ are the structure constants, we write the four-point function for super-Liouville theory as,\\
\begingroup
\begin{align}
\begin{split}
&\langle \Phi_1(Z_1,\bar Z_1)  \Phi_2(Z_2,\bar Z_2)\Phi_3(Z_3,\bar Z_3)\Phi_4(Z_4,\bar Z_4)\rangle_{\tau}\\&=\sum_{\tilde\Delta_1,\tilde\Delta_2,\tilde\Delta_3,\tilde\Delta_4}C_{\tilde\Delta_1\Delta_1\tilde\Delta_2} C_{\tilde\Delta_2\Delta_2\tilde\Delta_3}C_{\tilde\Delta_3\Delta_3\tilde\Delta_4}C_{\tilde\Delta_4\Delta_4\tilde\Delta_1}\\&
\times Z_1^{\tilde\Delta_1-\Delta_1-\tilde\Delta_2}Z_2^{\tilde\Delta_2-\Delta_2-\tilde\Delta_3}Z_3^{\tilde\Delta_3-\Delta_3-\tilde\Delta_4}Z_4^{\tilde\Delta_4-\Delta_4-\tilde\Delta_1}\Big[\mathcal{F}_c^{\{s\}\Delta_{1,2,3,4}\tilde\Delta_{1,2,3,4}}(q,Z_{1,2,3,4})\Big] \times (\textrm{anti-holomorphic})\label{4.6m}
\end{split}
\end{align}
\endgroup
where the necklace-channel conformal block is denoted by $\mathcal{F}^{\{s\}}$ and the expression is given by,
\begingroup
\begin{align}
\begin{split}\mathcal{F}^{\{s\}}_{c}&=q^{c/24-\tilde\Delta_1}\sum_{n,m,r,g=0}^{\infty}q^n \sum_{\substack{|M|+|K|=|N|+|L|=n\\|S|+|J|=|T|+|P|=m\\|O|+|Y|=|U|+|V|=r\\|E|+|W|=|C|+|X|=g }} [B^1_{\Delta }]^{MK,NL} \frac{\langle \tilde \Delta_1,MK|\Phi_1(Z_1)|SJ,\tilde{\Delta}_2\rangle}{\langle \tilde \Delta_1|\Phi_1(Z_1)|\tilde{\Delta}_2\rangle} [B^2_{\Delta }]^{SJ,TP}\\&\hspace{1 cm}\times \frac{\langle \tilde \Delta_2,TP|\Phi_2(Z_2)|OY,\tilde{\Delta}_3\rangle}{\langle \tilde \Delta_2|\Phi_2(Z_2)|\tilde{\Delta}_3\rangle} [B^3_{\Delta }]^{OY,UV} \frac{\langle \tilde \Delta_3,UV|\Phi_3(Z_3)|EW,\tilde{\Delta}_4\rangle}{\langle \tilde \Delta_3|\Phi_3(Z_3)|\tilde{\Delta}_4\rangle}\,\,[B^4_{\Delta }]^{EW,CX} \\&\hspace{5 cm}\times\frac{\langle \tilde \Delta_4,CX|\Phi_4(Z_4)|NL,\tilde{\Delta}_1\rangle}{\langle \tilde \Delta_4|\Phi_4(Z_4)|\tilde{\Delta}_1\rangle}\,.
\end{split}
\end{align}
\endgroup
We will be mainly focusing on the calculation of the Out-of-time-order (OTO) four-point function, which has the form of $\langle \mathcal{O}_A\mathcal{O}_B\mathcal{O}_A\mathcal{O}_B\rangle\,.$ We will set the conformal dimensions of the two primaries (super-Virasoro) inserted on the time circle in the torus to be equal while calculating the time-ordered (TO) super-Liouville four-point function. We can write the conformal block structure in the TO four-point function on the torus as, (defining $x=z_2/z_1$) 
\begingroup
\begin{align}
\begin{split}
&\mathcal{F}^{\{s\}}_{\langle \Phi_1(z_1,\bar z_1)  \Phi_1(z_2,\bar z_2)\Phi_3(z_3,\bar z_3)\Phi_3(z_4,\bar z_4)\rangle_{\tau}}={\bf \mathcal{A}}_c^{\Delta_{1,2},\tilde \Delta_{1,2,3,4}}(z_{1,2,3,4})+q^{1/2} {\bf \mathcal{B}}_c^{\Delta_{1,2,3,4},\tilde \Delta_{1,2,3,4}}(z_{1,2,3,4})+\mathcal{O}(q)+\cdots\,.
\end{split}
\end{align}
\endgroup
\noindent
Here $\mathcal{A}_c$ and $\mathcal{B}_c$ are given in (\ref{A.2 c}) and (\ref{A.3 d}) respectively.
In the next section, we use the four-point function given in \eqref{4.6m} to compute the OTOC using a suitable transformation and analyze its early-time behaviour. 
\section{A brief tour to OTOC} 
\label{sec 5}
The out-of-time-order correlator (OTOC) is a probe for early-time chaos. The OTOC is defined as the inner product between two states $A(t)B|0\rangle_{\beta}$ and $BA(t)| 0\rangle_{\beta}$ where the operators $A$ and $B$ are separated in space by $x$ and in Lorentzian time by $t$ at the inverse temperature $\beta\,.$
\noindent
The definition of the normalised OTOC is given by,
\begin{align}\mathbfcal{C}_{\beta}(x,t)=\frac{\langle B^{\dagger} A^{\dagger}(t) B\,A(t)\rangle_{\beta}}{\sqrt{\langle B^{\dagger} A^{\dagger}(t) A(t)\,B \rangle_{\beta}\langle A^{\dagger}(t) B^{\dagger} B\,A(t) \rangle_{\beta}}}\,.\end{align}

\noindent
 To compute the OTOC, we do the  analytic continuation of the following TO correlator,\\
$\langle B^{\dagger}(z_1,\bar z_1)B(z_2,\bar z_2)A^{\dagger}(z_3,\bar z_3)A(z_4,\bar z_4)\rangle_{\beta}$, so that the operator ordering becomes $\langle B^{\dagger} A^{\dagger}(t) B\,A(t) \rangle_{\beta}\,.$ \par
This can be obtained by the following choice of coordinates on the thermal cylinder, as \cite{Roberts:2014ifa},
\begin{align}
\begin{split}
&z_1=e^{\frac{2\pi}{\beta}(t+i\epsilon_1)},\hspace{2 cm} \bar z_1=e^{-\frac{2\pi}{\beta}(t+i\epsilon_1)}\,,\\&
z_2=e^{\frac{2\pi}{\beta}(t+i\epsilon_2)},\hspace{2 cm} \bar z_2=e^{-\frac{2\pi}{\beta}(t+i\epsilon_1)}\,,\\&z_3=e^{\frac{2\pi}{\beta}(x+i\epsilon_3)},\hspace{2 cm} \bar z_3=e^{\textcolor{black}{+}\frac{2\pi}{\beta}(x-i\epsilon_3)}\,,\\&z_4=e^{\frac{2\pi}{\beta}(x+i\epsilon_4)},\hspace{2 cm} \bar z_4=e^{\frac{2\pi}{\beta}(x-i\epsilon_4)}
\end{split}
\end{align}
where we analytically continued the correlator to the real-time considering $\epsilon_1<\epsilon_3<\epsilon_2<\epsilon_4$.\par
At late times, the OTOC is exponentially damped for truly chaotic systems, and it may decay polynomially for non-chaotic models. The time evolution of the cross-ratio $z$ is given by $(1-z)\rightarrow(1-z)e^{-2\pi i}$, which means, in turn, that the late time behaviour of the OTOC is given by the corresponding {\bf Regge limit} of the correlator.\\

In this case, the cross-ratio on the complex plane is given by,
\begin{align}
z\approx e^{\frac{-2\pi}{\beta}(x-t)}\epsilon_{12}^*\epsilon_{34}\,,\hspace{2 cm} \bar z\approx - e^{\frac{-2\pi}{\beta}(x+t)}\epsilon_{12}^*\epsilon_{34}\,,
\end{align}
where we use the abbreviation,
\begin{align}
\epsilon_{ij}=i\Big(e^{\frac{2\pi}{\beta}i\epsilon_i}-e^{\frac{2\pi}{\beta}i\epsilon_j}\Big)\,.
\end{align}
\noindent
Now, one needs to pick the monodromy around the branch cut in $z\sim[1,\infty)$.

\vspace{0.2 cm}
 \noindent
Furthermore, we will use the following limits of the fusion kernel to find the OTOC for our case.
\begin{align} 
\begin{split}
\mathbb{F}_{P_{3},P_{t}}\begin{bmatrix}
    P_3 & P_2 \\
    \mathbb{I}  & P_1
\end{bmatrix}&=\delta(P_1-P_{t})\ , \\
\mathbb{F}_{\mathbb{I},P_{t}}\begin{bmatrix}
    P_3 & P_2 \\
    P_3 & P_2
\end{bmatrix}&=C_{NS}(P_2,P_3,P_{t})\,\frac{W_{NS}(\alpha_{t})}{W_{NS}(Q-\alpha_t)}\ , \\
\mathbb{S}_{\mathbb{I},P}[\mathbb{I}]&= W_{NS}(\alpha)W_{NS}(Q-\alpha)\ .
\end{split}
\end{align}

\noindent
In case of finding the OTOC on the torus, one needs to remember that there are two cycles in this specific topology. One is called the spatial circle, and the other is called the time circle. We are interested in finding this probe of chaos (OTOC) for super-Liouville matter insertions.
For calculating the out-of-time-order correlator in this case, we need to perform an R-transformation on the time-ordered four-point function. Pictorially, this can be described as \cite{Mertens:2022irh} \footnote{\textcolor{black}{\bf Comment}: We should remember that if we set $\Delta_{\psi}=0$, we are no longer allowing four-point insertion. In this case, the four-point block will be reduced to a three-point necklace channel block. This is one of the main differences with the OPE channel block.}, 
\begin{align}
\begin{split}
\begin{minipage}[h]{0.17\linewidth}
	\vspace{4pt}
	\scalebox{0.85}{\includegraphics[width=\linewidth]{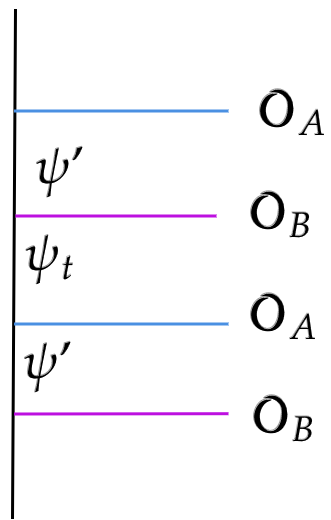}}
   \end{minipage} \hspace{0.3cm}=\int d\alpha_{s}\mathbb{R}^{(s)\gamma,\epsilon}_{\alpha_s \alpha_t} \begin{bmatrix}
\alpha_B & \alpha_{A} \\
\alpha_\psi & \alpha_{\psi'}
\end{bmatrix}
\begin{minipage}[h]{0.17\linewidth}
	\vspace{4pt}
	\scalebox{0.85}{\includegraphics[width=\linewidth]{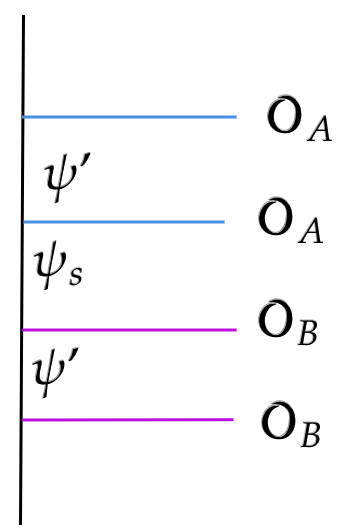}}
   \end{minipage}\label{5.3m}
  \end{split}
  \end{align}
  
  In the above equation, $\mathbb{R}$ matrix is related to the fusion matrix in the following fashion for \textit{Virasoro representation} as\footnote{Few basic moves are described pictorially in Appendix~(\ref{appd}).} \cite{Suchanek:2010kq},
  \begin{align}
  \mathbb{R}_{\alpha_s \alpha_t} \begin{bmatrix}
\alpha_3 & \alpha_2 \\
\alpha_4 & \alpha_1
\end{bmatrix}=e^{2\pi i (\Delta_2+\Delta_4-\Delta_s-\Delta_t)}\mathbb{F}_{\alpha_s \alpha_t} \begin{bmatrix}
\alpha_3 & \alpha_2 \\
\alpha_4 & \alpha_1
\end{bmatrix}\,.\end{align}
The $\mathbb{F}$-matrix can be written in terms of quantum 6j symbols via,
$$\mathbb{F}_{\alpha_s \alpha_t} \begin{bmatrix}
\alpha_3 & \alpha_2 \\
\alpha_4 & \alpha_1
\end{bmatrix}=|S_b(2\alpha_t)S_b(2\alpha_s)|\sqrt{\frac{C(\alpha_4,\alpha_t,\alpha_1)C(\bar\alpha_t,\alpha_3,\alpha_2)}{C(\alpha_3,\alpha_4,\alpha_s)C(\bar\alpha_s,\alpha_2,\alpha_1)}}\begin{Bmatrix}
\alpha_1 & \alpha_2& \alpha_s \\
\alpha_3 & \alpha_4 & \alpha_t
\end{Bmatrix}_b$$
where $S_b$ denotes the double sine function , $\alpha_j=\frac{Q}{2}+i\underbrace{bk_j}_{P_j}$ and $C(\alpha_1 , \alpha_2, \alpha_s)$ are the usual DOZZ coefficients in the Liouville theory. The 6j symbols can be found in \cite{Blommaert:2018oro}.
The generalization to the super-Liouville case is quite straightforward.

The necklace channel braiding $(\mathbb{R})$ matrix is given by,
\begingroup
\begin{align}
\mathbb{R}^{\gamma,\epsilon}_{\alpha_s \alpha_t}\begin{bmatrix}
\alpha_3 & \alpha_2 \\
\alpha_4 & \alpha_1
\end{bmatrix}=e^{\pi i \epsilon (\Delta_4 + \Delta_1 - \Delta_s-\Delta_t)}\,\mathbb{R}^{\gamma}_{\alpha_s \alpha_t}\begin{bmatrix}
\alpha_3 & \alpha_2 \\
\alpha_4 & \alpha_1
\end{bmatrix}\,\,\,\,\,\,\,\,\,\,\,\,\,\,\, \gamma=(e,o) \,\,\,\,\,\,\,\,\,\,\,\,\,\, \epsilon=\textrm{sign} (\textrm{Arg}\, 
 z_3-\textrm{Arg}\, z_2)
\end{align}
\endgroup
where,
\begin{align}
\begin{split}
\mathbb{R}^{\gamma}_{\alpha_s \alpha_t}\begin{bmatrix}
\alpha_3 & \alpha_2 \\
\alpha_4 & \alpha_1
\end{bmatrix}=M^{\gamma}_{\alpha_s,\alpha_t}
\begin{bmatrix}
\alpha_3 & \alpha_2 \\
\alpha_4 & \alpha_1
\end{bmatrix}
\frac{\Gamma_{NS}(2\alpha_s)\Gamma_{NS}(2Q-2\alpha_s)}{4\Gamma_{NS}(Q-2\alpha_t)\Gamma_{NS}(2\alpha_t-Q)}\int_{-i\infty }^{i\infty }\frac{d\tau}{i}J^{\gamma}_{\alpha_s,\alpha_t}
\end{split}
\end{align}
with,
\begin{align}
    \begin{split}
M^e_{\alpha_s,\alpha_t}
\begin{bmatrix}
\alpha_3 & \alpha_2 \\
\alpha_4 & \alpha_1
\end{bmatrix}=&\frac{4^{|\sigma|}\Gamma_{\sigma}(\alpha_t+\alpha_4-\alpha_2)\Gamma_{\sigma}(\bar\alpha_t+\alpha_4-\alpha_2)\Gamma_{\sigma}(\alpha_t+\bar\alpha_4-\alpha_2)\Gamma_{\sigma}(\bar\alpha_t+\bar\alpha_4-\alpha_2)}{4^{|\eta|}\Gamma_{\eta}(\alpha_s+\alpha_4-\alpha_3)\Gamma_{\eta}(\bar\alpha_s+\alpha_4-\alpha_3)\Gamma_{\eta}(\alpha_s+\bar\alpha_4-\alpha_3)\Gamma_{\eta}(\bar\alpha_s+\bar\alpha_4-\alpha_3)}\\ &\times\frac{ \Gamma_{\sigma}(\alpha_t+\alpha_1-\alpha_3)\Gamma_{\sigma}(\bar\alpha_t+\alpha_1-\alpha_3)\Gamma_{\sigma}(\alpha_t+\bar\alpha_1-\alpha_3)\Gamma_{\sigma}(\bar\alpha_t+\bar\alpha_1-\alpha_3)}{\Gamma_{\eta}(\alpha_s+\alpha_1-\alpha_2)\Gamma_{\eta}(\bar\alpha_s+\alpha_1-\alpha_2)\Gamma_{\eta}(\alpha_s+\bar\alpha_1-\alpha_2)\Gamma_{\eta}(\bar\alpha_s+\bar\alpha_1-\alpha_2)}
\end{split}
\end{align}
and for the even structure, the J($\tau$) matrix is given by \cite{Chorazkiewicz:2008es},

\begingroup\small
$$
J^{e}(\tau)=\begin{pmatrix}
\begin{bmatrix}N&N&N&N\\
N&N&N&N
\end{bmatrix}+\begin{bmatrix}R&R&R&R\\
R&R&R&R
\end{bmatrix} &\frac{1}{i}\begin{bmatrix}N&N&N&N\\
R&R&N&N
\end{bmatrix}+i\begin{bmatrix}R&R&R&R\\
N&N&R&R
\end{bmatrix}  \\
\frac{1}{i}\begin{bmatrix}N&N&N&N\\
N&N&R&R
\end{bmatrix}+i\begin{bmatrix}R&R&R&R\\
R&R&N &N
\end{bmatrix} & -\begin{bmatrix}N&N&N&N\\
R&R&R&R
\end{bmatrix}-\begin{bmatrix}R&R&R&R\\
N&N&N&N
\end{bmatrix}
\end{pmatrix}$$
\endgroup
where we use the abbreviations,
\begin{align}
\begin{bmatrix}N&N&N&N\\
N&N&N&N
\end{bmatrix}=\frac{S_{NS}(\bar \alpha_4-\alpha_3+\alpha_2+\tau)S_{NS}(\alpha_1+\tau)S_{NS}(\alpha_4-\alpha_3+\alpha_2+\tau)S_{NS}(\bar \alpha_1+\tau)}{S_{NS}(\bar \alpha_3+\bar\alpha_t+\tau)S_{NS}(\bar \alpha_3+\alpha_t+\tau)S_{NS}( \alpha_s+\alpha_2+\tau)S_{NS}(\bar \alpha_s+\alpha_2+\tau)}
\end{align}
similarly,
\begin{align}
\begin{bmatrix}N&R&N&R\\
N&N&N&N
\end{bmatrix}=\frac{S_{NS}(\bar \alpha_4-\alpha_3+\alpha_2+\tau)S_{R}(\alpha_1+\tau)S_{NS}(\alpha_4-\alpha_3+\alpha_2+\tau)S_{R}(\bar \alpha_1+\tau)}{S_{NS}(\bar \alpha_3+\bar\alpha_t+\tau)S_{NS}(\bar \alpha_3+\alpha_t+\tau)S_{NS}( \alpha_s+\alpha_2+\tau)S_{NS}(\bar \alpha_s+\alpha_2+\tau)}\,.
\end{align}
Here, $\bar\alpha_i=Q-\alpha_i$.
Again ,$$\lim_{\alpha_1\rightarrow0}\mathbb{R}^{e}_{\alpha_s\alpha_t}\begin{bmatrix}
    \alpha_3& \alpha_2\\
    \alpha_4& \alpha_1
\end{bmatrix}^{e}_{\lambda}={\bf \delta(\alpha_s-\alpha_2)}\delta^e_{\lambda}\delta(P_t-P_3)\,.$$
Before computing the exact form of the integral, we wish to extract the asymptotic behaviour of the 4-point necklace channel conformal block and see how it depends on the cross-ratio and the conformal dimension $\Delta_s$, which can easily be done using the shadow formalism \cite{Ferrara:1972kab, Iliesiu:2015qra,Alkalaev:2023evp}.
\section*{Torus blocks: Shadow formalism}

To calculate the OTOC in super-Liouville field theory on the torus, one should compute the {\bf asymptotic behaviour} of the necklace channel conformal block precisely \cite{Rosenhaus:2018zqn}. This computation will mainly help to fix the nature of OTOC with time. We show that, as there is a connection between \textit{comb channel} conformal blocks and necklace channel conformal blocks, we can extract the $n$-point necklace channel conformal block from the $n+2$-point \textit{comb channel} conformal block \cite{Rosenhaus:2018zqn}.
\begin{align}
\mathcal{F}_N^{(n)}(q,z|\Delta,\tilde {\Delta},\Delta_{\alpha})=\sum_{m=0}^\infty\frac{ q^{m+\Delta_{\alpha}}A_m^{(n+2)}(z|\Delta,\tilde {\Delta},\Delta_{\alpha})}{m!(2\Delta_{\alpha})_m},
\end{align}
where, 
\begin{align}A_m^{(n+2)}(z|\Delta,\tilde{ \Delta},\Delta_{\alpha})=\lim_{\substack{z_0\rightarrow 0\\ z_{n+1}\rightarrow 0}}\partial_0^m\partial_{n+1}^m\Big(z_0^{-2\Delta_{\alpha}}\mathcal{G}_{n+2}(1/z_{0},z,z_{n+1}|\Delta,\tilde{\Delta},\Delta_{\alpha})\Big)\,.\label{6.23 m}\end{align}
In (\ref{6.23 m}), $\mathcal{G}_{n+2}$ denotes the $n+2$-point \textit {comb channel }conformal block. We denote the external dimensions as $(\Delta,\Delta_1\cdots \Delta_{n})$ and the internal dimensions as $(\tilde \Delta_{0},\tilde{\Delta}_1\cdots \tilde{\Delta}_{n-2})$. In general, the computation of 6-point comb channel blocks yields the following compact form,
\begin{align}
\begin{split}
&\mathcal{G}_{\tilde \Delta_1,\tilde \Delta_2,\tilde \Delta_3}^{\Delta_1,\cdots \Delta_6}(z_0,z,z_{n+1})=\mathcal{L}^{(\Delta_1,\cdots \Delta_4) }(z_0,\cdots ,z_4)\,\chi_0^{\tilde{\Delta}_0}\chi_{1}^{\tilde{\Delta}_1}\chi_2^{\tilde{\Delta}_2}\\&\hspace{2 cm}\times{\bf F_{K}}\begin{bmatrix}
-\mathcal{S}_1&\tilde \Delta_0+\tilde \Delta_1-\Delta_2&\tilde \Delta_1+\tilde \Delta_2-\Delta_3&\,\,\,-\mathcal{S}_2  \\&\hspace{1cm} 2\tilde \Delta_0,\,2\tilde \Delta_1,\,2\tilde \Delta_2
\end{bmatrix}(\chi_0,\chi_{1},\chi_2)
\label{6.24 m}
\end{split}
\end{align}
where the leg-factor can be cast in a compact form as,

\begin{align}\mathcal{L}^{(\Delta_1,\cdots \Delta_4) }(z_0,\cdots ,z_4)=\Big(\frac{z_{12}z_{34}}{z_{01}z_{02}z_{45}z_{35}}\Big)^{\Delta_{\alpha}}\prod_{i=0}^{3}\Big(\frac{z_{i,i+2}}{z_{i,i+1}z_{i+1,i+2}}\Big)^{\Delta_{i+1}}\end{align}
and the \textit{comb function} is given by,
\begingroup
\begin{align}
\begin{split}
&{\bf F_{K}}\begin{bmatrix}
    a_1,& b_1,\cdots b_{k-1},&a_2\\&
    c_1, \cdots ,c_k
\end{bmatrix}(x_1,\cdots x_k)\\&=\sum_{n_1,\cdots n_k=0}^{\infty}\frac{(a_1)_{n_1}(b_1)_{n_1+n_2}(b_2)_{n_2+n_3}\cdots {(b_{k-1})_{n_{k-1}+n_k}(a_2)_{n_k}}}{{(c_1)_{n_1}\cdots (c_k)_{n_k}}}\frac{x_1^{n_1}}{n_1!}\cdots \frac{x_k^{n_k}}{n_k!}\,.
\end{split}
\end{align}
\endgroup
In (\ref{6.24 m}), we define the cross-ratios as,
\begin{align}
    \chi_0=\frac{z_{01}z_{23}}{z_{02}z_{13}};\,\,\,\,\,\,\chi_{1}=\frac{z_{12}z_{34}}{z_{13}z_{24}};\,\,\,\,\,\,\chi_2=\frac{z_{23}z_{45}}{z_{24}z_{35}}
\end{align}
and we also have used the short notation of $\Delta_{ij}=\Delta_i-\Delta_j$.
\noindent
${\bf F_{K}}$ denotes the comb function. Its properties and a few identities are defined in \cite{Rosenhaus:2018zqn}.
In (\ref{6.24 m}) we define, 
\begin{align}
\mathcal{S}_1=\Delta_1-\Delta_{\alpha}-\tilde \Delta_{0},\hspace{2 cm} \mathcal{S}_2=\Delta_4-\Delta_{\alpha}-\tilde \Delta_{2}\,.
\end{align}
\noindent
Now the quantity $A^{(6)}_m{(z|\Delta,\tilde \Delta,\Delta_{\alpha})}$, which is related to the 4-point necklace channel block, becomes,
\begingroup
\begin{align}
\begin{split}
A^{(6)}_m{(z|\Delta,\tilde \Delta,\Delta_{\alpha})}=&\lim_{\substack{z_0\rightarrow 0\\ z_{5}\rightarrow 0}}(\Delta_1-\Delta_{\alpha }-n_1)_m (1-z_0){}^{-\Delta_{\alpha }+\Delta_1-m-n_1}(-\Delta_4-\Delta_{\alpha }+n_3)_m\\&\hspace{4 cm}\times(\chi_{1}-z_5){}^{-\Delta_{\alpha }-\Delta_4-m+n_3}\times \Xi({\bf \rho})\,,\\&
=(\Delta_1-\Delta_{\alpha }-n_1)_m(-\Delta_4-\Delta_{\alpha }+n_3)_m(\chi_{1}){}^{-\Delta_{\alpha }-\Delta_4-m+n_3}\, \Xi({\bf \rho})\,.\label{6.30 m}
\end{split}
\end{align}
\endgroup
Here, ${\bf \rho}$ denotes the other insertion points except $z_0$ and $z_5\,.$ $\Xi({\bf \rho})$ refers to the part of the correlation function independent of the cross-ratio. We need mainly the dependence on the cross ratio in leading behaviour to find the out-of-time-order correlator. We will use this in Sec.~(\ref{sec g}).
\subsection{OTOC in plane for super-Liouville CFT}
In superplane, the super-Liouville four-point function can be written as \cite{Belavin:2007eq,Eden:1998hh,Eden:2000qp,Khorrami:1998kw},

\begin{align}
\langle\Phi_1\Phi_2\Phi_3\Phi_4\rangle=Y(a+b_1 W_{234}+b_4W_{123}+cV),
\end{align}
where \begin{align}Y=\prod_{i<j}(z_{ij})^{\Delta/3-\Delta_i-\Delta_j},\,\,\,\,\,\,\,\,\,\,\,\,\,\,\, W_{ijk}= \frac{\theta_i z_{jk}-\theta_j z_{ik}+\theta_k z_{ij}+\theta_i\theta_j\theta_k}{\sqrt{z_{ij}z_{ik}z_{jk}}},
\end{align}
\begin{align}
  V=\frac{\theta_1\theta_2 z_{34}}{z_{13}z_{24}}+\frac{\theta_3\theta_4 z_{12}}{z_{13}z_{24}}+\frac{\theta_1\theta_4 z_{23}}{z_{13}z_{24}}+\frac{\theta_2\theta_3 z_{14}}{z_{13}z_{24}}-\frac{\theta_1\theta_3 }{z_{13}}-\frac{\theta_2\theta_4} {z_{24}}+3\frac{\theta_1\theta_2\theta_3\theta_4}{z_{13}z_{24}},\,\,\,\,\,\,\,\,\,\Delta=\sum_i\Delta_i,
\end{align}

and $a,b_1,b_4 \text{ and } c$ are functions of cross ratio. The leading behaviour of the $4$-point comb channel conformal block (each of $a,b_1,b_4,c$) has the following form for our case as \cite{Belavin:2024mzd},
\begin{align}\mathcal{V}_c^{\Delta_{1,2,3,4},\tilde\Delta_{1,2,3,4}}\sim\chi_{1}^{\alpha_s(Q-\alpha_s)}\,.
\end{align}
Now, to compute the OTOC we need to compute the following integral as given in \eqref{5.3m}\footnote{ The fermionic contribution to the leading order dependence on cross-ratio is the same as the bosonic case.},
\begingroup
\begin{align}
\begin{split}
\textrm{OTOC}
&=\mathcal{F}_r\int_{Q/2}^{Q/2+i\infty} d\alpha_s e^{\pi i \epsilon(\tilde\Delta_2+\tilde\Delta_4-\Delta_s-\Delta_t)}\frac{W_{NS}(Q)W_{NS}(\alpha_s)}{\pi W_{NS}(Q-\alpha_A)W_{NS}(Q-\alpha_B)}\\&\hspace{4 cm}\times C_{NS}(\alpha_s,\alpha_A,\alpha_B)\mathcal{V}_c^{\Delta_{1,2,3,4},\tilde\Delta_{1,2,3,4}}\mathbb{M}^p_{\alpha_t,\alpha_s}\Big[\alpha_A,\alpha_B\Big]\,,\\
\\
&\hspace{0 cm}=\underbrace{\lim_{P_s\rightarrow0} \partial_{P_s}^2\Bigg[{\bf \Psi(\alpha_A,\alpha_B,Q,P_s)\bf\bar{\Psi}(\bar\alpha_A,\bar\alpha_B,Q,\bar P_s)}\mathbb{M}^p_{\alpha_t,\alpha_s}\Big[\alpha_A,\alpha_B\Big]\Bigg]}_{\bf \tilde \Theta(\alpha_A,\alpha_B,Q)}\mathcal{F}_r\int_{0}^\infty dP_s P_s^2 \chi_{1}^{2 P_s^2}\,,\\&
={\bf\tilde{\Theta}}(\alpha_A,\alpha_B,Q)\frac{\chi_{1}^{Q^2/4}}{(-\log(\chi_{1}))^{3/2}}\mathcal{F}_r\sim \chi_{1}^{Q^2/4} \Big(-\log(\chi_{1})\Big)^{-3/2}\mathcal{F}_r,
\hspace{1 cm}\alpha_A , \alpha_B > \frac{Q}{4}\,,\label{5.19u}
\end{split}
\end{align}
\endgroup

where $\mathbb{M}^p_{\alpha_t,\alpha_s}\Big[\alpha_A,\alpha_B\Big]$ is the monodromy matrix which is defined to be, \begin{align}
\mathcal{F}^{AA}_{BB}(h_p|\chi_1)\xrightarrow{(1-\chi_1)\to (1-\chi_1)e^{-2\pi i}}\int_{\mathbb{S}'} d\alpha ~\mathbb{M}^p_{\alpha_t,\alpha_s}\Big[\alpha_A,\alpha_B\Big]\,\mathcal{F}^{AA}_{BB}(h_\alpha|\chi_1).\label{5.22b}
\end{align}
In (\ref{5.22b}) the contour $\mathbb{S}'$ runs from $Q/2\to Q/2+i \infty$ and in fact it can be expressed in terms of fusion matrix as \cite{Kusuki:2019gjs}, 
\begin{align}
\mathbb{M}^p_{\alpha_p,\alpha}
\begin{bmatrix}
    \alpha_A & \alpha_A\\
    \alpha_B & \alpha_B
\end{bmatrix}=\int_{\mathbb{S}'} d\beta e^{-2\pi i(h_{\beta}-h_A-h_B)}F_{\alpha_p,\beta}\begin{bmatrix}
    \alpha_A & \alpha_B\\
    \alpha_A & \alpha_B
\end{bmatrix}F_{\beta,\alpha}\begin{bmatrix}
    \alpha_A & \alpha_A\\
    \alpha_B & \alpha_B
\end{bmatrix}.
\end{align}
In \eqref{5.19u} the function ${\bf \Psi(\alpha_A,\alpha_B,Q,P_s)}$ is defined in (\ref{B.18 m}) of { Appendix}~(\ref{App}).
This behaviour of OTOC in the super-Liouville theory is exactly similar to the behaviour of OTOC obtained in \cite{Kusuki:2019gjs} for the case of Liouville theory. To compute the integral, we used the {\bf saddle-point approach}, and we have taken $C_{NS}$ approximated for two light conformal dimensions and one heavy conformal dimension. In deriving its asymptotic form, we used the expression of $\Gamma_b(x)$ defined in Appendix~(\ref{AppB}).
 $\mathcal{F}_r$ is independent of $\alpha_s$, and it comes from the other remaining part of the conformal block.




\subsection{OTOC on even torus}
\label{sec g}
The two-point bosonic and fermionic super-Virasoro blocks in torus at leading order in $q$ is given by\footnote{only $\Delta_s$ dependent part is written here.},
\hfsetfillcolor{gray!5}
\hfsetbordercolor{white}
\begin{align}
\begin{split}\tikzmarkin[disable rounded corners=true]{werp}(0.5,-0.5)(-0.1,1)&\textcolor{black}{\mathbfcal{B}_B^{(1)}}=\,\frac{\Big[(\Delta_s+\tilde\Delta_2-\Delta_2)(\tilde\Delta_4+\Delta_s-\Delta_3)\Big]}{{2\Delta_s}}\chi_{1}+\mathcal{O}(q)\,,\\&
\textcolor{black}{\mathbfcal{B}_F^{(1)}}=\Big[\frac{(-\tilde\Delta_2-\Delta_s-\Delta_2+1/2)(-\Delta_s-\tilde\Delta_4-\Delta_3+1/2)}{2\Delta_s}\Big]\chi_{1}+\mathcal{O}(q)\,.\tikzmarkend{werp}
\end{split}
\end{align}
In the following part, we focus on the {\bf "classical" torus conformal block}, which is given by,
\begin{align}
\mathcal{V}_c^{\Delta_{1,2,3,4},\tilde\Delta_{1,2,3,4}}(q,z_{1,2,3,4})
=\exp (\frac{c}{6}f^{\epsilon_{1,2,3,4},\tilde\epsilon_{1,2,3,4}}) \,\,\,\,\, \text{as}\,\,\,\ c\rightarrow \infty\,,
\end{align}
where the function f is the classical conformal block and\footnote{Heavy operators scales as $\Delta_H \sim c \epsilon$ and light operators scales as $\Delta_L \sim \epsilon$. $\epsilon$ denotes the classical dimension.},
$$\epsilon_i=\frac{\Delta_i}{k},\,\,\,\,\,\,\,\, \tilde\epsilon_i=\frac{\tilde\Delta_i}{k}, \,\,\,\,\,\,\,\,\text{where}\,\,\,\,\,\,\,\, k=\frac{c}{6}.$$
Now, using (\ref{A.1 m}) the $s$-channel block function is given by,
\begin{align}
f^{\epsilon_{1,2,3,4},\tilde\epsilon_{1,2,3,4}}=(\tilde\epsilon_1-1/4)\log(q)+\sum_{n=0}^\infty q^nf_{n,B/F}^{(1)}(\epsilon,\tilde\epsilon|x,y),
\end{align}
where the first few expansion coefficients are,
\begingroup
\begin{align}
\begin{split}
&f^{(1)}_{B0}=\frac{1}{2\tilde\epsilon_3}\Big[\frac{(\tilde\epsilon_1+\tilde\epsilon_2-\epsilon_1)(\tilde\epsilon_2+\tilde\epsilon_3-\epsilon_2)}{2\tilde\epsilon_2}\chi_{1}\frac{(\tilde\epsilon_3+\tilde\epsilon_4-\epsilon_3)(\tilde\epsilon_4+\tilde\epsilon_1-\epsilon_4)}{2\tilde\epsilon_4}\Big]\,,\\&
f^{(1)}_{F0}=\frac{1}{2\tilde\epsilon_3}\Big[\frac{(-\tilde\epsilon_1-\tilde\epsilon_2-\epsilon_1+1/2)(-\tilde\epsilon_2-\tilde\epsilon_3-\epsilon_2+1/2)}{2\tilde\epsilon_2}\chi_{1}\frac{(-\tilde\epsilon_3-\tilde\epsilon_4-\epsilon_3+1/2)(-\tilde\epsilon_4-\tilde\epsilon_1-\epsilon_4+1/2)}{2\tilde\epsilon_4}\Big]
\end{split}
\end{align}
\endgroup
with\footnote{ This definition of $\chi_1$ is valid for torus one-point superblocks.}, \textcolor{black}{$\chi_{1}:=\frac{z_{12}z_{34}}{z_{13}z_{24}}$}\,. Also, B/F denotes the bosonic or fermionic counterpart.\par
The out-of-time-order OPE channel conformal block can be obtained from the necklace channel conformal block by composition of three fusion($\mathbb{F}$) and one $\mathbb{S}$-move \cite{deBoer:2024kat}.
\begin{figure}
\begin{center}
\scalebox{1.0}{\includegraphics[width=\linewidth]{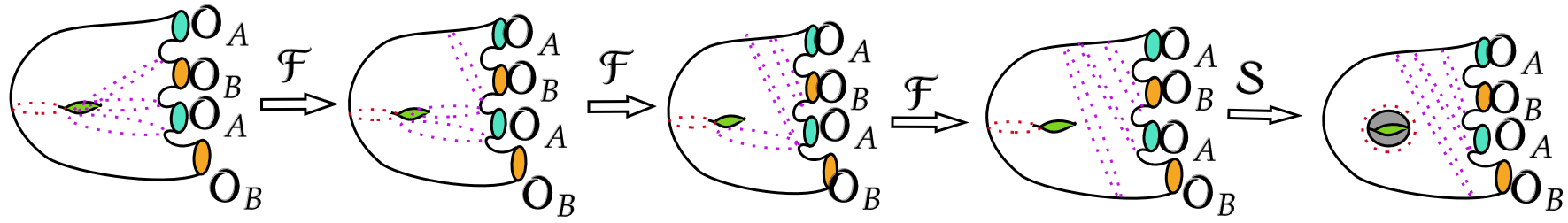}}
  \caption{Figure depicting the transformation needed to get the OTO OPE block from the Necklace channel out of time order conformal block by modular-$\mathbb{S}$ transformation and the fusion-$\mathbb{F}$ transformation.}
  \label{Fig2}
  \end{center}
 \end{figure}

\begin{align}
\begin{split}
  & \textcolor{black}{ \mathcal{F}_{OPE}=\int \frac{d\tilde\alpha_2}{2}\frac{d\tilde\alpha_3}{2}\frac{d\tilde\alpha_4}{2}\frac{d\tilde\alpha_1}{2}\mathbb{F}_{\alpha_2'\tilde\alpha_2}\begin{bmatrix}
    \alpha_B & \tilde\alpha_3 \\
    \alpha_A & \tilde\alpha_1
    \end{bmatrix}
\mathbb{F}_{\alpha_3'\tilde\alpha_3}\begin{bmatrix}
    \alpha_A & \tilde\alpha_4 \\
    \alpha_2' & \tilde\alpha_1
    \end{bmatrix}\mathbb{F}_{\alpha_4'\tilde\alpha_4}\begin{bmatrix}
    \alpha_B & \tilde\alpha_1\\
    \alpha_3'& \tilde\alpha_1 
    \end{bmatrix}\mathbb{S}_{\alpha_1'\tilde\alpha_1}[\alpha_4']\mathcal{F}_N^{OTO}}\,.
\end{split}
\end{align}
\noindent
Now in the limit, $\alpha_3'=\alpha_2'=\alpha_B$ and $\alpha_A= \epsilon\rightarrow0$ we get,\footnote{To get a simplified expression for OTOC in the OPE channel, we took the $\alpha_A\to 0$ limit which should be contrasted with the one in (\ref{5.19u}) for plane. }
\begin{align}
\begin{split}&\mathbb{F}_{\alpha_3'\tilde\alpha_3}\begin{bmatrix}
    \epsilon & \tilde\alpha_4 \\
    \alpha_2' & \tilde\alpha_1
\end{bmatrix}=\delta(\tilde\alpha_3-\alpha_2')+\mathcal{O}(\epsilon)\,,\\&
\mathbb{F}_{\alpha_2'\tilde\alpha_2}\begin{bmatrix}
   \alpha_B & \tilde\alpha_3 \\
   \epsilon & \tilde\alpha_1
\end{bmatrix}=\delta(\tilde\alpha_2-\alpha_B)+\mathcal{O}(\epsilon)\,.\end{split}
\end{align}
The transformations are schematically sketched in Fig.~\eqref{Fig2}.
Now considering $\alpha_4',\alpha_1'\rightarrow 0$, the integral simplifies to,

\begingroup
\begin{align}
\begin{split} \mathcal{F}_{OPE}&=\int \frac{d\tilde\alpha_2}{2}\frac{d\tilde\alpha_3}{2}\frac{d\tilde\alpha_4}{2}\frac{d\tilde\alpha_1}{2}
\delta(\tilde\alpha_3-\alpha_2')\delta(\tilde\alpha_2-\alpha_B)C_{{NS}}(\tilde\alpha_4,\tilde\alpha_1,\alpha_B)\frac{W_{NS}(Q)W_{NS}(\tilde\alpha_4)}{\pi W_{NS}(Q-\tilde\alpha_1)W_{NS}(Q-\alpha_B)}\\&\hspace{9 cm}\times
\mathbb{S}_{\alpha_1'\tilde\alpha_1}[\alpha_4']\textcolor{black}{\mathcal{F}_N^{OTO}}\,,\\&
=\int \frac{d\tilde\alpha_4}{2}\frac{d\tilde\alpha_1}{2}C_{{NS}}(\tilde\alpha_4,\tilde\alpha_1,\alpha_B)\frac{W_{NS}(Q)W_{NS}(\tilde\alpha_4)}{\pi W_{NS}(Q-\tilde\alpha_1)W_{NS}(Q-\alpha_B)}
\mathbb{S}_{0\tilde\alpha_1}[0]\times \mathcal{F}_N^{OTO}\Big|_{\tilde\alpha_3=\alpha_2',\tilde\alpha_2=\alpha_B}+(\cdots)
\label{6.17 m}
\end{split}
\end{align}
\endgroup
where in (\ref{6.17 m}) we define ,
\begin{align}
\begin{split}
&W_{NS}(\alpha)=\frac{2(\pi\mu\textcolor{black}{\gamma(\frac{bQ}{2}))}^{-\frac{Q-2\alpha}{2b}}\pi(\alpha-Q/2)}{\Gamma(1+b(\alpha-Q/2))\Gamma(1+\frac{1}{b}(\alpha-Q/2))}\label{5.31y}
\end{split}
\end{align}
where $\gamma(x)=\frac{\Gamma(x)}{\Gamma(1-x)}$, and the structure constant is given by \footnote{We thank Junkai Wang for pointing out a typo.},
\begingroup
\begin{align}
\begin{split}
C_{NS}(\alpha_1,\alpha_2,\alpha_3)&=\lambda^{(Q-\sum_{i=1}^3\alpha_i)/b}\\&\times\frac{\Upsilon'_{NS}(0)\Upsilon_{NS}(2\alpha_1)\Upsilon_{NS}(2\alpha_2)\Upsilon_{NS}(2\alpha_3)}{\Upsilon_{NS}(\alpha_1+\alpha_2+\alpha_3-Q)\Upsilon_{NS}(\alpha_1+\alpha_2-\alpha_3)\Upsilon_{NS}(-\alpha_1+\alpha_2+\alpha_3)\Upsilon_{NS}(\alpha_3+\alpha_1-\alpha_2)}
\end{split}
\end{align}
\endgroup
with  $\lambda=\pi \mu \gamma(bQ/2)b^{1-b^2}\,.$ $\mathcal{F}^{OTO}_N$ is the out-of-time-order conformal block. Also we need the following  modular-$\mathbb{S}$ transformation matrix elements,
\begin{align}
&\mathbb{S}_{0,\alpha_1}[0]=\textrm{sin}(\pi b^{-1}(\alpha_1-Q/2))\textrm{sin}(\pi b(\alpha_1-Q/2))\,,\\&
\mathbb{S}_{0,\alpha_3}[0]=\textrm{sin}(\pi b^{-1}(\alpha_3-Q/2))\textrm{sin}(\pi b(\alpha_3-Q/2))\,.
\end{align}

\section*{Behaviour of OTOC with time:}

Finally, we have all the ingredients to analyze the behaviour of OTOC as a function of time. To do that, we mainly focus on the bosonic part ($\phi_a$) of the super-Virasoro primary operators (NS sector) as defined in (\ref{2.25nm}), leaving the more general study involving a fermionic part, as defined in (\ref{2.25nnm}), for future work. The bare $n$-point necklace channel torus block for this scalar primary is given by \cite{Gerbershagen:2021yma},\footnote{$\tilde{\Delta}_i$ and ${\Delta}_i$ are intermediate and external conformal dimensions, respectively. Exponentiation of the block is not used. }
\begingroup
\begin{align}
\begin{split}
\mathcal{V}^h_{\Delta}(\rho)&=\, \prod_{i=1}^{n}\rho_i^{\tilde\Delta_i}\,\,\\&\times{\bf F_{N}}\begin{bmatrix}
\tilde\Delta_1+\tilde\Delta_2-\Delta_1&\tilde \Delta_2+\tilde \Delta_3-\Delta_2&\cdots &\,\,\,\tilde\Delta_n+\tilde\Delta_1-\Delta_n  \\&\hspace{1cm} 2\tilde \Delta_1,\,2\tilde \Delta_2,\,2\tilde \Delta_3,\cdots,2\tilde{\Delta}_n
\end{bmatrix}(\rho_1,\rho_2,\cdots ,\rho_n)
\end{split}
\end{align}
\endgroup
where the cross-ratios on the torus are defined as,
\begingroup
\begin{align}
\begin{split}
    &\rho_1=\frac{q z_{12}z_{n-1,n}}{(z_{n-1}-qz_1)(z_n-q z_2)}=q\chi_{1},\hspace{2.5 cm}\rho_2=\frac{z_{23}(z_n-qz_1)}{z_{13}(z_{n}-qz_2)}=1-\chi_{1}\,,\\&
\rho_n=\frac{z_{n-2,n-1}(z_n-qz_1)}{z_{n-2,n}(z_{n-1}-qz_1)}=1-\chi_{1}, \hspace{2.5 cm}\rho_i=\frac{z_{i-2,i-1}z_{i,i+1}}{z_{i-2,i}z_{i-1,i+1}}=\chi_{1}\,,\hspace{1 cm} \mbox{for} \,\,\,\, 3\leq i\leq n-1\,.
    \end{split}
\end{align}
\endgroup
The above cross-ratios are obtained for the configuration $z_1=0 , z_2 =\chi_{1}, z_3=1 ,z_4=\infty$.
The monodromy ($\mathbb{M}_1$) around $\chi_{1}=1$ is given by\footnote{For our case, the $_2F_1$ has branch cut in $z\in [1,\infty)\,.$},
\begingroup
\begin{equation}
\mathbb{M}_{1}^{T}
\!\left[\mathcal{V}_{\Delta}^{h}\right]
\underset{}{\sim}
(q\chi_1)^{\widetilde{\Delta}_{1}}
\chi_1^{\widetilde{\Delta}_{3}}\mathcal{N}
\label{new}
\end{equation}\\
where, the coefficients are given $\mathcal{N}$ is given in \eqref{C.14t}. 
The leading behaviour with respect to time can be easily extracted by substituting\footnote{Here $\chi_{1}$ means cross ratio in the super-plane and ultimately $(q\chi_1)^{\widetilde{\Delta}_{1}}$ governs the leading time dependence of the OTOC.} \cite{Kudler-Flam:2019kxq},
\begin{align}\tau=i\,\frac{K_1(1-\chi_{1})}{K_1(\chi_{1})},~~~~~ \chi_{1}=\left(\frac{\Theta_{2}(e^{i\pi \tau})}{\Theta_{3}(e^{i\pi \tau})}\right)^4.
\end{align} in \eqref{5.19u}. $ K_1(z)$ is the elliptic integral of the first kind. The monodromy around $\chi_{1}=1$ is mapped to torus modular parameter via ,
\begin{align}
\tau \to \frac{\tau}{1+2\tau},~~~~ \bar\tau \to \bar\tau.
\end{align}
The behaviour of the holomorphic part of (non-normalized) OTOC in time is given by\footnote{We used the conformal transformation to obtain the OTOC as a function of time from the cross-ratio space $\chi_{1}$. },
\begin{align}
\textrm{OTOC}\xrightarrow{\chi_{1}\sim 0} \frac{(-i e^{-\frac{2 \pi  t}{\beta }})^{\frac{Q^2}{4}} \Bigg(i \exp \Big[-\frac{2 \pi \Big (t \,K_1(-i e^{-\frac{2 \pi  t}{\beta }})+(\beta +2 i t) \,K_1(1+i e^{-\frac{2 \pi  t}{\beta }})\Big)}{\beta  \Big(K_1(-i e^{-\frac{2 \pi  t}{\beta }})+2 i\, K_1(1+i \,e^{-\frac{2 \pi  t}{\beta }})\Big)}\Big]\Bigg)^{\tilde\Delta_1}}{\Big(-\log (-i e^{-\frac{2 \pi  t}{\beta }})\Big)^{3/2}}. \label{OTOC}
\end{align}
\begin{figure}[htb!]
    \centering
    \scalebox{0.40}{\includegraphics{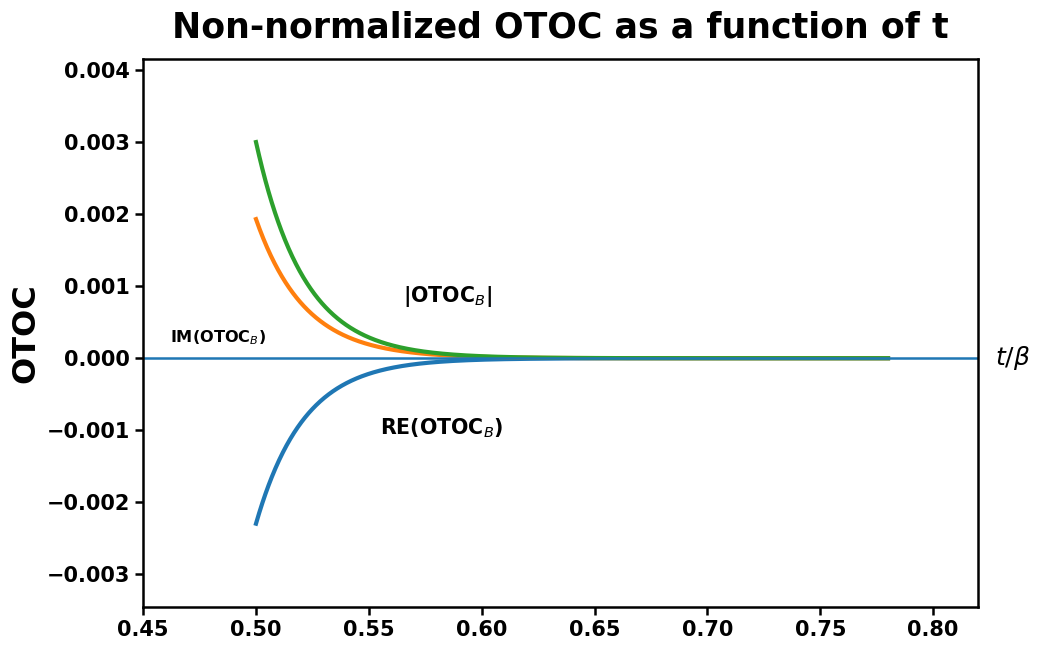}}
\caption{\textcolor{black}{Figure depicting the plot of non-normalized bosonic part (upto an overall time-independent rescaling factor) of the out-of-time-order correlator (in necklace channel) for the case of super-Liouville OTOC (for $Q=3.5\,,\tilde \Delta_1=4$ and $\beta=10$). The \textcolor{black}{{\bf blue curve}} depicts the real part of OTOC, and the {\bf orange curve} denotes the imaginary part of the OTOC. {\bf Green curve} denotes the absolute value of the OTOC.  Both the real and imaginary parts approach zero at late time.}}
    \label{fig:sgp}
\end{figure}

\textcolor{black}{It's evident from (non-normalized) (\ref{OTOC}) that the OTOC is not of the single exponential form, i.e., it is (the real part of it) is not of the form $\alpha-\beta e^{\lambda\,t}\,.$ Fig.~(\ref{fig:sgp}) shows the plot of the real, imaginary, and absolute value of the non-normalized OTOC.\footnote{In Fig.~(\ref{fig:sgp}), the subscript $B$ denotes the bosonic part and the expression in (\ref{OTOC}) has been multiplied by an overall real, time-independent factor.} This is distinct from what is known for the Schwarzian theory dual to JT gravity in the leading semiclassical regime \cite{Maldacena:2016hyu}. Further studies are required, including subleading corrections from the geometry, to better understand its implications for quantum chaos.}


\section{Conclusion and Outlook}
\label{sec 6}
Mainly motivated by the aspects of Topological QFT and Gravity, in this paper, we discuss how to compute the partition function of $\mathcal{N}=1$ super-Liouville theory using super-Virasoro TQFT. Due to different spin structures related to the super-torus, the bosonic or fermionic fields may or may not have zero modes. Correspondingly, the path integral and the modular sum change. Apart from calculating partition functions for different topologies \textcolor{black}{with sphere and torus boundaries}, we also compute the OTOC on the torus. As a probe matter, we include super-Liouville primaries in the boundary of the torus. Below, we summarize the main findings of the paper,

\begin{itemize}
\item  First, we show that the claim of \cite{Eberhardt:2023mrq}, that the gravitational partition function is equal to the sum of the modulus square of the Virasoro partition function from the {\bf Teichmuller} $\times$ {\bf Teichmuller} correspondence also holds for the $\mathcal{N}=1$ super-Liouville theory. We claimed and showed that the (ungraded) inner product of the superconformal blocks with two different Liouville momenta satisfies delta function type normalization. The bulk mapping class group (MCG) becomes a subset of the boundary mapping class group for Riemann surfaces with boundaries. Due to this, we needed to perform an extra modular sum to compute the gravitational partition function. The modular sum depends on the spin structure corresponding to the different boundary conditions. 

\item Second, we compute the partition function of supergravity by using the super-Virasoro TQFT. \textit{Apart from the spherical boundary, we also consider torus wormholes. Different spin structures lead to different partition functions.} Odd spin structures change the boundary condition so that the zero modes contribute, and we also had to sum over odd modular parameter transformations. 

\item Eventually, we calculate the OTOC on the super-torus with super-Liouville matter insertions (restricting to NS insertions). For this purpose, we first compute the necklace channel out-of-time-order conformal block in the large charge ($c\rightarrow \infty$) limit. In this semiclassical limit, one can see that the conformal blocks are exponentiated. One should be able to verify this using the monodromy method. However, in the torus, it is a bit tedious to calculate the exponential function using superconformal Ward identities on the super-torus. Hence, we use braiding transformation to compute the out-of-time order necklace channel conformal block from the time-ordered one. Then, we find a simplied expression for the OPE channel conformal block using fusion and modular transformations on the torus. \textcolor{black}{ As the leading order behaviour of the super-torus four-point conformal blocks in $\chi_1\to 0$ limit is uniform i.e $\chi_1^{\Delta_s}$ for all the pieces of super-blocks, we consider only that for the computation of the saddle point integral for super-torus OTOC in necklace channel \cite{Belavin:2024mzd}. We leave the more general analysis for future studies. }

\item \textcolor{black}{The result of the off-shell torus wormhole partition function in VTQFT differs from the result obtained in \cite{Cotler:2020ugk}, and the difference lies in the fact of the bulk mapping class group. When we take two trumpets (with one asymptotic $T^2$ boundary and one geodesic $T^2$ boundary) and try to glue them along some bulk torus slice, one needs to perform a non-trivial gauging of the bulk mapping class group (corresponding to the twist angle upto $2\pi$). This makes the inner product different from what is achieved in VTQFT and makes the inner product non-orthogonal. One can verify this by calculating the $\langle \rho \rho\rangle$ correlator (using ensemble averaging) and finding the Vandermonde determinant to compute the overlap between the states of different Liouville momenta as shown in \cite{Jafferis:2024jkb}. In Virasoro or super-Virasoro TQFT, the problem is universal because of the orthogonality between Liouville momenta states. The torus wormhole itself is an off-shell geometry.
 Now other off-shell geometries ($\Sigma_{g,n}\times S^1$) can also be calculated in $\mathcal{N}=1$ supergravity using super-Virasoro TQFT or Index theorems following \cite{Eberhardt:2022wlc}, and a detailed investigation of that is indeed important, which we leave for future investigations.}

\item \textcolor{black}{It is worth investigating the nature of OTOC after including higher genus Riemann surfaces in the boundary. Though the corrections are expected to be subleading, one should still compute them explicitly to figure out any notable changes in the nature of the OTOC.} 

\end{itemize}

\noindent
Now, we end by discussing some possible future directions. One can, in principle, generalize this study for $\mathcal{N}=2$ super-Liouville theory and test whether the normalization (\ref{eqclaim}) of the conformal block holds. However, one has to first study the fusion and modular transformation matrices for $\mathcal{N}=2$ as they are not well-known.   Furthermore, one should also consider non-Gaussian contractions for multi-boundary OPE statistics as well as test the \textit{Eigenstate-Thermalization Hypothesis} (ETH). Another extension is to calculate the Handlebody partition function for the super-Liouville case generalizing \cite{Eberhardt:2023mrq} as well as to incorporate grading structure inside the inner product for multi-boundary geometries. Currently, there has been a resurgence of studying field theories on null manifolds with Carrollian symmetry \cite{Bagchi:2019xfx,Bagchi:2019clu} and its supersymmetric extensions \cite{Bagchi:2022owq}. It will be interesting to generalize the computations presented in Sec.~\eqref{sec 4} and Sec.~\eqref{sec 5} for BMS$_3$ super-torus generalizing \cite{Bagchi:2020rwb}.
\section*{Acknowledgement}
 AB thanks the speakers and participants of the workshop “Quantum Information in QFT and AdS/CFT-III" organized at IIT Hyderabad between 16-18th September,
2022 and funded by SERB through a Seminar Symposia (SSY) grant (SSY/2022/000446),
“Quantum Information Theory in Quantum Field Theory and Cosmology" held in 4-9th
June, 2023 hosted by Banff International Research Centre at Canada 
AB would also like to thank the Department of Physics of BITS Pilani, Goa Campus, for hospitality during the course of this work. S.G (PMRF ID: 1702711) and S.P (PMRF ID: 1703278)
are supported by the Prime Minister’s Research Fellowship of the Government of India. S.G would like to thank Prof. Jan Plefka at Humboldt University, Prof. Niels Emil Bjerrum-Bohr at Niels Bohr Institute and Prof. Rafael Porto at DESY for their kind hospitality during the course of the work.  PN is supported by a Fulbright-Nehru Postdoctoral Research Fellowship. She would like to thank Vijay Balasubramanian and the hospitality of the University of Pennsylvania during the course of this work.  AB is supported by the Core Research Grant
(CRG/2023/ 001120) and Mathematical Research Impact Centric Support Grant (MTR/2021/000490) by the Department of Science and Technology Science 
and Engineering Research Board (India), India 
AB also acknowledges the associateship program of the Indian Academy of Science, Bengaluru. 
\appendix
\section{Sketching the derivation of the necklace channel conformal block} 
\label{App}
\begin{align}
\mathcal{F}^{\{s\}}= {\bf \mathcal{A}}_c^{\Delta_{1,2,3,4},\tilde \Delta_{1,2,3,4}}(z_{1,2,3,4})+q^{1/2} {\bf \mathcal{B}}_c^{\Delta_{1,2,3,4},\tilde \Delta_{1,2,3,4}}(z_{1,2,3,4})+\mathcal{O}(q),
\label{A.1 m}\end{align}
The coefficients can be defined as ,
\begin{align}
\begin{split}\mathcal{A}_c&=\sum_{m,r,g=0}^\infty\sum_{\substack{|S|+|J|=|T|+|P|=m\\|O|+|Y|=|U|+|V|=r\\|E|+|W|=|C|+|X|=g }}   \frac{\langle \tilde \Delta_1|\Phi_1(Z_1)|SJ,\tilde{\Delta}_2\rangle}{\langle \tilde \Delta_1|\Phi_1(Z_1)|\tilde{\Delta}_2\rangle} [B^2_{\Delta }]^{SJ,TP} \frac{\langle \tilde \Delta_2,TP|\Phi_2(Z_2)|OY,\tilde{\Delta}_3\rangle}{\langle \tilde \Delta_2|\Phi_2(Z_2)|\tilde{\Delta}_3\rangle}\\&\times [B^3_{\Delta }]^{OY,UV} \frac{\langle \tilde \Delta_3,UV|\Phi_3(Z_3)|EW,\tilde{\Delta}_4\rangle}{\langle \tilde \Delta_3|\Phi_3(Z_3)|\tilde{\Delta}_4\rangle}\,\,[B^4_{\Delta }]^{EW,CX} \frac{\langle \tilde \Delta_4,CX|\Phi_4(Z_4)|\tilde{\Delta}_1\rangle}{\langle \tilde \Delta_4|\Phi_4(Z_4)|\tilde{\Delta}_1\rangle}\label{A.2 c}
\end{split}
\end{align}
and similarly the other coefficient can be determined as ,

\begin{align}
\begin{split}\mathcal{B}_c&=\sum_{m,r,g=0}^\infty\sum_{\substack{|S|+|J|=|T|+|P|=m\\|O|+|Y|=|U|+|V|=r\\|E|+|W|=|C|+|X|=g }}   \frac{\langle \tilde \Delta_1|G_{\frac{1}{2}}\Phi_1(Z_1)|SJ,\tilde{\Delta}_2\rangle}{\langle \tilde \Delta_1|\Phi_1(Z_1)|\tilde{\Delta}_2\rangle} [B^2_{\Delta }]^{SJ,TP} \frac{\langle \tilde \Delta_2,TP|\Phi_2(Z_2)|OY,\tilde{\Delta}_3\rangle}{\langle \tilde \Delta_2|\Phi_2(Z_2)|\tilde{\Delta}_3\rangle}\\&\times [B^3_{\Delta }]^{OY,UV} \frac{\langle \tilde \Delta_3,UV|\Phi_3(Z_3)|EW,\tilde{\Delta}_4\rangle}{\langle \tilde \Delta_3|\Phi_3(Z_3)|\tilde{\Delta}_4\rangle}\,\,[B^4_{\Delta }]^{EW,CX} \frac{\langle \tilde \Delta_4,CX|\Phi_4(Z_4)G_{-\frac{1}{2}}|\tilde{\Delta}_1\rangle}{\langle \tilde \Delta_4|\Phi_4(Z_4)|\tilde{\Delta}_1\rangle}\,.\label{A.3 d}
\end{split}
\end{align}
Using (\ref{B.13}) one achieves,
\begingroup
\begin{align}
   \begin{split} 
\frac{\langle \tilde \Delta_2,TP|\Phi(z,\theta)|OY,\tilde{\Delta}_3\rangle}{\langle \tilde \Delta_2|\Phi(z,\theta)|\tilde{\Delta}_3\rangle}\sim\frac{\langle\tilde \Delta_2|G_{\frac{1}{2}}\Phi(z,\theta)G_{-\frac{1}{2}}|\tilde\Delta_3\rangle}{C_{\Phi_1\Phi_2\Phi_3}}=&\frac{(\tilde \Delta_3+\tilde \Delta_2-\Delta)\,C_{\Phi\phi\Phi}\,z^{\tilde \Delta_2-\Delta-\tilde \Delta_3}}{C_{\Phi_1\Phi_2\Phi_3}}+\\&\frac{(-\tilde \Delta_2-\tilde \Delta_3-\Delta+1/2)\,C_{\Phi\psi\Phi}\,\theta\, z^{\tilde \Delta_2-\Delta-1/2-\tilde \Delta_3}}{C_{\Phi_1\Phi_2\Phi_3}}\,.
\end{split}
\end{align}
\endgroup
\textbullet
{\bf Non-vanishing structure constants:}
\begin{align}
C_{\Phi_1\Phi_2\Phi_3}=\Big( \pi \mu \, \gamma\Big(\frac{Q b}{2}\Big)b^{1-b^2}\Big)^{\frac{Q-a}{b}}\frac{\Upsilon'_{NS}(0)\Upsilon_{NS}(2a_1)\Upsilon_{NS}(2a_2)\Upsilon_{NS}(2a_3)}{\Upsilon_{NS}(a-Q)\Upsilon_{NS}(a_{1+2-3})\Upsilon_{NS}(a_{3+2-1})\Upsilon_{NS}(a_{3+1-2})}\,, \label{eq5}
\end{align}

\begin{align}
\tilde C_{\Phi_1\Phi_2\Phi_3}=\Big( \pi \mu \, \gamma\Big(\frac{Q b}{2}\Big)b^{1-b^2}\Big)^{\frac{Q-a}{b}}\frac{2i \Upsilon'_{NS}(0)\Upsilon_{NS}(2a_1)\Upsilon_{NS}(2a_2)\Upsilon_{NS}(2a_3)}{\Upsilon_{R}(a-Q)\Upsilon_{R}(a_{1+2-3})\Upsilon_{R}(a_{3+2-1})\Upsilon_{R}(a_{3+1-2})} \label{eq6}
\end{align}
where $\Upsilon_{NS}$and $\Upsilon_{R}$ are defined in (\ref{8.8p}).
\noindent
Now we also have,
\begin{align}C_{\Phi_1\psi_2\Phi_3}=x_{31}\tilde C_{\Phi_1\Phi_2\Phi_3}\hspace{2cm}C_{\Phi_1\psi_2\psi_3}=-\frac{\Delta_2+\Delta_3-\Delta_1}{x_{23}} C_{\Phi_1\Phi_2\Phi_3}\,.\end{align}
\section*{NS algebra and Superfields}

The mode expansion of the superfield of conformal dimension $\Delta $ is given by,
\begin{align}
\Phi_{\Delta}(Z)=\sum_{m\in \mathbb{N}_0}z^{-m-\Delta}\Phi_m+\sum_{2m\in\mathbb{N}_0 }z^{-m-1/2-\Delta}\theta\,\Phi_{m}\label{B.1m}
\end{align}
where $\mathbb{N}_0$ denotes non-negative integers and Z denotes the super-coordinate $Z\equiv Z(z,\theta)$.The OPE of G(z) with the superfield $\Phi$ is given by\footnote{In the subsequent calculation, the `$SL$' subscript in labelling the super-primary is implicit. },

$$ G(z)\Phi_{NS}(0,0)=\sum_{k\in \mathbb{Z}+1/2}z^{k-3/2}G_{-k}\Phi_{NS}(0,0)\,.$$
\noindent
Again the OPE with Virasoro generators $L_n$ is defined  as,
$$T(z)\Phi_{NS}=\sum_{k\in \mathbb{Z}}z^{n-2}L_{-n}\Phi_{NS}(0,0)\,.$$
\noindent
The generator algebra takes the following form,

\begin{align}
\begin{split}
[L_m,L_n]&=(m-n)L_{m+n}+\frac{c}{12}m(m^2-1)\delta_{m+n,0}\,,\\&
[L_m,G_k]=\frac{m-2k}{2}G_{m+k}\,,\\&
\{G_k,G_l\}=2L_{k+l}+\frac{c}{3}(k^2-\frac{1}{4})\delta_{k+l,0}\,.
\end{split}
\end{align}
\noindent
The superfields satisfy the algebra,
\begin{align}
\begin{split}&[L_n,\Phi_{\Delta,\bar\Delta}(z,\theta,\bar z ,\bar \theta)]=z^n[z\partial_z+(n+1)(\Delta+\frac{1}{2}\theta \partial_{\theta})]\Phi_{\Delta,\bar \Delta}\,,\\&
[\bar L_n,\Phi_{\Delta,\bar\Delta}(z,\theta,\bar z ,\bar \theta)]=\bar z^n[\bar z\partial_{\bar z}+(n+1)(\bar \Delta+\frac{1}{2}\bar \theta \partial_{\bar \theta})]\Phi_{\Delta,\bar \Delta}\,,\\&
[G_k,\Phi_{\Delta,\bar\Delta}(z,\theta,\bar z ,\bar \theta)]=z^{k-1/2}[z\partial_\theta-\theta z \partial_z -\theta(2k+1)\Delta]\Phi_{\Delta,\bar \Delta}\,,\\&
[\bar G_k,\Phi_{\Delta,\bar\Delta}(z,\theta,\bar z ,\bar \theta)]=\bar z^{k-1/2}[\bar z\partial_{\bar\theta}-\bar\theta \bar z \partial_{\bar z} -\bar\theta(2k+1)\bar \Delta]\Phi_{\Delta,\bar \Delta}\,.
\end{split}
\end{align}
\noindent
The three-point function of the super-primary fields is given by 
\begin{align}
\begin{split}
&\langle \Phi(z_3,\theta_3;\bar z_3,\bar \theta_3)\Phi(z_2,\theta_2;\bar z_2,\bar \theta_2)\Phi(z_1,\theta_1;\bar z_1,\bar \theta_1)\rangle=\\&\,Z_{32}^{\beta_1}Z_{31}^{\beta_{2}}Z_{21}^{\beta_{3}}\langle\Phi(\infty,0;\infty,0)\Phi(1,\Theta;1,\bar \Theta)\Phi(0,0;0,0)\rangle\label{B.4 m1}
\end{split}
\end{align}
where $\beta_i=\Delta_i-\Delta_j-\Delta_k, Z_{12}=z_1-z_2-\theta_1\theta_2, $ along with$$\Theta=\frac{1}{\sqrt{z_{12}z_{13}z_{23}}}\Big( \theta_1z_{23}+\theta_2z_{31}+\theta_3z_{12}-\frac{1}{2}\theta_1\theta_2\theta_3\Big)\,.$$ 
\\
\noindent
We need to compute the following quantity\footnote{We have taken the definition of the superfield a bit different from \eqref{2.25nm}. },
\begin{align}
    \begin{split}
        \langle\tilde \Delta_2|G_{\frac{1}{2}}\Phi(z)G_{-\frac{1}{2}}|\tilde\Delta_3\rangle\equiv   \langle\tilde \Delta_2|G_{\frac{1}{2}}(\phi(z)+\theta \psi(z))G_{-\frac{1}{2}}|\tilde\Delta_3\rangle\,.\label{B.5}
    \end{split}
\end{align}
To do this, we need to have the mode expansion of the scalar and the fermionic field as mentioned in \eqref{B.1m}. We have,
\begin{align}
    \begin{split}
        \psi_{\Delta+1/2}(z)=\sum_{m\in \mathbb Z+1/2}z^{-m-1/2-\Delta}\,\psi_m\,.\label{B.6m}
    \end{split}
\end{align}
The mode coefficient can be extracted using the Cauchy residue theorem and is given by,
\begin{align}
    \begin{split}
        \psi_m=\oint_0 \frac{dz}{2\pi i}\,z^{m+\Delta-1/2}\,\psi_{\Delta+1/2}(z)\,.\label{B.8a}
    \end{split}
\end{align}
Now, in \eqref{B.5}, let's first concentrate on the Fermionic part and do the algebra,
\begin{align}
     \langle\tilde \Delta_2|G_{\frac{1}{2}}\psi(z)G_{-\frac{1}{2}}|\tilde\Delta_3\rangle=\sum_{m\in \mathbb Z+1/2}z^{-m-1/2-\Delta}  \langle\tilde \Delta_2|G_{\frac{1}{2}}\psi_m G_{-\frac{1}{2}}|\tilde\Delta_3\rangle\,.
\end{align}
Our goal is to express the correlation function in terms of the three-point function $ \langle\tilde \Delta_2|\psi(z)|\tilde\Delta_3\rangle$. To do this, one has to know the following algebra $\{G_{r},\psi_m\}$ and $[G_{r},\phi_m]$.
\begin{align}
    \begin{split}
        \{G_{r},\psi_m\}&=\oint_0 \frac{dz}{2\pi i}\,z^{m+\Delta-1/2}\{G_r,\psi_{\Delta+q/2}(z)\}\,,\\ &
        =\oint_0 \frac{dz}{2\pi i}\,z^{m+\Delta+r-1}(z\partial_z+(2r+1)\Delta)\phi_{\Delta}(z)\,,\\ &
        =\oint_0 \frac{dz}{2\pi i}z^{m+r+\Delta-1}[-(m+r+\Delta)+(2r+1)\Delta]\phi_{\Delta}(z)\,,\\ &
        =[-(m+r+\Delta)+(2r+1)\Delta]\phi_{m+r}\,.
    \end{split}
\end{align} \\\\
Similarly, we can have the algebra for the second one,
\begin{align}
    \begin{split}
        [G_r,\phi_m]&=\oint_0 \frac{dz}{2\pi i}z^{m+\Delta-1}[G_r,\phi_\Delta(z)]\,,\\ &
        =\oint_0 \frac{dz}{2\pi i}z^{m+r+\Delta-1/2}\psi(z)=\psi_{m+r}\,.
    \end{split}
\end{align}
Now, \eqref{B.8a} can be evaluated as follows,
\begingroup\small
\begin{align}
    \begin{split}
         \langle\tilde \Delta_2|G_{\frac{1}{2}}\psi(z)G_{-\frac{1}{2}}|\tilde\Delta_3\rangle&=\sum_{m\in \mathbb Z+1/2}z^{-m-1/2-\Delta}  \langle\tilde \Delta_3|G_{\frac{1}{2}}\psi_{-m} G_{-\frac{1}{2}}|\tilde\Delta_2\rangle^*\,,\\ &
         =\sum_{m\in \mathbb Z+1/2}z^{-m-1/2-\Delta}  \langle\tilde \Delta_3|\big(\{G_{\frac{1}{2}},\psi_{-m}\}-\psi_{-m}G_{\frac{1}{2}}\big)G_{-\frac{1}{2}}|\tilde \Delta_2\rangle^*\,,\\ &
        = \sum_{m\in \mathbb Z+1/2}z^{-m-1/2-\Delta}  \langle\tilde \Delta_3|(m-1/2\textcolor{black}{+\Delta})\phi_{-m+1/2}G_{-\frac{1}{2}}-\psi_{-m}G_{\frac{1}{2}}G_{-\frac{1}{2}}|\tilde \Delta_2\rangle^*\,,\\ &
        =\sum_{m\in \mathbb Z+1/2}z^{\textcolor{black}{-m-1/2-\Delta}} (m-1/2+\Delta)\langle \tilde \Delta_3|\phi_{-m+1/2}G_{-\frac{1}{2}}|\tilde\Delta_2\rangle^*\\ &
        \hspace{2 cm}-\sum_{m\in \mathbb Z+1/2}z^{m-1/2+\Delta}\langle \tilde \Delta_3|\psi_{-m}2L_0|\tilde\Delta_2\rangle^*\,,\\&
       =(-\tilde \Delta_2-\tilde \Delta_3-\Delta+1/2)\langle \tilde \Delta_2|\psi_{\Delta+1/2}(z)|\tilde \Delta_3\rangle\,.
    \end{split}
\end{align}
\endgroup
In a similar fashion, one can compute the following correlator,
\begin{align}
    \begin{split}
       \langle\tilde \Delta_2|G_{\frac{1}{2}}\phi(z)G_{-\frac{1}{2}}|\tilde\Delta_3\rangle & =\sum_{m\in \mathbb Z}z^{-m-\Delta} \langle\tilde \Delta_3|G_{\frac{1}{2}}\phi_{-m}G_{-\frac{1}{2}}|\tilde \Delta_2\rangle^*\,,\\ &
       =\sum_{m\in \mathbb Z}z^{-m-\Delta}\langle \tilde \Delta_3|\psi_{-m+1/2}G_{-\frac{1}{2}}|\tilde \Delta_2\rangle ^*-2\sum_{m\in \mathbb Z}z^{-m-\Delta}\langle \tilde \Delta_3|\phi_{-m }L_0|\tilde \Delta_2\rangle^*\,,\\ &
       =(\tilde \Delta_3+\tilde \Delta_2-\Delta)\langle \tilde \Delta_2|\phi_{\Delta}(z)|\tilde \Delta_3\rangle\,.\label{B.13 r}
    \end{split}
\end{align}
Therefore the correlator involving the superfield $\Phi$ is given by,
\begingroup
\begin{align}
    \begin{split}
         \langle\tilde \Delta_2|G_{\frac{1}{2}}\Phi(z,\theta)G_{-\frac{1}{2}}|\tilde\Delta_3\rangle=(\tilde \Delta_3+\tilde \Delta_2-\Delta)\,\,z^{\tilde \Delta_2-\Delta-\tilde \Delta_3}+(-\tilde \Delta_2-\tilde \Delta_3-\Delta+1/2)\,C_{\Phi\psi\Phi}\,\theta\, z^{\tilde \Delta_2-\Delta-1/2-\tilde \Delta_3}\label{B.13}
    \end{split}
\end{align}
\endgroup
where the structure constants take the following non-vanishing contributions,
\begin{align}
    C_{\Phi\phi\Phi}=  C_{\phi\phi\phi}-\tilde\theta_2\tilde\theta_3 C_{\psi\phi\psi},\, C_{\Phi\psi\Phi}= \tilde\theta_2 C_{\psi\psi\phi}-\tilde \theta_3 C_{\phi\psi\psi}\,.
\end{align}
In the last line, we used the relation,
\begin{align}
\begin{split}\sum_m(\tilde\Delta_2-\tilde\Delta_3)\langle \tilde \Delta_2|\psi_m|\tilde \Delta_3\rangle z^{-m-\Delta-1/2}=-\sum_m \langle\tilde\Delta_2|m \psi_m |\tilde\Delta_3\rangle z^{-m-1/2-\Delta}
\end{split}
\end{align}
and 
\begin{align}
\begin{split}\sum_m(\tilde\Delta_2-\tilde\Delta_3)\langle \tilde \Delta_2|\phi_m|\tilde \Delta_3\rangle z^{-m-\Delta}=-\sum_m \langle\tilde\Delta_2|m \phi_m |\tilde\Delta_3\rangle z^{-m-\Delta}\,.
\end{split}
\end{align}
\noindent
In a similar way, we also have,
\textcolor{black}{ 
\begin{align}
    \begin{split}
&\langle\tilde\Delta_2|G_{\frac{1}{2}}\Phi(z)|\tilde\Delta_3\rangle=\sqrt{z}C_{\Phi \psi \Phi }+\theta (\tilde \Delta_2-\tilde\Delta_{3}+\Delta )C_{\Phi \phi \Phi }\,,\\&
\langle\tilde\Delta_2|\Phi(z)G_{-\frac{1}{2}}|\tilde\Delta_3\rangle=-C_{\Phi \psi \Phi }+\theta z^{-1}(\tilde \Delta_2-\tilde\Delta_{3}+\Delta )C_{\Phi \phi \Phi }\,.\label{B.17 m}
\end{split}
\end{align}
}
 We decompose the super-primary field into the Liouville primary field and the fermionic field. The non-vanishing matrix elements are given by (\ref{B.13 r}) and (\ref{B.17 m}). In principle, one should compute the matrix elements to all orders in $q$.
 \section*{Asymptotic formulas}
\textbullet $\,\,$ We define the Upsilon function as,
\begin{align}
    \Upsilon_b(x)=\frac{1}{\Gamma_b(x)\Gamma_b(Q-x)}\,.
\end{align}
This function is an entire function of x with zeroes located at $x=-mb-nb^{-1}$ and $x=Q+mb+nb^{-1}$ and it has the following integral representation,
\begin{align}
\log\Upsilon_b(x):=\int_{0}^{\infty}\frac{dt}{t}\Bigg[(Q/2-x)^2e^{-t}-\frac{\sinh^2((Q/2-x)t/2)}{\sinh(bt/2)\sinh(t/2b)}\Bigg]\,\,\,\,\,\,\,\,\, \text{for } 0<Re(x)<Q\,.
\end{align}
The asymptotics of the Upsilon function for large x in the upper half plane is given by\,\cite{Collier:2018exn},

\begin{align}
\begin{split}\log \Upsilon_b(x)\sim &x^2 \log(x)-\Bigg(3/2+\frac{i\pi}{2}\Bigg)x^2-Q x\,\log x+(1+\frac{i\pi}{2})\,Qx+(\frac{Q^2+1}{6})\,\log x-\\&-\frac{i\pi}{12}(Q^2+1)-2\log\Gamma_0(b)+\mathcal{O}(x^{-1})\,.\label{B.23 m}\end{split}\end{align}

In (\ref{B.23 m}) the function $\Gamma_0(b)$ independent of $x$ is given by,
\begin{align}
    \log\Gamma_0(b)=-\int_{0}^{\infty}\frac{dt}{t}\Bigg(\frac{e^{-Qt/2}}{(1-e^{-bt})(1-e^{-t/b})}-\frac{1}{t^2}-\frac{Q^2-2}{24}e^{-t}\Bigg)\,.
\end{align}\\\\
\textbullet \,The quantity $\Psi$ is defined below comes from the asymptotic expansions of $C_{NS}$.
\begin{align}
    {\bf \Psi}(\alpha_A,\alpha_B,Q,P)\equiv \frac{\Psi_1}{\Psi_2}\label{B.18 m}
\end{align}
where,
\begingroup
\begin{align}
    \begin{split}
      \Psi_1=&\Bigg(\alpha_A \alpha_B \text{$\Upsilon_1 $}(0) e^{-i \pi  (\frac{Q}{2}-i P) (\frac{Q}{2}+i P)} (\frac{1}{2} (-2 \alpha_A-Q)+Q) (Q-\alpha_A) (2 \alpha_A+Q) (\frac{1}{2} (-2 \alpha_B-Q)+Q) (Q-\alpha_B) (2 \alpha_B+Q) \\ &(C_1+\frac{1}{2} (Q^2+1) \log (\frac{Q}{2}+i P)-((\frac{3}{2}+\frac{i \pi }{2}) (\frac{Q}{2}+i P)^2)+(1+\frac{i \pi }{2}) Q (\frac{Q}{2}+i P)+(\frac{Q}{2}+i P)^2 \log (\frac{Q}{2}+i P)\\ &-Q (\frac{Q}{2}+i P) \log (\frac{Q}{2}+i P)-\frac{1}{12} i \pi  (Q^2+1)) (C_1+\frac{1}{2} (Q^2+1) \log (\frac{1}{2} (Q+2 (\frac{Q}{2}+i P)))-\frac{1}{4} (\frac{3}{2}+\frac{i \pi }{2})\\ & (Q+2 (\frac{Q}{2}+i P))^2+\frac{1}{2} (1+\frac{i \pi }{2}) Q (Q+2 (\frac{Q}{2}+i P))+\frac{1}{4} (Q+2 (\frac{Q}{2}+i P))^2 \log (\frac{1}{2} (Q+2 (\frac{Q}{2}+i P)))\\ &-\frac{1}{2} Q (Q+2 (\frac{Q}{2}+i P)) \log (\frac{1}{2} (Q+2 (\frac{Q}{2}+i P)))-\frac{1}{12} i \pi  (Q^2+1))\Bigg)\frac{W_{NS}(Q)W_{NS}(\alpha)}{\pi W_{NS}(Q-\alpha_A)W_{NS}(Q-\alpha_B)}\,,\nonumber
     \end{split}
     \end{align}
\endgroup
\begingroup
     \begin{align}
     \begin{split}
   \hspace{-4 cm}  \Psi_2&=4 C_2^8 (-\frac{1}{4} (\frac{3}{2}+\frac{i \pi }{2}) (i P+\frac{Q}{2}-\alpha_A-\alpha_B)^2+\frac{1}{4} \log (\frac{1}{2} (i P+\frac{Q}{2}-\alpha_A-\alpha_B)) (i P+\frac{Q}{2}-\alpha_A-\alpha_B)^2\\ &+\frac{1}{2} (1+\frac{i \pi }{2}) Q (i P+\frac{Q}{2}-\alpha_A-\alpha_B)-\frac{1}{2} Q \log (\frac{1}{2} (i P+\frac{Q}{2}-\alpha_A-\alpha_B)) (i P+\frac{Q}{2}-\alpha_A-\alpha_B)+C_1\\ &-\frac{1}{12} i \pi  (Q^2+1)+\frac{1}{2} (Q^2+1) \log (\frac{1}{2} (i P+\frac{Q}{2}-\alpha_A-\alpha_B))) \Big(-\frac{1}{4} (\frac{3}{2}+\frac{i \pi }{2}) (i P+\frac{3 Q}{2}-\alpha_A-\alpha_B)^2\\ &+\frac{1}{4} \log (\frac{1}{2} (i P+\frac{3 Q}{2}-\alpha_A-\alpha_B)) (i P+\frac{3 Q}{2}-\alpha_A-\alpha_B)^2+\frac{1}{2} (1+\frac{i \pi }{2}) Q (i P+\frac{3 Q}{2}-\alpha_A-\alpha_B)\\ &-\frac{1}{2} Q \log (\frac{1}{2} (i P+\frac{3 Q}{2}-\alpha_A-\alpha_B)) (i P+\frac{3 Q}{2}-\alpha_A-\alpha_B)+C_1-\frac{1}{12} i \pi  (Q^2+1)\\ &+\frac{1}{2} (Q^2+1) \log (\frac{1}{2} (i P+\frac{3 Q}{2}-\alpha_A-\alpha_B))\Big) (-\frac{1}{4} (\frac{3}{2}+\frac{i \pi }{2}) (i P+\frac{Q}{2}+\alpha_A-\alpha_B)^2\nonumber \end{split}
\end{align}
\begin{align}
     \begin{split}&+\frac{1}{4} \log (\frac{1}{2} (i P+\frac{Q}{2}+\alpha_A-\alpha_B)) (i P+\frac{Q}{2}+\alpha_A-\alpha_B)^2+\frac{1}{2} (1+\frac{i \pi }{2}) Q (i P+\frac{Q}{2}+\alpha_A-\alpha_B)\\ &-\frac{1}{2} Q \log (\frac{1}{2} (i P+\frac{Q}{2}+\alpha_A-\alpha_B)) (i P+\frac{Q}{2}+\alpha_A-\alpha_B)+C_1-\frac{1}{12} i \pi  (Q^2+1)\\ &+\frac{1}{2} (Q^2+1) \log (\frac{1}{2} (i P+\frac{Q}{2}+\alpha_A-\alpha_B))) (-\frac{1}{4} (\frac{3}{2}+\frac{i \pi }{2}) (i P+\frac{3 Q}{2}+\alpha_A-\alpha_B)^2\\ &+\frac{1}{4} \log (\frac{1}{2} (i P+\frac{3 Q}{2}+\alpha_A-\alpha_B)) (i P+\frac{3 Q}{2}+\alpha_A-\alpha_B)^2+\frac{1}{2} (1+\frac{i \pi }{2}) Q (i P+\frac{3 Q}{2}+\alpha_A-\alpha_B)\\ &-\frac{1}{2} Q \log (\frac{1}{2} (i P+\frac{3 Q}{2}+\alpha_A-\alpha_B)) (i P+\frac{3 Q}{2}+\alpha_A-\alpha_B)+C_1-\frac{1}{12} i \pi  (Q^2+1)\\ &+\frac{1}{2} (Q^2+1) \log (\frac{1}{2} (i P+\frac{3 Q}{2}+\alpha_A-\alpha_B))) (-\frac{1}{4} (\frac{3}{2}+\frac{i \pi }{2}) (-i P+\alpha_A+\alpha_B-\frac{Q}{2})^2\\ &+\frac{1}{4} \log (\frac{1}{2} (-i P+\alpha_A+\alpha_B-\frac{Q}{2})) (-i P+\alpha_A+\alpha_B-\frac{Q}{2})^2+\frac{1}{2} (1+\frac{i \pi }{2}) Q (-i P+\alpha_A+\alpha_B-\frac{Q}{2})\\ &-\frac{1}{2} Q \log (\frac{1}{2} (-i P+\alpha_A+\alpha_B-\frac{Q}{2})) (-i P+\alpha_A+\alpha_B-\frac{Q}{2})+C_1-\frac{1}{12} i \pi  (Q^2+1)+\frac{1}{2} (Q^2+1)\\ & \times \log (\frac{1}{2} (-i P+\alpha_A+\alpha_B-\frac{Q}{2}))) (-\frac{1}{4} (\frac{3}{2}+\frac{i \pi }{2}) (i P+\alpha_A+\alpha_B-\frac{Q}{2})^2+\frac{1}{4} \log (\frac{1}{2} (i P+\alpha_A+\alpha_B-\frac{Q}{2})) \\ & \times(i P+\alpha_A+\alpha_B-\frac{Q}{2})^2+\frac{1}{2} (1+\frac{i \pi }{2}) Q (i P+\alpha_A+\alpha_B-\frac{Q}{2})-\frac{1}{2} Q \log (\frac{1}{2} (i P+\alpha_A+\alpha_B-\frac{Q}{2})) \\ & \times(i P+\alpha_A+\alpha_B-\frac{Q}{2})+C_1-\frac{i \pi}{12}   (Q^2+1)+\frac{Q^2+1}{2}  \log (\frac{1}{2} (i P+\alpha_A+\alpha_B-\frac{Q}{2}))) (-\frac{1}{4} (\frac{3}{2}+\frac{i \pi }{2}) (-i P+\frac{Q}{2}+\alpha_A+\alpha_B)^2\\ &+\frac{1}{4} \log (\frac{1}{2} (-i P+\frac{Q}{2}+\alpha_A+\alpha_B))(-i P+\frac{Q}{2}+\alpha_A+\alpha_B)^2+\frac{1}{2} (1+\frac{i \pi }{2}) Q (-i P+\frac{Q}{2}+\alpha_A+\alpha_B)\\ &-\frac{1}{2} Q \log (\frac{1}{2} (-i P+\frac{Q}{2}+\alpha_A+\alpha_B)) (-i P+\frac{Q}{2}+\alpha_A+\alpha_B)+C_1-\frac{1}{12} i \pi  (Q^2+1)\\ &+\frac{1}{2} (Q^2+1) \log (\frac{1}{2} (-i P+\frac{Q}{2}+\alpha_A+\alpha_B))) (-\frac{1}{4} (\frac{3}{2}+\frac{i \pi }{2}) \times(i P+\frac{Q}{2}+\alpha_A+\alpha_B)^2\\ &+\frac{1}{4} \log (\frac{1}{2} (i P+\frac{Q}{2}+\alpha_A+\alpha_B)) (i P+\frac{Q}{2}+\alpha_A+\alpha_B)^2+\frac{1}{2} (1+\frac{i \pi }{2}) Q (i P+\frac{Q}{2}+\alpha_A+\alpha_B)\\ &-\frac{1}{2} Q \log (\frac{1}{2} (i P+\frac{Q}{2}+\alpha_A+\alpha_B)) (i P+\frac{Q}{2}+\alpha_A+\alpha_B)+C_1-\frac{1}{12} i \pi  (Q^2+1)+\frac{1}{2} (Q^2+1) \log (\frac{1}{2} (i P+\frac{Q}{2}+\alpha_A+\alpha_B)))\,.
    \end{split}
\end{align}
\endgroup
\section{Definition of different functions}
\label{AppB}
In this appendix, we will review some special functions that we heavily used in the main text.\\\\
\textbullet{\bf Barnes double gamma function}

For Re(x)>0 the function $\Gamma_b(x)$ is given by the integral representation,
\begin{align}
\log\Gamma_b(x)=\int \frac{dt}{t}\Big[\frac{e^{-xt}-e^{-Q/2 t}}{(1-e^{-tb})(1-e^{-t/b})} -\frac{\Big(Q/2-x\Big)^2}{2e^t} -\frac{\Big(Q/2-x\Big)}{t}       \Big]\,.
\end{align}
$\Gamma_b(x)$ is a special case of \textit{Barnes double gamma} $\Gamma_{\nu_1\nu_2}(x)$ with $\nu_1=\nu_2^{-1}=b$, being an analytic continuation of the function,
$$\log\Gamma_{\nu_1\nu_2}=\frac{\partial}{\partial s}\sum_{m,n=0}^{\infty}(m\nu_1+n\nu_2+z)^{-s}$$ to $s=0$, showing a self duality property,
\begin{align}
    \Gamma_b(x)=\Gamma_{b^{-1}}(x).
\end{align}
$\Gamma_b(x)$ satisfies also,
$$\Gamma_b(x+b)=\frac{\sqrt{2\pi }b^{bx-1/2}\Gamma_b(x)}{\Gamma(bx)}\,,$$
\begin{align}
S_{NS}(x)=S_b(\frac{x}{2})S_b(\frac{Q+x}{2})\,.
\end{align}
\textbullet 
{\bf Reflection Properties:}
\begin{align}
S_{NS}(x)S_{NS}(Q-x)=S_R(x)S_R(Q-x)=1\,.
\end{align}
\textbullet 
{\bf Location of the Zeros and Poles :}
\begin{align}
\begin{split}
&S_{NS}(x)=0 \iff x=Q+mb+nb^{-1},\,\,\,\,\,\, m,n\in \mathbb{Z}_{\geq0}\,,\,\,\,\,\,\,m+n\in 2\mathbb{Z}\\&
S_{R}(x)=0 \iff x=Q+mb+nb^{-1},\,\,\,\,\,\, m,n\in \mathbb{Z}_{\geq0}\,,\,\,\,\,\,\,m+n\in 2\mathbb{Z}+1\\&
S_{NS}^{-1}(x)=0 \iff x=-mb-nb^{-1},\,\,\,\,\,\, m,n\in \mathbb{Z}_{\geq0}\,,\,\,\,\,\,\,m+n\in 2\mathbb{Z}\\&
S_{R}^{-1}(x)=0 \iff x=-mb-nb^{-1},\,\,\,\,\,\, m,n\in \mathbb{Z}_{\geq0}\,,\,\,\,\,\,\,m+n\in 2\mathbb{Z}+1
\end{split}
\end{align}
\textbullet 
{\bf Basic Residue:}
\begin{align}\lim_{x\rightarrow 0}x\,S_{NS}(x)=\frac{1}{\pi}\,.
\end{align}
\noindent
\textbullet 
{\bf Few important functions:}
\begin{align}
  \frac{\Gamma_{NS}(2\alpha) }{\Gamma_{NS}(2\alpha-Q) }=W_{NS}(\alpha)\lambda^{\frac{Q-2\alpha}{2b}}\,.
\end{align}
Where,
\begin{align}
\begin{split}
&W_{NS}(\alpha)=\frac{2(\pi\mu\textcolor{black}{\gamma(\frac{bQ}{2}))}^{-\frac{Q-2\alpha}{2b}}\pi(\alpha-Q/2)}{\Gamma(1+b(\alpha-Q/2))\Gamma(1+\frac{1}{b}(\alpha-Q/2))}\label{5.31y}
\end{split}
\end{align}
\textbullet \, $\Gamma_{NS}(x)$ has a pole at 0, the behaviour near the pole is given by
\begin{align}
    \Gamma_{NS}(x)\xrightarrow[]{x\to 0} \frac{\Gamma_{NS}(Q)}{\pi x},\hspace{2 cm} \Upsilon'_{NS}(0)=\frac{\pi}{\Gamma^2_{NS}(Q)}.
\end{align}
\noindent
\textbullet{\bf Upsilon functions}:
\begin{align}
   \frac{ \Upsilon_{NS}(2x)}{\Upsilon_{NS}(2x-Q)}=\mathcal{G}_{NS}(x)\lambda^{-\frac{Q-2x}{b}},
\end{align}
where,
\begin{align}
    \mathcal{G}_{NS}(x)=\frac{W_{NS}(Q-x)}{W_{NS}(x)}.
\end{align}

\section{Torus two-point function (Liouville CFT) with general left and right moving temperature}\label{C.1m}

We first calculate the torus two-point function in 2d CFT where we have some intermediate left and right moving temperature $\beta_L$ and $\beta_R$, respectively.
We insert the two operators $\mathcal{O}_1$ and $\mathcal{O}_2$ in the \textit{Euclidean time circle}. To distinguish them from the insertion in the spatial circle, we use a \textit{tilde} symbol. Now we use the modular crossing kernels and Virasoro fusion kernel to calculate the torus two-point function in the following fashion Fig.~(\ref{fig4}),  considering $\alpha=\frac{Q}{2}+i P$ (where P is the Liouville Momenta),

\begin{figure}
\begin{center}
\scalebox{0.6}{\includegraphics{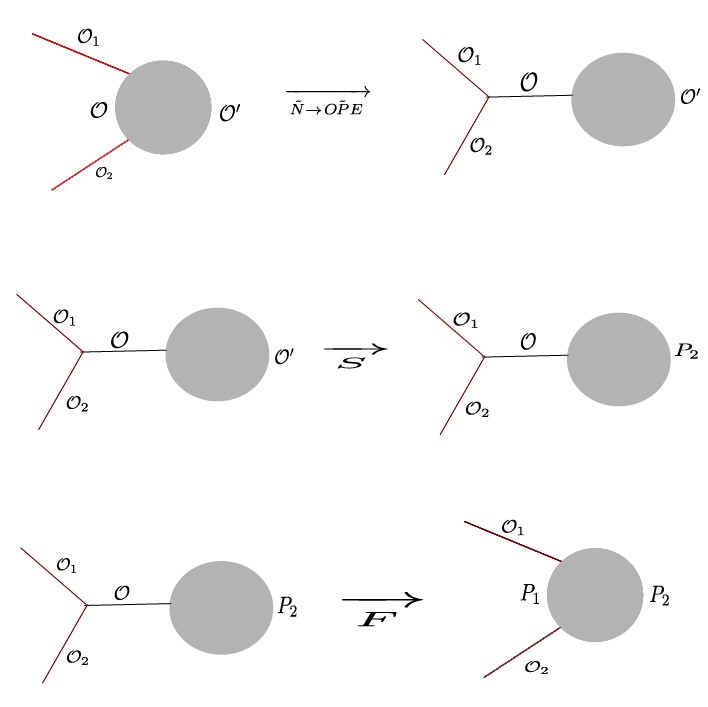}}
\caption{In the first step we transform from the $\tilde{N}$ channel to the $\widetilde{OPE}$ channel. Then we use the modular crossing transformation. At last, we use the Virasoro fusion kernel to write the OPE channel in terms of the Necklace channel.  }\label{fig4}
\end{center}
\end{figure}

\begin{align}
\begin{split}
\mathcal{F}^{\tilde{N}}_{h h'}&=\widetilde{\mathcal{F}^{OPE}_{P' P}}\,,\\&
=\int  \frac{dP_2}{2}  \frac{dP_1}{2} \mathbb{S}_{P' P_2}[P] \,\mathbb{F}_{P,P_1}\begin{bmatrix}
\tilde{P} & P_2\\
\tilde{P} & P_2
\end{bmatrix}\mathcal{F}^N_{P_1,P_2}\,.
\label{3.1p}
\end{split}  
\end{align}


\noindent
Now we know that the fusion kernel can be written in the following integral form \cite{Ghosh:2019rcj},
\begin{equation}
   \mathbb{F}_{P'', P}\begin{bmatrix}
P^* & \tilde{P} \\
P^* & \tilde{P}\end{bmatrix}=P(\alpha_i;\alpha'',\alpha)P(\alpha_i;Q-\alpha'',Q-\alpha)\int_{\mathcal{C}'} \frac{ds}{i}\prod_{k=1}^4\frac{\mathcal{S}_b(s+U_k)}{\mathcal{S}_b(s+V_k)}  
\end{equation} 
where, 
\begin{equation}
    P(\alpha_i;\alpha'',\alpha)=\frac{\Gamma_b^2(\alpha''+\alpha^*-\tilde{\alpha})\Gamma(\alpha''+Q-\tilde{\alpha}-\alpha^*)\Gamma(\alpha''-Q+\tilde{\alpha}+\alpha^*)\Gamma_b(2\alpha)}
{\Gamma_b^2(\alpha)\Gamma(\alpha+Q-2\tilde{\alpha})\Gamma(\alpha-Q+2\alpha^*)\Gamma_b(2\alpha-Q)}
\end{equation}
along with, 
\begin{equation}
    \mathcal{S}_b(x)=\frac{\Gamma_b(x)}{\Gamma_b(Q-x)}.
\end{equation}
 \noindent
\begin{align}
\begin{split}
&U_1=i(P_1-P_4)=0\,,\hspace{3 cm} V_1=Q/2+i(-P''+P^*-\tilde P)\,,\\&
U_2=-i(P_1+P_4)=-2i\tilde P \,,\hspace{2.1 cm} V_2=Q/2+i(P''+P^*-\tilde P)\,,\\&
U_3=i(P_2+P_3)=2iP^*\,,\hspace{2.3 cm} V_3=Q/2+iP\,,\\&
U_4=i(P_2-P_3)=0\,,\hspace{3 cm} V_4=Q/2-iP\,.
\end{split}
\end{align}
Various limits of the fusion kernel with heavy operator insertions are discussed in \cite{Collier:2019weq}.\\
Now we will proceed to show the formalism to compute the Virasoro torus two-point function. If we insert the operator in the dual Necklace channel (Euclidean time circle), this  can be written as,\newpage
\begin{align}
\begin{split}
 \hspace{-3 cm}\langle\mathcal{O}_1^s(v,\bar{v})\mathcal{O}_2^s(0)\rangle_{T^2(\tau,\bar{\tau})}
&=\begin{minipage}[h]{0.05\linewidth}
	\vspace{-3 pt}\scalebox{0.3}{\includegraphics{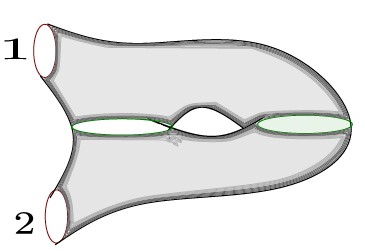}}\end{minipage}\\&
=\sum_{p,q}\begin{minipage}[h]{0.05\linewidth}
	\vspace{-9 pt}\scalebox{0.3}{\includegraphics{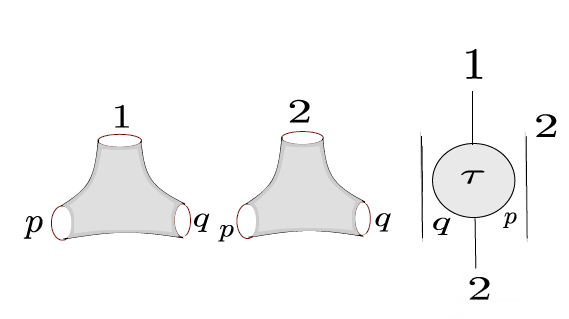}}\end{minipage}\\&
=\sum_{p,q}C_{\mathcal{O}_1\mathcal{O}_p\mathcal{O}_q}C_{\mathcal{O}_2\mathcal{O}_p\mathcal{O}_q}|\mathcal{F}^{g=1}_{12}(h_p,h_q;\tau,v)|^2\,.
\label{4.1w}
\end{split}
\end{align}
 \noindent
Considering the probe operators $\mathcal{O}_1=\mathcal{O}_2=\tilde{\mathcal{O}}$ to be heavy, we can write (\ref{4.1w}) as,
$$=\sum_{p,q}|C_{\tilde{\mathcal{O}}\mathcal{O}_p\mathcal{O}_q}|^2|\mathcal{F}^{g=1}_{12}(h_p,h_q;\tau,v)|^2\,.$$

\noindent
Now the average over the product of two torus  two-point function is given by \cite{Yan:2023rjh}:
\begin{align}
\begin{split}
&\overline{\langle\mathcal{O}_1(v,\bar{v})\mathcal{O}_2(0)\rangle_{T^2(\tau,\bar{\tau})}{\langle\mathcal{O}_1(v',\bar{v'})\mathcal{O}_2(0)\rangle_{T^2(\tau',\bar{\tau'})}}}\,,\\&=\sum_{p,q,g,h}\overline{|C_{\mathcal{O}_1\mathcal{O}_p\mathcal{O}_q}|^2|C_{\mathcal{O}_1\mathcal{O}_g\mathcal{O}_h}|^2}|\mathcal{F}^{\tilde{N},g=1}_{12}(h_p,h_q;\tau,v)|^2|\mathcal{F}^{\tilde{N},g=1}_{12}(h_g,h_h;\tau,v)|^2\,,\\&
=2\Bigg|\int dh_p dh_q \rho_0(h_p) \rho_0(h_q) C_0(\tilde{h},h_p,h_q)C_0(\bar{\tilde{h}},\bar{h}_p,\bar{h}_q)|\mathcal{F}^{\tilde{N},g=1}_{12}(h_p,h_q;\tau,v)|^2\Bigg|^2\label{5.5 m}
\end{split}
\end{align}
where $$\mathcal{F}^{\tilde{N},g=1}_{1 2}(h,h';\tau,v)=
\int  \frac{dP_2}{2}  \frac{dP_1}{2} \mathbb{S}_{P' P_2}[P] \,\mathbb{F}_{P,P_1}\begin{bmatrix}
\tilde{P} & P_2\\
\tilde{P} & P_2
\end{bmatrix}\mathcal{F}^N_{P_1,P_2}.$$
and 

\begin{align}
    C_0(P_1,P_2,P_3)= \frac{\Gamma_b(2Q)\Gamma_b(\frac{Q}{2} \pm i P_1\pm i P_2 \pm i P_3)}{\sqrt{2}\Gamma_b(Q)^3\prod_{k=1}^3\Gamma_b(Q\pm 2i P_k)}.
\end{align}
In (\ref{5.5 m}) the `bar' denotes the averaging over the dimensions of the heavy operators while the other conformal dimensions are held fixed. The Liouville structure constants show a universal behavior in this limit \cite{Collier:2018exn}.
\subsection*{Fusion matrix in super Liouville field theory}
We describe the fusion matrix in SLFT in this appendix.
Generally, the fusion matrices take the following form $\{i,j\}=1,2$ which correspond to parity of even and odd respectively ,
\begingroup
\begin{align}
    \begin{split}
&\mathbb{F}_{\alpha_s,\alpha_t}\left[\begin{array}{cc}
\alpha_3& \alpha_2\\
\alpha_4& \alpha_1 \end{array}\right]^i_j
={\Gamma_i(2Q-\alpha_t-\alpha_2-\alpha_3)\Gamma_i(Q-\alpha_t+\alpha_3-\alpha_2)\Gamma_i(Q+\alpha_t-\alpha_2-\alpha_3)
\Gamma_i(\alpha_3+\alpha_t-\alpha_2)\over
\Gamma_j(2Q-\alpha_1-\alpha_s-\alpha_2)\Gamma_{j}(Q-\alpha_s-\alpha_2+\alpha_1)\Gamma_j(Q-\alpha_1-\alpha_2+\alpha_s)
\Gamma_{j}(\alpha_s+\alpha_1-\alpha_2)}\\& \times{\Gamma_i(Q-\alpha_t-\alpha_1+\alpha_4)\Gamma_i(\alpha_1+\alpha_4-\alpha_t)\Gamma_i(\alpha_t+\alpha_4-\alpha_1)
\Gamma_i(\alpha_t+\alpha_1+\alpha_4-Q)\over
\Gamma_{j}(Q-\alpha_s-\alpha_3+\alpha_4)\Gamma_j(\alpha_3+\alpha_4-\alpha_s)\Gamma_{j}(\alpha_s+\alpha_4-\alpha_3)
\Gamma_j(\alpha_s+\alpha_3+\alpha_4-Q)}\\& \times {\Gamma_{\rm NS}(2Q-2\alpha_s)\Gamma_{\rm NS}(2\alpha_s)\over \Gamma_{\rm NS}(Q-2\alpha_t)\Gamma_{\rm NS}(2\alpha_t-Q)}
{1\over i}\int_{-i\infty}^{i\infty}d\tau J_{\alpha_s,\alpha_t}\left[\begin{array}{cc}
\alpha_3& \alpha_2\\
\alpha_4& \alpha_1 \end{array}\right]^i_j\, ,
\end{split}
\end{align} \label{eq4}
\endgroup
We will consider $i=j=1$, for the fusion matrix in NS sector. we have,
\begingroup
\begin{align}
    \begin{split}
&J_{\alpha_s,\alpha_t}\left[\begin{array}{cc}
\alpha_3& \alpha_2\\
\alpha_4& \alpha_1 \end{array}\right]^1_1
= {S_{NS}(Q+\tau-\alpha_1)S_{\rm NS}(\tau+\alpha_4+\alpha_2-\alpha_3)S_{NS}(\tau+\alpha_1)S_{NS}(\tau+\alpha_4+\alpha_2+\alpha_3-Q)\over
S_{NS}(Q+\tau+\alpha_4-\alpha_t)S_{NS}(\tau+\alpha_4+\alpha_t)S_{NS}(Q+\tau+\alpha_2-\alpha_s)S_{NS}(\tau+\alpha_2+\alpha_s)}\\& +{S_{R}(Q+\tau-\alpha_1)S_{R}(\tau+\alpha_4+\alpha_2-\alpha_3)S_{R}(\tau+\alpha_1)S_{R}(\tau+\alpha_4+\alpha_2+\alpha_3-Q)\over
S_{R}(Q+\tau+\alpha_4-\alpha_t)S_{R}(\tau+\alpha_4+\alpha_t)S_{R}(Q+\tau+\alpha_2-\alpha_s)S_{R}(\tau+\alpha_2+\alpha_s)}\, , \label{eq3}
\end{split}
\end{align}
\endgroup
where $S_{NS,R}$ are defined by,

\begingroup
\begin{align}
    S_{NS}(x)=\frac{\Gamma_{NS}(x)}{\Gamma_{NS}(Q-x)},\hspace{2 cm}S_{R}(x)=\frac{\Gamma_{R}(x)}{\Gamma_{R}(Q-x)} ~~~~~~\textrm{with}~~~~ \Gamma_{R}(x)=\Gamma_b(\frac{x+b}{2})\Gamma_b(\frac{x+b^{-1}}{2})
\end{align}
\endgroup

\noindent
Alternatively, it can be shown that for $ \alpha_s = 0$, the fusion matrix takes the following form \cite{Poghosyan:2016kvd},
\begin{align}
\begin{split}
\label{Fusionmatrixpole}
F_{0,\alpha_t}\left[\begin{array}{cc}
\alpha_3& \alpha_1\\
\alpha_3& \alpha_1 \end{array}\right]^1_1=C_{NS}(\alpha_t,\alpha_1,\alpha_3){W_{NS}(Q)W_{NS}(\alpha_t)
\over \pi W_{NS}(Q-\alpha_1)W_{NS}(Q-\alpha_3)}\, .
\end{split}
\end{align}
One can also verify that,
\begin{align}W_{NS}(x)W_{NS}(Q-x) = -4\sin\pi b(x-Q/2)\sin\pi {1\over b}(x-Q/2)\,.\end{align}
Also, the monodromy coefficient in (\ref{new}) in are,
\begin{align}
\begin{split}
&\mathcal{N}=\left[
e^{2\pi i(\widetilde{\Delta}_{2}+\widetilde{\Delta}_{4})}
C_{1}
(1-\chi_1)^{\widetilde{\Delta}_{2}+\widetilde{\Delta}_{4}}
+
e^{2\pi i(\Delta_{2}+\Delta_{3})}
C_{2}
(1-\chi_1)^{\Delta_{2}+\Delta_{3}}
\right]\,,\\&
C_{1}
=
\frac{
\Gamma\!\left(2\widetilde{\Delta}_3\right)
\Gamma\!\left(
\Delta_2+\Delta_3
-\widetilde{\Delta}_2-\widetilde{\Delta}_4
\right)
}{
\Gamma\!\left(
\Delta_2-\widetilde{\Delta}_2+\widetilde{\Delta}_3
\right)
\Gamma\!\left(
\widetilde{\Delta}_3+\Delta_3-\widetilde{\Delta}_4
\right)
}\,,\\&
C_{2}
=
\frac{
\Gamma\!\left(2\widetilde{\Delta}_3\right)
\Gamma\!\left(
\widetilde{\Delta}_2+\widetilde{\Delta}_4
-\Delta_2-\Delta_3
\right)
}{\Gamma\!\left(\widetilde{\Delta}_2+\widetilde{\Delta}_3-\Delta_2
\right)
\Gamma\!\left(
\widetilde{\Delta}_3+\widetilde{\Delta}_4-\Delta_3
\right)
}\\
&\times
F_{2}^{\mathrm{Appell}}
\left(
\widetilde{\Delta}_2+\widetilde{\Delta}_4-\Delta_2-\Delta_3;
\widetilde{\Delta}_1+\widetilde{\Delta}_2-\Delta_1,
\widetilde{\Delta}_4+\widetilde{\Delta}_1-\Delta_4;
\right.\\
&\hspace{4cm}\left.
2\widetilde{\Delta}_2,
2\widetilde{\Delta}_4;
1,1
\right)\,.\label{C.14t}
\end{split}
\end{align}

\section{Different transformations}
\label{appd}
We have the following basic moves \cite{Collier:2023fwi}: 
\begin{itemize} 
\item Braiding:
\begin{align}
     \label{eq:braiding}
    \begin{tikzpicture}[baseline={([yshift=-.5ex]current bounding box.center)}, scale=0.9]
        \draw[very thick, blue] (0,-1.2) to (0,-.2);
        \fill[blue] (0,-.2) circle (.07);
        \draw[very thick, blue, bend left=20] (0,-.2) to (-1,.8);
        \draw[very thick, blue, bend right=20] (0,-.2) to (1,.8);
        \node at (-.4,.4) {1};
        \node at (.4,.4) {2};
        \node at (.17,-.55) {3};
        \draw[very thick, out=270, in=90] (1.5,1) to (.5,-1);
        \draw[very thick, out=270, in=90] (-1.5,1) to (-.5,-1);
        \draw[very thick, out=270, in=270, looseness=1.5] (-.5,1) to (.5,1);
        \draw[very thick] (-1,1) circle (.5 and .2); 
        \draw[very thick] (1,1) circle (.5 and .2);
        \begin{scope}[shift={(0,-1)}, yscale=.4]
            \draw[very thick] (-.5,0) arc (-180:0:.5);
            \draw[dashed, very thick] (-.5,0) arc (180:0:.5);
        \end{scope}
    \end{tikzpicture}
    \ = \mathbb{B}_{12}^3 \ 
    \begin{tikzpicture}[baseline={([yshift=-.5ex]current bounding box.center)}, scale=0.9]
        \draw[very thick, blue] (0,-1.2) to (0,-.2);
        \fill[blue] (0,-.2) circle (.07);
        \draw[very thick, blue, bend left=20] (0,-.2) to (-1,.8);
        \draw[very thick, blue, bend left=20] (0,-.2) to (-1.1,.1);
        \draw[very thick, dashed, blue, bend left=30] (-1.1,.1) to (-.4,.7);
        \draw[very thick, blue] (-.4,.7) .. controls (0,.1) and (.5,.1) .. (1,.8);
        \node at (-.27,.2) {1};
        \node at (.6,.1) {2};
        \node at (.17,-.55) {3};
        \draw[very thick, out=270, in=90] (1.5,1) to (.5,-1);
        \draw[very thick, out=270, in=90] (-1.5,1) to (-.5,-1);
        \draw[very thick, out=270, in=270, looseness=1.5] (-.5,1) to (.5,1);
        \draw[very thick] (-1,1) circle (.5 and .2); 
        \draw[very thick] (1,1) circle (.5 and .2);
        \begin{scope}[shift={(0,-1)}, yscale=.4]
            \draw[very thick] (-.5,0) arc (-180:0:.5);
            \draw[dashed, very thick] (-.5,0) arc (180:0:.5);
        \end{scope}
    \end{tikzpicture}
\end{align} 
\item Fusion:
\begin{align}
    \begin{tikzpicture}[baseline={([yshift=-.5ex]current bounding box.center)}, scale=0.9]
        \draw[very thick, blue, bend left=30] (-1.8,1.2) to (-.6,0);
        \draw[very thick, blue, bend right=30] (-1.8,-1.2) to (-.6,0);
        \draw[very thick, blue, bend right=30] (2.2,1.2) to (.6,0);
        \draw[very thick, blue, bend left=30] (2.2,-1.2) to (.6,0);
        \draw[very thick, blue] (-.6,0) to (.6,0);
        \fill[blue] (-.6,0) circle (.07);
        \fill[blue] (.6,0) circle (.07);
        \draw[very thick] (-2,1.2) circle (.2 and .5);
        \draw[very thick] (-2,-1.2) circle (.2 and .5);
        \begin{scope}[shift={(2,1.2)}, xscale=.4]
            \draw[very thick] (0,-.5) arc (-90:90:.5);
            \draw[very thick, dashed] (0,.5) arc (90:270:.5);
        \end{scope}
        \begin{scope}[shift={(2,-1.2)}, xscale=.4]
            \draw[very thick] (0,-.5) arc (-90:90:.5);
            \draw[very thick, dashed] (0,.5) arc (90:270:.5);
        \end{scope}
        \begin{scope}[xscale=.4]
            \draw[very thick] (0,-.9) arc (-90:90:.9);
            \draw[very thick, dashed] (0,.9) arc (90:270:.9);
        \end{scope}
        \draw[very thick, out=0, in=180] (-2,1.7) to (0,.9) to (2,1.7);
        \draw[very thick, out=0, in=180] (-2,-1.7) to (0,-.9) to (2,-1.7);
        \draw[very thick, out=0, in=0, looseness=2.5] (-2,-.7) to (-2,.7);
        \draw[very thick, out=180, in=180, looseness=2.5] (2,-.7) to (2,.7);
        \node at (-.7,.8) {3};
        \node at (.8,.8) {2};            
        \node at (.8,-.8) {1};
        \node at (-.8,-.8) {4}; 
        \node at (0,.3) {$s$};
        \end{tikzpicture}        
    \ =\sum_s\  \mathbb{F}_{st} \begin{bmatrix}
                3 & 2 \\
                4 & 1
    \end{bmatrix}
    \ 
    \begin{tikzpicture}[baseline={([yshift=-.5ex]current bounding box.center)}, scale=0.9]
        \draw[very thick, blue, bend left=10] (-1.8,1.2) to (0,.4);
        \draw[very thick, blue, bend right=10] (-1.8,-1.2) to (0,-.4);
        \draw[very thick, blue, bend right=10] (2.2,1.2) to (0,.4);
        \draw[very thick, blue, bend left=10] (2.2,-1.2) to (0,-.4);
        \draw[very thick, blue] (0,-.4) to (0,.4);
        \fill[blue] (0,.4) circle (.07);
        \fill[blue] (0,-.4) circle (.07);
        \draw[very thick] (-2,1.2) circle (.2 and .5);
        \draw[very thick] (-2,-1.2) circle (.2 and .5);
        \begin{scope}[shift={(2,1.2)}, xscale=.4]
            \draw[very thick] (0,-.5) arc (-90:90:.5);
            \draw[very thick, dashed] (0,.5) arc (90:270:.5);
        \end{scope}
        \begin{scope}[shift={(2,-1.2)}, xscale=.4]
            \draw[very thick] (0,-.5) arc (-90:90:.5);
            \draw[very thick, dashed] (0,.5) arc (90:270:.5);
        \end{scope}
        \begin{scope}[yscale=.3]
            \draw[very thick] (-.97,0) arc (-180:0:.97);
            \draw[very thick, dashed] (.97,0) arc (0:180:.97);
        \end{scope}
        \draw[very thick, out=0, in=180] (-2,1.7) to (0,.9) to (2,1.7);
        \draw[very thick, out=0, in=180] (-2,-1.7) to (0,-.9) to (2,-1.7);
        \draw[very thick, out=0, in=0, looseness=2.5] (-2,-.7) to (-2,.7);
        \draw[very thick, out=180, in=180, looseness=2.5] (2,-.7) to (2,.7);
        \node at (-.9,.5) {3};
        \node at (.8,.5) {2};            
        \node at (.8,-.5) {1};
        \node at (-.9,-.5) {4}; 
        \node at (.2,0) {$t$};
    \end{tikzpicture}
\end{align}
\item  Modular S-transformation:
\begin{align}
    \begin{tikzpicture}[baseline={([yshift=-.5ex]current bounding box.center)}, scale=0.9]
        \draw[very thick,blue] (0,-1.2) to (0,-.2);
        \fill[blue] (0,-.2) circle (.07);
        \draw[very thick, blue, out=140, in=-60] (0,-.2) to (-1,.8);
        \draw[very thick, blue, out=120, in=180] (-1,.8) to (0,1.8);
        \draw[very thick, blue, out=40, in=240] (0,-.2) to (1,.8);
        \draw[very thick, blue, out=60, in=0] (1,.8) to (0,1.8);
        \node at (-.32,.4) {1};
        \node at (.25,-.5) {0};
        \draw[very thick, out=270, in=90] (1.5,1) to (.5,-1);
        \draw[very thick, out=270, in=90] (-1.5,1) to (-.5,-1);
        \draw[very thick, out=90, in=180] (-1.5,1) to (0,2.2);
        \draw[very thick, out=0,in=90] (0,2.2) to (1.5,1);
        \draw[very thick, bend left=70] (-.7,1) to (.7,1);
        \draw[very thick, bend right=70] (-.8,1.2) to (.8,1.2);
        \begin{scope}[shift={(0,-1)}, yscale=.4]
            \draw[very thick] (-.5,0) arc (-180:0:.5);
            \draw[dashed, very thick] (-.5,0) arc (180:0:.5);
        \end{scope}
        \begin{scope}[shift={(1.1,1)}, yscale=.4]
            \draw[very thick] (-.4,0) arc (-180:0:.4);
            \draw[dashed, very thick] (-.4,0) arc (180:0:.4);
        \end{scope}
    \end{tikzpicture} 
    \ =
    \sum_{2} \mathbb{S}_{12}[0]\ 
    \begin{tikzpicture}[baseline={([yshift=-.5ex]current bounding box.center)}, scale=0.9]
        \draw[very thick, blue] (0,-1.2) to (0,-.2);
        \fill[blue] (0,-.2) circle (.07);
        \draw[very thick, blue, bend left=20] (0,-.2) to (.3,.8);
        \draw[very thick, blue, dashed, bend left=50] (.3,.8) to (.85,-.2);
        \draw[very thick, blue, bend left=20] (.85,-.2) to (0,-.2);
        \node at (-.13,.5) {2};
        \node at (-.25,-.5) {0};
        \draw[very thick, out=270, in=90] (1.5,1) to (.5,-1);
        \draw[very thick, out=270, in=90] (-1.5,1) to (-.5,-1);
        \draw[very thick, out=90, in=180] (-1.5,1) to (0,2.2);
        \draw[very thick, out=0,in=90] (0,2.2) to (1.5,1);
        \draw[very thick, bend left=70] (-.7,1) to (.7,1);
        \draw[very thick, bend right=70] (-.8,1.2) to (.8,1.2);
        \begin{scope}[shift={(0,-1)}, yscale=.4]
            \draw[very thick] (-.5,0) arc (-180:0:.5);
            \draw[dashed, very thick] (-.5,0) arc (180:0:.5);
        \end{scope}
        \draw[very thick] (0,1) circle (1.1 and .8);
    \end{tikzpicture} 
\end{align}
\end{itemize}
$\mathbb{B}$ is easy to determine,
\begin{align}
   \mathbb{B}_{12}^3=\pm \mathrm{e}^{\pi i (\Delta_3-\Delta_1-\Delta_2)}\,.
\end{align}  
\section{Computation of semiclassical Liouville torus partition function}\label{AppE}
In this section we will compute the semiclassical Liouville partition function at large central charge and will show how the the cosmological constant can be treated as an integral regulator while integrating over zero mode. We start with the Liouville theory on torus,
\begin{align}
    \begin{split}
        S=\int_{\mathcal{T}}d^2 z \sqrt{g}\,(\frac{1}{4\pi}g^{ab} \partial_a \phi \partial_b \phi+\frac{1}{4\pi}Q R \phi+  \mu e^{2b\phi})\,.
    \end{split}
\end{align}
where, $Q=b+1/b$ and central charge is, $c=1+6Q^2$.
Now the torus metric is conformally flat (even any metric in 2D is conformally flat) and can be written as,
\begin{align}
ds^2=e^{2\varphi(z,\bar{z})}|dz|^2,\,\textrm{det}{(g)}=e^{4\varphi}
\end{align}
Therefore, the Ricci scalar can be written as,
\begin{align}
    R=-2e\,^{-2\varphi(z,\bar{z})}\hat g^{ab}\,\nabla_a \nabla_b \varphi.
\end{align}
Note that, for our case, the transformed metric is flat, so $\hat g^{ab}=\delta^{ab}$. Hence, it follows that the second term in the action can be written as,
\begin{align}
    Q R\phi\to -2Q e^{-2\varphi(z,\bar{z})}\delta^{ab}\phi\, \partial_a \partial_b\varphi\,.
\end{align}
Therefore, the Liouville action takes the form,
\begin{align}
    \begin{split}
        S&=\int d^2 z  \,e^{2\varphi(z,\bar{z})}\,\left(\frac{1}{4\pi}e^{-2\varphi}\delta^{ab}\hat{\partial}_a\phi\, \hat{\partial}_b \phi-\frac{2Q}{4\pi}e^{-2\varphi}\delta^{ab}\,\phi\,\hat\partial_b\hat{\partial}_a\varphi\, + \mu e^{2b\phi}\right)\,,\\ &
        \to \int d^2 z  \,\,\left(\frac{1}{4\pi}\delta^{ab}\hat{\partial}_a\phi\, \hat{\partial}_b \phi+\frac{2Q}{4\pi}\,\hat\partial_b\varphi\,\hat{\partial}_a\varphi\, +\mu e^{2b\phi+2\varphi}\right)\,.
        \label{3}
    \end{split}
\end{align}
Now, the classical limit corresponds to $b\to 0$. In this limit, one can identify the field $\varphi=b\,\phi$ \cite{Zamolodchikov:2001ah}. The e.o.m for the Liouville field is,
\begin{align}
    \frac{1+2bQ}{4\pi}\Delta\phi=4b\mu e^{4 b \phi},\,\,\textrm{(Note that in classical limit $\phi$ scales as $\varphi/b$)}\,.
\end{align}
In the classical limit, the Liouville field satisfies the zero mode condition (for torus) \cite{Nakayama:2004vk}:
\begin{align}
\Delta \phi=0\implies \tau_2 \partial_{\bar{z}}\partial_z=0
\end{align}
which admits only constant solutions (say $\phi_0$) i.e. 
 Ker($\Delta$)=constant (independent of $z,\bar{z}$) on torus. Now we compute the semiclassical partition function by considering only the contribution from the classical saddle to the Liouville interaction. Decomposing the Liouville field as $\phi=\phi_0+\delta\phi$,
 \begin{align}
     \begin{split}
         Z&=\int d\phi_0\,e^{-\mu \,e^{4b \phi_0}\int d^2z}\int \mathcal{D}\delta\phi \,e^{-\int d^2z\,\left(\frac{1+2bQ}{4\pi}\,\delta^{ab}\partial_a\delta\phi\,\partial_b\delta\phi\right)+\cdots}\,,\\ &
         \to -\frac{\log(\mu)}{4b\int d^2z}\,Z_{boson}(\tau)\,. \label{8}
     \end{split}
 \end{align}
In the absence of $\mu$, the zero mode integral ($\int d\phi_0$) gives a divergent result. However, as shown in \eqref{8}, in the presence of $\mu$ the integral gives a convergent result, and that is how $\mu$ acts as an integral regulator. All in all, ignoring the overall constants, we will get,
\begin{align}
    Z_{L}\sim Z_{\textrm{boson}}(\tau)\,.
\end{align}
Extrapolating the same argument, the super-Liouville partition function at large central charge takes the form:
\begin{align}
    Z_{SL}\sim Z_{\textrm{boson}}\times Z_{\textrm{fermion}}.
\end{align}
\bibliography{ref}
\bibliographystyle{utphysmodb}
\end{document}